\documentclass[aps,pra,superscriptaddress]{revtex4}
\pdfoutput=1
\usepackage[intlimits]{amsmath}
\usepackage{amsfonts}

\usepackage{comment}
\usepackage{subfigure}
\usepackage{enumitem}
\usepackage[usenames]{color}
\usepackage{braket}

\usepackage{graphicx}
\usepackage{epstopdf}

\advance\voffset by -1cm
\advance\hoffset by -0.5cm
\newcommand\be            {\begin{equation}}
\newcommand\ee            {\end{equation}}
\newcommand\bes           {\begin{subequations}}
\newcommand\esu           {\end{subequations}}

\newcommand{\bigx}[1]{\bBigg@{#1}}

\newcommand\mc            {\mathcal}

\newcommand\p            {\partial}

\renewcommand\th         {\theta}

\newcommand{\spz}{\hspace{0.7cm}}

\def\3pt#1#2#3{{\langle{#1}\vert{#2}\vert{#3}\rangle}}

\newcommand\doi[2]        {\href{http://dx.doi.org/#1}{#2}}

\newcommand{\EQ}{\begin{equation}}
\newcommand{\EN}{\end{equation}}
\usepackage{epsfig}
\usepackage{color}

\usepackage{amsmath}
\usepackage{graphicx}
\usepackage{amsfonts}
\usepackage{amssymb}
\usepackage{slashed}
\graphicspath{{Figures/}}

\begin{document}
\bibliographystyle{plainnat}

\title{{\Large {\bf Non Relativistic Limit of Integrable QFT \\ and Lieb-Liniger Models}}}
\author{Alvise Bastianello}
\affiliation{SISSA and INFN, Sezione di Trieste, via Bonomea 265, I-34136, 
Trieste, Italy}
\author{Andrea De Luca} 
\affiliation{LPTMS, CNRS, Univ. Paris-Sud, Universit\'{e} Paris-Saclay, 91405 Orsay, France}
\author{Giuseppe Mussardo}
\affiliation{SISSA and INFN, Sezione di Trieste, via Bonomea 265, I-34136, 
Trieste, Italy}
\affiliation{International Centre for Theoretical Physics (ICTP), 
I-34151, Trieste, Italy}

\begin{abstract}
\noindent
In this paper we study a suitable limit of integrable QFT with the aim to identify continuous non-relativistic integrable models with local interactions. This limit amounts to  
sending to infinity the speed of light $c$ but simultaneously adjusting the coupling constant $g$ of the quantum field theories in such a way to keep finite the energies of the various excitations. The QFT considered here are Toda Field Theories and the $O(N)$ non-linear sigma model. In both cases the resulting non-relativistic integrable models consist only of Lieb-Liniger models, which are fully decoupled for the Toda theories while symmetrically coupled for the $O(N)$ model. These examples provide explicit evidence of the universality and ubiquity of the Lieb-Liniger models and, at the same time, 
suggest that these models may exhaust the list of possible non-relativistic integrable theories of bosonic particles with local interactions.

\vspace{3mm}
\noindent
Pacs numbers: 11.10.St, 11.15.Kc, 11.30.Pb

\end{abstract}
\maketitle

\maketitle

\section{Introduction}

Nowadays there are known many integrable quantum field theories (QFT) which have relativistic invariance as their distinguished feature: the list includes 
purely bosonic models, as the Sinh-Gordon, or more generally Toda models \cite{Zamo1989,AFZ1983,BCDS1990,CM1989,ChM1990,M1992,Mbook,D1997,Oota,Delius}, fermionic systems, as the Gross-Neveu model \cite{ZZ}, as well as supersymmetric versions of all these examples (see, for instance, \cite{Witten-Shankar,Sc1990,A1994,Hollo}). This list increases even further if one also adds to it theories with soliton excitations, as the Sine Gordon model, or various sigma models based on group manifolds \cite{ZZ,Zamsigma,Ogievetsky}. 

Compared to this extraordinary richness of examples concerning relativistic integrable models, the paucity of non-relativistic integrable models (NRIM) with interactions strictly local and galilean invariant is rather dazzling, their list being essentially given by the Lieb-Liniger model!  The fact that the Lieb-Liniger model \cite{LL,yang-yang,yang,suth} is, under certain general conditions discussed below, probably the only non-trivial example of NRIM may not be accidental and the aim of this paper is indeed to argue about the universal ubiquity of this model. Let's also add that the only other known NRIM are, to the best of our knowledge, variations of the same theme, alias systems made of coupled Lieb-Liniger models relative to different species of particles \cite{yang,suth}. Moreover, it is important to stress, the integrability of these coupled Lieb-Liniger models is only realised under special conditions: when the particle species have the same mass and all 
interactions -- both among the particles of the same species or interspecies -- have the same coupling constant. As shown below, these conditions for the integrability of the coupled Lieb-Liniger models can be simply derived by using an elementary application of the Yang-Baxter equations although for a more detailed discussion of the problem one may look at the ref.~\cite{Lamacraft}. 

In this paper we have tried to pinpoint possible new NRIM making use of the richness of the relativistic ones. Namely, we have explored the possibility of identifying a new class of NRIM in terms of the non-relativistic limit (NR) of known integrable QFT. Let's comment more precisely the terms of the problem: given the relativistic invariance of the starting QFT and the absence in their Hamiltonian of higher derivative terms than $(\partial_x \phi)^2$, the Hamiltonian of the corresponding non-relativistic integrable models obtained with our procedure will have the general form 
\begin{equation}
H \,=\,\int \,\left[\left(\sum_{k=1}^r \frac{1}{2 m_k} \partial_x \psi^\dagger_k \partial_x \psi_k
\right)  + V\left(\{\psi^\dagger_k,\psi_l \}\right)\right] \, dx \,\,\,, 
\label{generalHamiltoniannr1}
\end{equation}
where $V(\{\psi^\dagger_k,\psi_l \})$ is a {\em local} potential term, function of the complex bosonic fields $\psi^\dagger_k$ and $\psi_l$ but not of their derivatives  ($m_k$ is the mass of these fields). This implies that the general form of the equations of motion of the non-relativistic models sought in this paper 
will be  
\begin{equation}
i \partial_t \psi^\dagger_k \,=\,\frac{1}{2 m_k} \partial_x^2 \psi_k^\dagger + \frac{\delta V}{\delta \psi_k}\left(\{\psi^\dagger_k,\psi_l \}\right) \,\,\,. 
\label{generaleqm1}
\end{equation} 
These equations of motion, in particular, are invariant under a parity transformation $x \rightarrow - x$ and possess galilean invariance. Hence, a non-relativistic integrable model as KdV, for instance, will not appear in our screening since its equation of motion reads 
\begin{equation}
u_t = 6 u u_x - u_{xxx} \,\,\,,
\label{kdv}
\end{equation}
and therefore does not belong to the family of equations of motion (\ref{generaleqm1}) (in particularly, it is not even invariant under $x \rightarrow - x$). Moreover, 
our analysis will not include either models as the Calogero-Sutherland or Haldane-Shastry \cite{CSHS} (since they are non-local) or XXZ spin chain or alike \cite{XXZ} (simply because these are models inherently defined on a lattice and also because, ironically, their continuous limit is usually given by relativistic invariant theories). 

These considerations help in clarifying that the non-relativistic integrable models we are looking for are those entirely defined by the local potential term $V(\{\psi^\dagger_k,\psi_l \})$. In other words, the question posed in this paper is the following: given a non-relativistic model with general Hamiltonian given by eq.\,(\ref{generalHamiltoniannr1}), what are the possible expressions of the local potential $V(\{\psi^\dagger_k,\psi_l \})$ that ensure the integrability of the theory? 

\vspace{1mm}

Previously the idea to take the NR limit of a known integrable QFT to identify a non-relativistic integrable model has been successfully applied to the Sinh-Gordon model \cite{KMT}, and the NR limit of this model was indeed shown to be the Lieb-Liniger model. This identification, in particular, was the key point that permitted to compute previously unknown correlation functions of the Lieb-Liniger model in terms of Form Factors \cite{FMS1993} and Thermodynamics Bethe Ansatz \cite{ZamTBA} of the Sinh-Gordon model (see refs. \cite{KMT} for a comprehensive analysis of these points). On the basis of this precedent, it has been quite natural to explore further whether it would be possible to define new NRIM using models which are generalisation of the Sinh-Gordon model, namely the affine simply-laced Toda field theories 
\cite{Zamo1989,AFZ1983,BCDS1990,CM1989,ChM1990,M1992,Mbook,D1997,Oota,Delius}: these theories are build up in terms of the simple roots of the ADE Lie algebras, and the Sinh-Gordon model (corresponding to the algebra $A_1$) is indeed their simplest representative. 

Apart from being a generalisation of the Sinh-Gordon model, there are many other good reasons for considering the affine Toda field theories: (i) these theories have a remarkable pattern of mass spectrum and bound states; (ii) their elastic $S$-matrix is relatively simple, being made of pure phases; (iii) moreover, from a more general point of view concerning the classification of integrable quantum field theories, one may also argue in favour of them as the only integrable interacting relativistic models for a given number $r$ of bosonic fields. For all these pleasant properties presented by the Toda field theories, one could have expected a-priori that their non-relativistic limit be an attractive way for spotting new classes of models which are simultaneously local, integrable and non-relativistic (and, moreover, with different masses). Unfortunately this is not what happens: indeed, the NR limit of the affine simply-laced Toda QFT gives, as a final result, theories made of a set of {\em decoupled} Lieb-Liniger models (with different masses), their number $r$ being the rank of the corresponding Lie algebra. In other words, the attempt to identify new NRIM  through the route of the Toda field theories proves fruitless but for an interesting reason: the purely transmissive nature of their $S$-matrix amplitudes,  which in non-relativistic theories  acts as a main obstacle for building integrable models other than the decoupled Lieb-Liniger ones. 

Once learnt this lesson, given that the absence of reflection amplitudes in the $S$-matrix is directly related to the distinguishability of the particles, our next attempt to identify new NRIM has been to consider the NR limit of QFT whose particles cannot be distinguished in terms of the eigenvalues of the conserved quantities \cite{Mbook}. These theories must necessarily have particles of the same mass and the simplest example analysed in this paper is given by the $O(N)$ non-linear sigma model. For the presence of the constraint on the $N$ fields, the non-relativistic limit of this model cannot be pursued along the operatorial formalism adopted for the Toda field theories but needs instead a path integral approach. In any case, the final result is morally not much different from the previous ones, in the sense that we end up with the symmetrically coupled Lieb-Liniger models, which are presently the other known NRIM. In other words, Lieb-Liniger models are ubiquitous and it seems difficult to find other examples of NRIM, as we will claim below.  

The paper is organised as follows. In Section \ref{integrability} we argue in favour of the context of the relativistic quantum field theories (versus the non-relativistic ones) as the most encouraging playground for answering questions relative to the integrability or not of a given model. This discussion provides 
the starting point for the rest of the paper. In this Section we will also recall the main argument that points to the Toda field theories as possible exhaustive class of relativistic integrable models involving a set $r$ of bosonic gaussian fields. The main properties of these theories are summarised in Section \ref{TODAFT}.  
In Section \ref{LLSect} we introduce and discuss the Lieb-Liniger model and its generalization in terms of various species. 
In Section \ref{SinhGordon} we study in detail the NR limit of the simplest Toda field theory, alias the Sinh-Gordon model, and prove that 
in this limit the model reduces to a Lieb-Liniger model. This section helps in clarifying that the NR limit of a model not only consists of  
sending the speed of light $c$ to infinity but it is also necessary to fine tune the coupling constant of the theory. This discussion provides then a gentle introduction to the remaining parts of the paper, where we will analyse the NR limit for the richer phenomenology presented by the other Toda theories. In Section \ref{BD} we present the analysis relative to the closest companion of the Sinh-Gordon model, namely the Bullough-Dodd model. The comparison between the two models is an important step in our analysis because, even though both models are made of only one bosonic particle, their relativistic dynamics is completely different: while in the Bullough-Dodd the particle is a bound state of itself, in the Sinh-Gordon model on the contrary there are no bound states. Moreover, while the Sinh-Gordon model is invariant under a ${\mathcal Z}_2$ symmetry, the Bullough-Dodd does not have any internal symmetry at all. Despite all these qualitative differences, surprisingly enough the 
NR limit of both models is exactly the same and consists of a single Lieb-Liniger model. In Section \ref{Todanon} we show that this result applies as well to the most general cases of Toda Field Theories, in the sense that the NR limit of all these theories simply consists of decoupled Lieb-Liniger models with different masses but the same coupling constant. In order to establish the existence or not of other possible NRIM, one necessarily has to consider theories with indistinguishable particles and the simplest example discussed in Section \ref{secsigma} is the $O(N)$ non-linear sigma model. In this case the NR limit consists of coupled Lieb-Liniger models at their integrable point.  Our conclusions are finally gathered in Section \ref{Conclusions}. 
The paper also contains few appendices: in Appendix \ref{DynkinToda} we collect some useful data of the Toda Field Theories; in Appendix \ref{AppendixA}
we present the simple argument that leads to the condition of integrability for coupled Lieb-Liniger models; in Appendix \ref{A}  and Appendix \ref{gammaprop}
 we collect some technical steps in the derivation of the non-relativistic limit of the Bullough-Dodd and $O(N)$ models respectively.  

\section{Integrability in relativistic field theories} \label{integrability}

The quest of defining integrability has a long story, starting of course from the familiar definition given by Liouville in classical mechanics, where he stated that 
a system made of $N$ degrees of freedom is integrable if it possesses $N$ independent first integrals of motion in involution. Despite this transparent 
definition of integrable systems in classical mechanics, to define what is an integrable system in many-body quantum mechanics -- and possibly reaching  
a classification of all such models -- turns out to be a non trivial problem (see for instance \cite{M1992,Sutherland,CauxMossel,BootKM}). Here we are concerned with quantum many-body systems only in $(1+1)$ dimensions. Without entering into a long and detailed examination of all pros and cons of possible alternative definitions of integrability for such quantum systems (see ref.\cite{CauxMossel}), in the following we adopt as a guiding line the simplest and the most basic one, 
alias the request that a quantum integrable system must have non-trivial set of charges which commute with its Hamiltonian (this set must be infinite if we are dealing with many body systems). In particular, we demand each charge to be expressible as spatial integral of an appropriate local density.

Although this definition essentially embodies the physical concept of integrability in the quantum realm, it has a practical weakness: given an Hamiltonian, no general procedure is known to determine whether an ensemble of conserved charges effectively exists (and in such a case how to construct it). 
Even for all known integrable models, such a task usually requires a certain amount of ingenuity. Therefore, it is in general a hard question to decide 
whether a given model is integrable or not. It is useful therefore to specialize more the problem to the case of QFT describing interacting particles. In this case, in the presence of an infinite set of local conserved charges ${\cal Q}_n$ (where the index $n$ can be put in correspondence, for instance, with the higher derivative term $\partial_x^n {\cal O}$ present in the expression of the associate density current of this charge, with ${\cal O}$ a local field of the theory), a QFT has an $S$-matrix that is completely elastic and factorizable and all scattering processes consists of two-body elastic amplitudes \cite{ZZ}.

\vspace{2mm}
\noindent
{\bf Integrability in quantum many-body systems: QFT vs NR models}. 
This conclusion has an immediate consequence for relativistic QFT: if the scattering has to be elastic, all production and decay processes must be forbidden as a consequence of the peculiar values of the parameters of the model. Viceversa, if for a given model we are able to show the existence of production processes, this fact alone automatically provides an explicit proof that such a model is {\em not} integrable.  As we will see, this criterion has the advantage of being 
checkable for any process involving a finite number of particles in a {\em finite} number of steps, as originally shown in a seminal paper by P. Dorey \cite{D1997}. We will see explicitly how this algorithmic procedure can be put in action in few simple examples, leading to the construction of explicit integrable Hamiltonians. Moreover, the factorization of the scattering leads to the Yang-Baxter equation, a consistency equation
coming from three-body processes, that the scattering matrix has to satisfy. The combination of this constraint with the demand of relativistic invariance gives rise to a powerful machinery: the scattering matrix is forced to a well-defined functional form and the spectrum of masses can be iteratively constructed employing the bootstrap approach. 

On the other hand, the number of particles is always conserved in a NR theory. Therefore, the absence of particle production cannot be used as a distinctive feature
of integrability, but  integrability could be in principle checked perturbatively by mean of the factorization of the scattering matrix. Nevertheless, passing from relativistic to non relativistic models, some important tools that are fundamental in the relativistic construction (for instance the crossing symmetry)
are lost and so is the bootstrap program \cite{smirnov}. This kind of technical difficulties are only partially overcome with the use of the quantum inverse scattering method and the algebraic Bethe Ansatz, which nonetheless are inherently built to deal with lattice models.

In light of this technical robustness, the approach of this paper is to start from integrable QFT and then take their non-relativistic limit in order to identify integrable non-relativistic models.

\vspace{2mm}
\noindent
{\bf Classes of QFT considered in this paper}. 
In this paper we focus the attention on two classes of QFT: purely Lagrangian models and the $O(N)$ non-linear sigma model. While the $O(N)$ non-linear sigma model will be discussed in detail in the second part of the paper, let's concentrate here after our attention on the Lagrangian models, which seem well suited to 
present in simple terms some important issues concerning integrability. The Lagrangian theories we have in mind can be characterised in terms of the approach stated by 't Hooft and Veltman in their famous preprint {\em Diagrammar} \cite{Diagrammar}. Namely, the QFT considered here are only those which can be uniquely defined in terms of the set of their Feynman diagrams, made of standard propagators and vertices relative to a given number $r$ of bosonic fields $\varphi_a(x)$. In other words, in the first part of this paper we will exclude from our considerations models, as for instance the Sine-Gordon,  where the non-perturbative effects are the predominant ones, not because these models are non interesting (quite the contrary!) but simply because, to make any progress on a subject that is already subtle, we prefer first to focus the attention on a class of QFT where any computation can be done -- in principle -- on the basis of the explicit set of rules that defines them. For these theories, which are nevertheless non-trivial, quoting \cite{Diagrammar}, "{\em diagrams form the basis from which everything must be derived. They define the operational rules, and tell us when to worry about Schwinger terms, subtractions, and whatever other mythological objects need to be introduced} ". The Feynman diagrams are those succinctly given by a local Lagrangian density which therefore contains all and only the fields associated to the particle excitations of the model. Moreover, we will assume that a judicial choice of masses and coupling constants can make everything finite. 

In the second part of the paper, we will consider instead the simplest QFT with a constraint, i.e. the $O(N)$ non-linear sigma model. 

\begin{figure}[t]
\begin{center}
\includegraphics[scale=0.55]{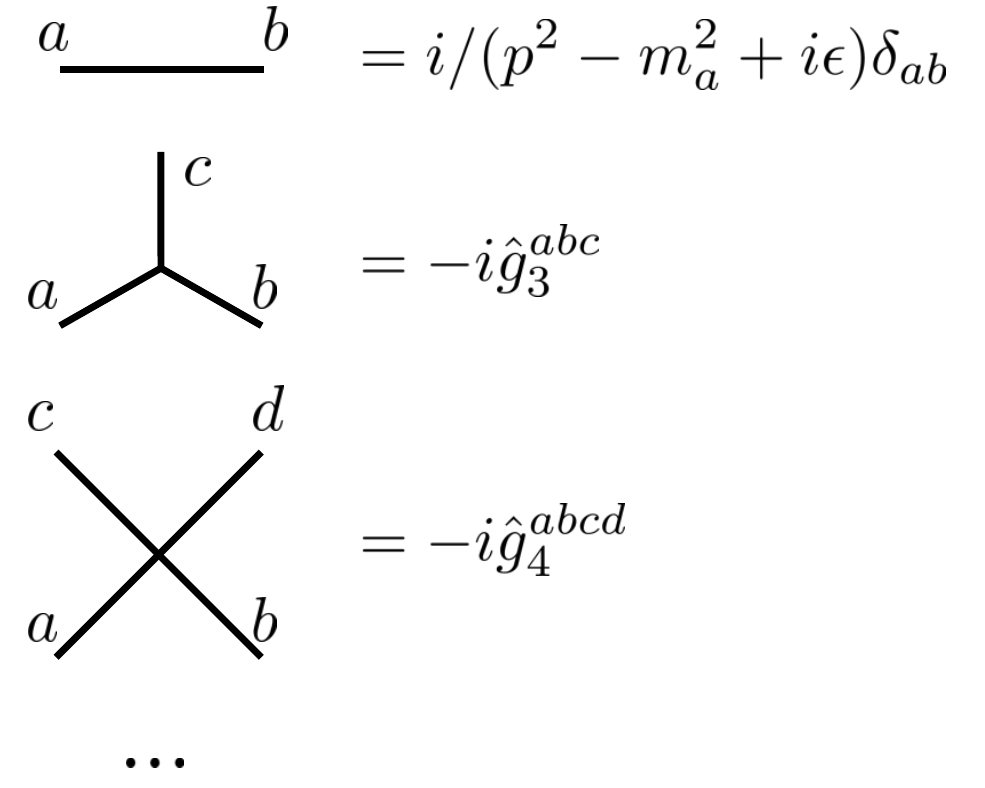}
\caption{\emph{Propagator and interactions vertexes associated with the Lagrangian (\ref{LagrangianGeneral}).}}\label{propvert}
\end{center}
\end{figure}

\vspace{2mm}
\noindent
{\bf Lagrangian densities}. 
The Lagrangian densities we will initially consider in this paper involve $r$ bosonic fields with bare masses $m_k$ and coupling constants $g_n$ associated to the $n$-th polynomial interactions   
\EQ
{\mathcal L} \,=\,\sum_{k=1}^r \frac{1}{2} \left[(\partial_{ \mu}\phi_k)^2 - m_k^2 \phi_k^2\right] - \frac{\hat g_3^{abc}}{3!} \phi_a\phi_b\phi_c - 
\frac{\hat g_4^{abcd}}{4!} \phi_a\phi_b\phi_c\phi_d - 
\frac{\hat g_5^{abcde}}{5!} \phi_a\phi_b\phi_c\phi_d\phi_e  - \cdots 
\label{LagrangianGeneral}
\EN 
Notice that, for dimensional analysis, all coupling constants $g_n$ have dimension $(mass)^2$ and therefore, without losing generality, can be chosen 
to be all proportional to an overall mass scale $m^2$. We also use the same scale to express the masses, hence 
\begin{equation}
m_k \,=\, m\, \mu_k
\,\,\,\,\,\,\,\,
,
\,\,\,\,\,\,\,\,
\hat g_n \,=\, m^2 \,  g_n \,\,\,.
\end{equation}
Clearly the number $r$ of particles may not be sufficient to uniquely identify a theory and, to do so, sometimes we also need to specify the symmetries of the Lagrangian (this becomes clear from the examples discussed below). The Feynman rules associated to the Lagrangian (\ref{LagrangianGeneral}) are given in Figure \ref{propvert}.

\vspace{2mm}
\noindent
{\bf Conditions for integrability}. To select which models are integrable in the class of theories (\ref{LagrangianGeneral}), we can proceed by implementing two conditions: 
\begin{enumerate}
\item to find, if they exist, special values of the masses $\mu_k$ and the couplings $g_n$ such that there are no production processes at the tree level:  
namely all processes in which $n$ particles turns into $m$ particles (with $ n \neq m$) have vanishing amplitude.  
\item to check that such values (or better, some particular combination thereof) are stable under renormalization (both finite and infinite) of the masses and the couplings, $m_i \rightarrow \tilde{m}_i$ and $g_n \rightarrow \tilde g_n$. It is worth to remind that in the $(1+1)$ dimensional theories given by the Feynman rules above, the only divergencies come from the tadpole diagrams that can be cured by an appropriate normal order prescription of the various fields, while all other 
diagrams give rise instead to a finite renormalization of the masses and the couplings. Although the infinite renormalization poses a problem too, 
typically it is this finite renormalization of the masses and the couplings that may be particularly dangerous, because may spoil their fine-tuned balance 
which is at the basis of the integrability of the model at the tree level. Note that the tree level is already enough to grant integrability at the classical level \cite{AM} so, what this second step of the procedure is doing, is to check whether the quantum fluctuations and the associated renormalization flows  spoil or not the integrable structure. 
\end{enumerate}

\begin{figure}[b]
\begin{center}
\includegraphics[scale=0.4]{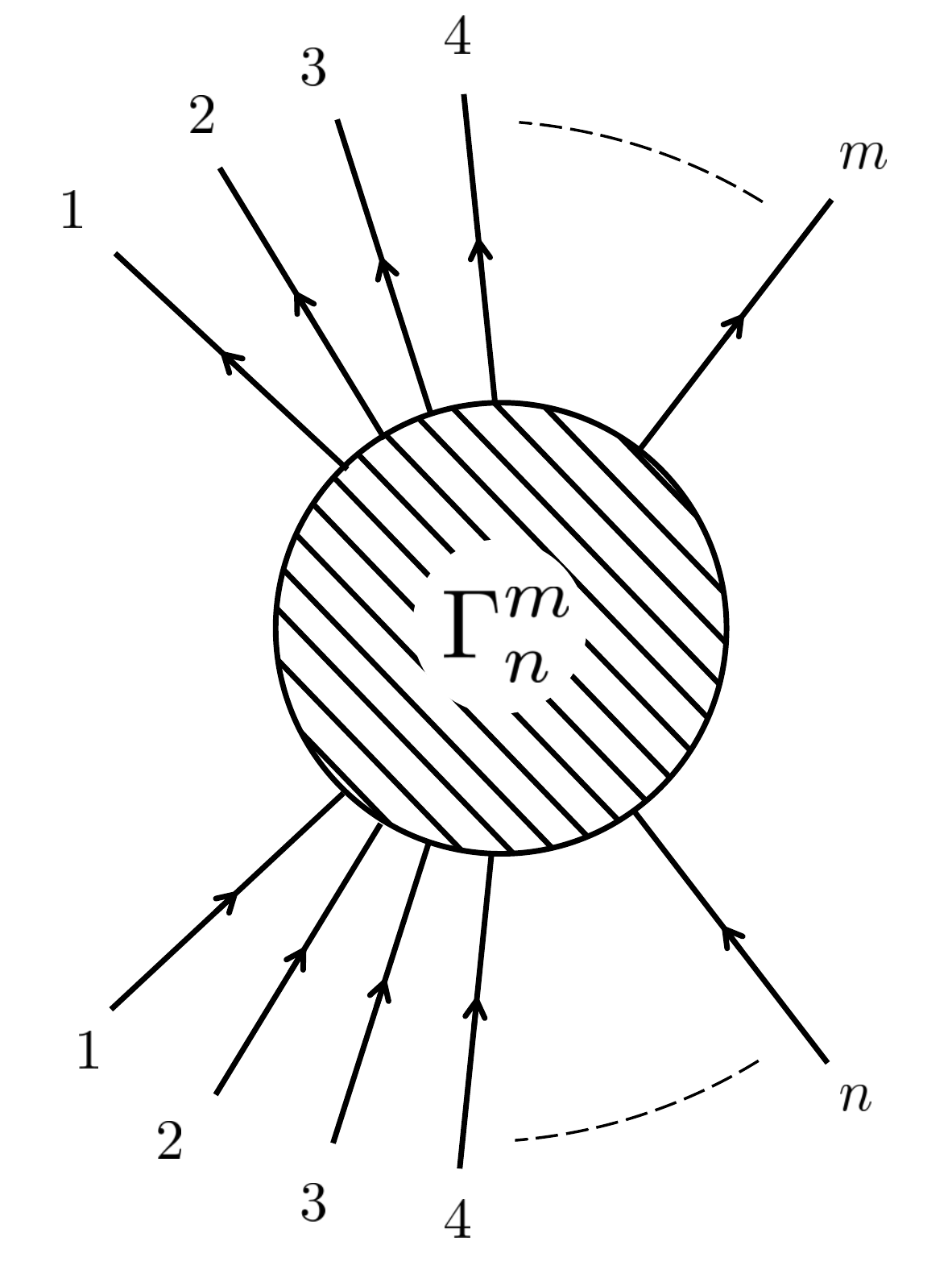}
\caption{{\em Vertex function $\Gamma_n^m$ relevant for the production process $n \rightarrow m$.}}
\label{productiongraph}
\end{center}
\end{figure}

The first condition consists in imposing that {\em all} the vertex functions $\Gamma_{n}^{m}(p_{a_1},\ldots,p_{a_n} | q_{a_1},\ldots,q_{a_m})$ shown in 
Figure \ref{productiongraph} (with the number of particles $n$ of the {\em in}-state different from the number of particles $m$ of the {\em out}-state), 
once computed at the tree level and on-shell of all particles, must identically vanish
\begin{equation}
\Gamma_n^m|_{on-shell} \,=\, 0 \,\,\, 
\,\,\,
.
\label{vertexproduction}
\end{equation}
The on-shell conditions means that the momenta of both in-coming and out-going particles satisfy the on-shell equations $((p_{a_i})^\mu (p_{a_i})_\mu= m_{a_i}^2;  (q_{a_k})^\mu(q_{a_k})_\mu = m_{a_k}^2)$ and 
moreover that the overall momenta involved in the process must satisfy the conservation law $\sum_i p_{a_i} =\sum_k q_{a_k}$.  

The vertex functions $\Gamma_n^m$ are given by the sum of all Feynman diagrams entering these amplitudes, and these diagrams can be ordered, say, in terms of their perturbative degree. 
At the tree level, these vertex functions consist of Feynman diagrams without internal loops. 
Given that the set of masses is finite, it is pretty obvious that it is impossible to identically satisfy the {\em infinite} number of conditions (\ref{vertexproduction}) with only a {\em finite} number of coupling constants $g_n$. This suggests that the Lagrangian densities (\ref{LagrangianGeneral}) for models which are candidates to be integrable must necessarily be expressed by an infinite series of interaction terms rather than finite polynomials in the various fields $\phi_k$. 

\begin{figure}
\begin{center}
\includegraphics[scale=0.5]{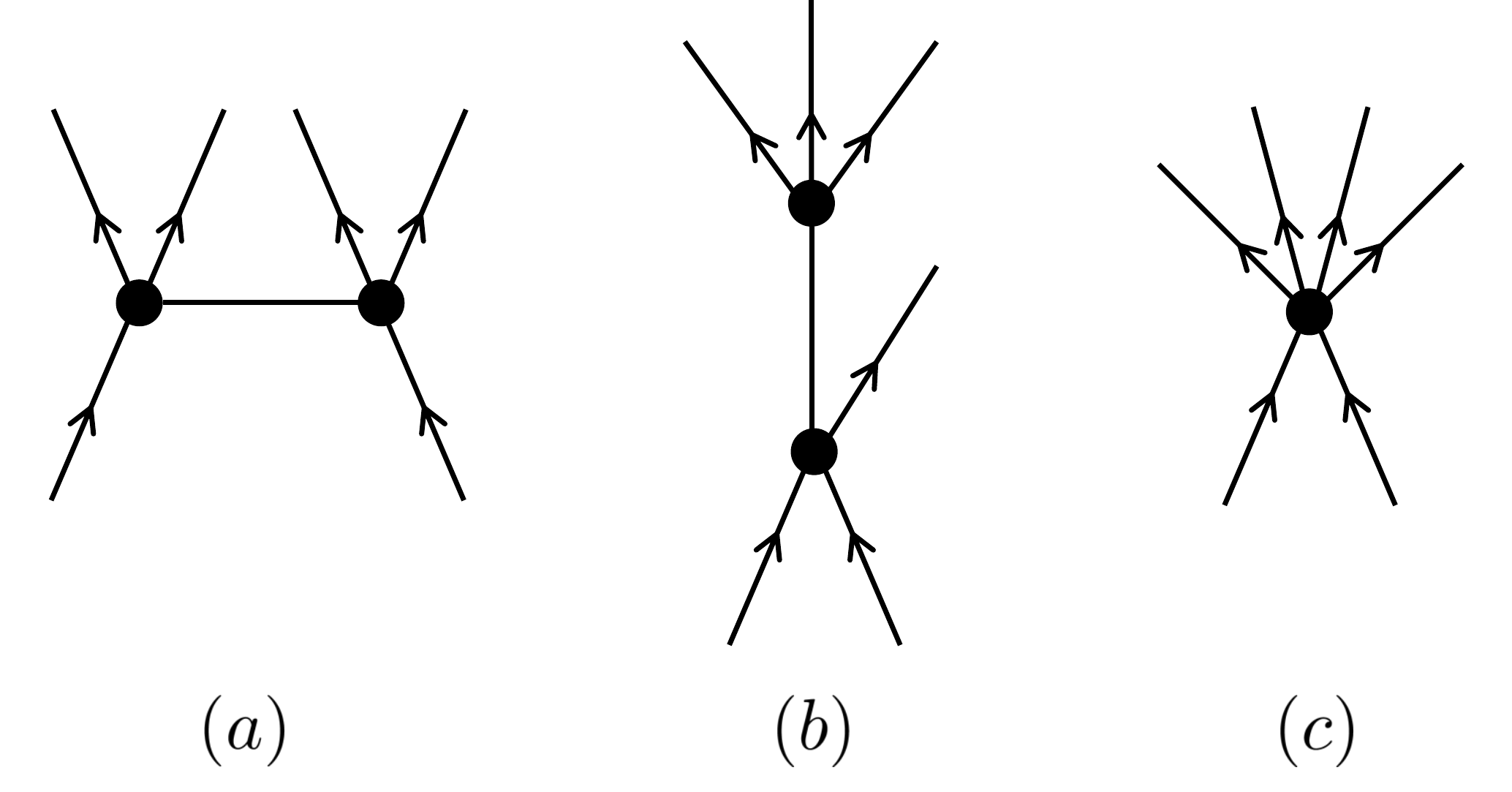}
\caption{{\em Feynman graphs at the tree level for the production process $2 \rightarrow 4$.}}
\label{produzione}
\end{center}
\end{figure}

\vspace{2mm}
\noindent
{\bf Simplest $Z_2$ integrable model: Sinh-Gordon}. 
This expectation is explicitly confirmed by the analysis of the simplest theory of the class (\ref{LagrangianGeneral}): this consists of a $Z_2$ symmetric theory built on only one field $\phi(x)$ and therefore the general expression of its Lagrangian is given in this case by  
\begin{equation} 
{\cal L} \,=\,\frac{1}{2} \left[(\partial_{\mu}\phi)^2 - m^2 \phi^2\right] - \frac{\hat g_4}{4!} \phi^4 - \frac{\hat g_6}{6!} \phi^6 - \cdots 
\label{putativeShG}
\end{equation}
This kind of theory has been originally analysed by Dorey \cite{D1997} and, closely following his analysis, let's consider the diagrams that, at the tree level, 
contribute to the production process $2 \rightarrow 4$. It is sufficient to consider the case where the initial particles have just the energy to create the four out-coming particles: for the momenta $(p^{(0)},p^{(1)})$ of the on-shell initial particles in the center of mass reference frame, we have $(2 m,\pm \sqrt{3} m)$. The total energy is then $E_t = 4 m$ and therefore the four final particles are all at rest, their common value of the momenta being $(m,0)$. Hence, the value of each graphs of Figure \ref{produzione} is given by  

\begin{eqnarray*}
(a) & \rightarrow & \,\,i\,m^2\, \frac{g_4^2}{32} \,\,\,\nonumber \\
(b) & \rightarrow & -i\,\,m^2\, \frac{g_4^2}{96 } \,\,\,\\
(c) & \rightarrow & -i \,m^2 \, \frac{g_6}{48 } \,\,\,.\nonumber 
\end{eqnarray*}
The total sum of these graphs is then  
\EQ
(a) + (b) + (c) \,=\,\frac{i}{48 }m^2 \,(g_4^2 - g_6) \,\,\,. 
\label{totsum3}
\EN 
Notice that this quantity has physical dimension $m^2$, a result that below we prove to hold in general for any tree level diagram.   
Eq.\,(\ref{totsum3}) shows that, in order to suppress this production process, we need to have necessarily a $\phi^6$ vertex in the Lagrangian (\ref{putativeShG}) and moreover to set the value of its coupling constant to 
\begin{equation}
g_6 \,=\, g_4^2\,\,\,.
\label{g_6}
\end{equation} 
In the absence of the $\phi^6$ vertex, the Feynman graphs built only on $\Phi^4$ Landau-Ginzburg theory would have given a non-zero value of the production amplitude $2 \rightarrow 4$, already at the tree level. With the value (\ref{g_6}) of the $\phi^6$ vertex, one can proceed further and compute at the tree-level the amplitude $\Gamma_2^6$ for the production process $2 \rightarrow 6$: it turns out that this amplitude is a constant and therefore it can be cancelled only if it exists a $\phi^8$ vertex in (\ref{putativeShG}), with the relative coupling constant satisfying the condition $g_8 = g_4^3$. Generalising this analysis to the higher 
on-shell production amplitudes $2 \rightarrow n$, one finds that the conditions that ensure their cancellation requires the existence of arbitrarily higher couplings whose values are fixed by the equation 
\begin{equation}
g_{2 n} \,= \, (g_4)^{n-1} \,\,\,. 
\end{equation}
Notice that choosing $g_4 > 0$, all the other infinitely many coupling constants are positive as well, and rescaling all of them by the same factor, the series can be resummed, with the final result   
\begin{equation}
{\cal L} \,=\, \frac{1}{2} (\partial_{\mu}\phi)^2 - \frac{m^2}{g^2} \left(\cosh(g \phi) -1\right) \,\,\,. 
\label{SinhGordonLagrangian}
\end{equation}
This is the Sinh-Gordon model, which is also the simplest example of the Toda Field Theories \footnote{Choosing instead $g_4 < 0$, the higher couplings have an alternate sign: in this case the final Lagrangian is given by the Sine-Gordon model, which however we decide not to consider here since this model has 
an hidden part of its spectrum related to its solitons.}. 

\begin{figure}
\begin{center}
\includegraphics[scale=0.3]{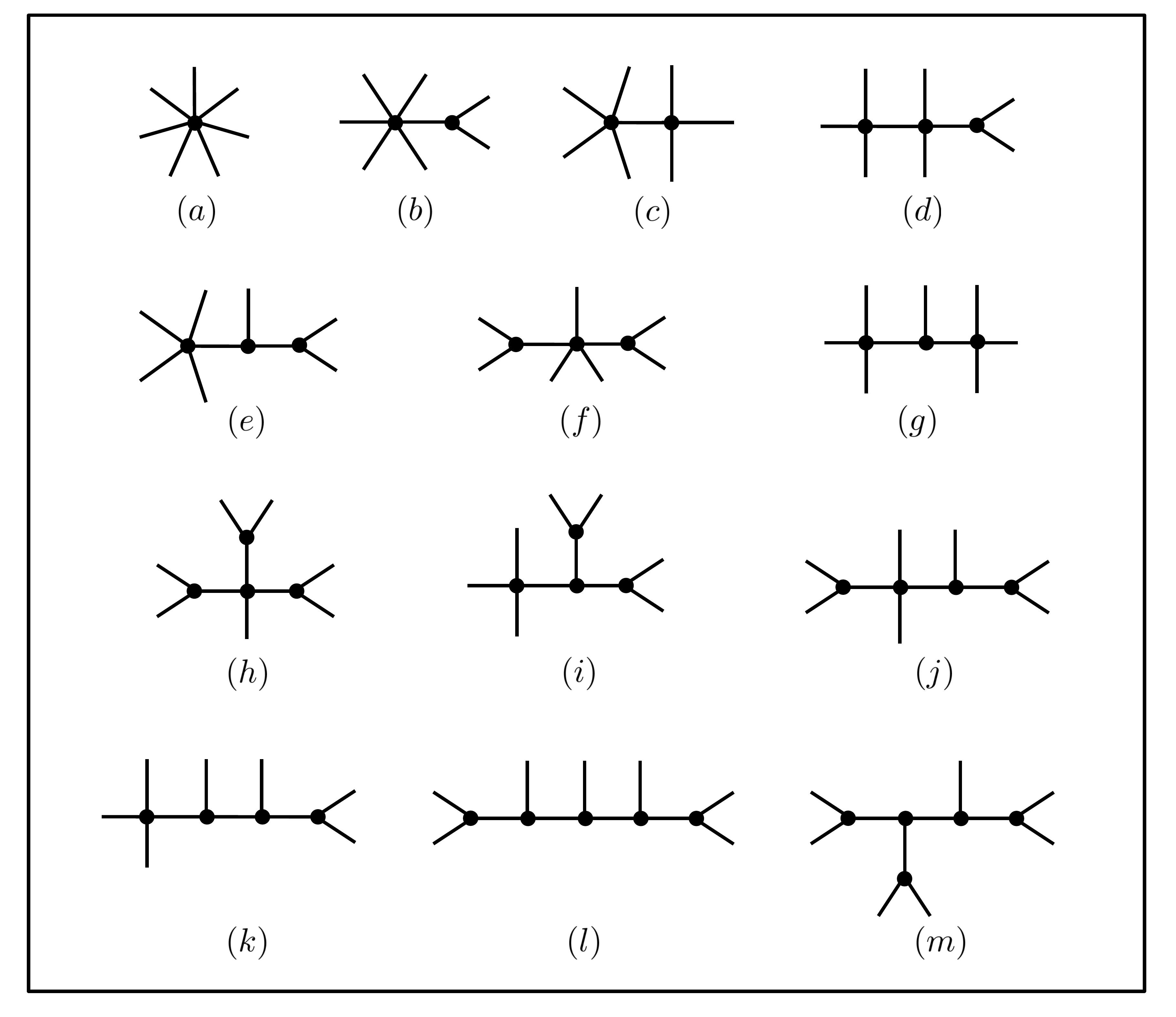}
\caption{{\em Feynman graphs at the tree level entering the amplitudes $\Gamma_n^m$, with $n+m = 7$.}}
\label{manyfeynmans}
\end{center}
\end{figure}

\begin{figure}[t]
\begin{center}
\includegraphics[scale=0.4]{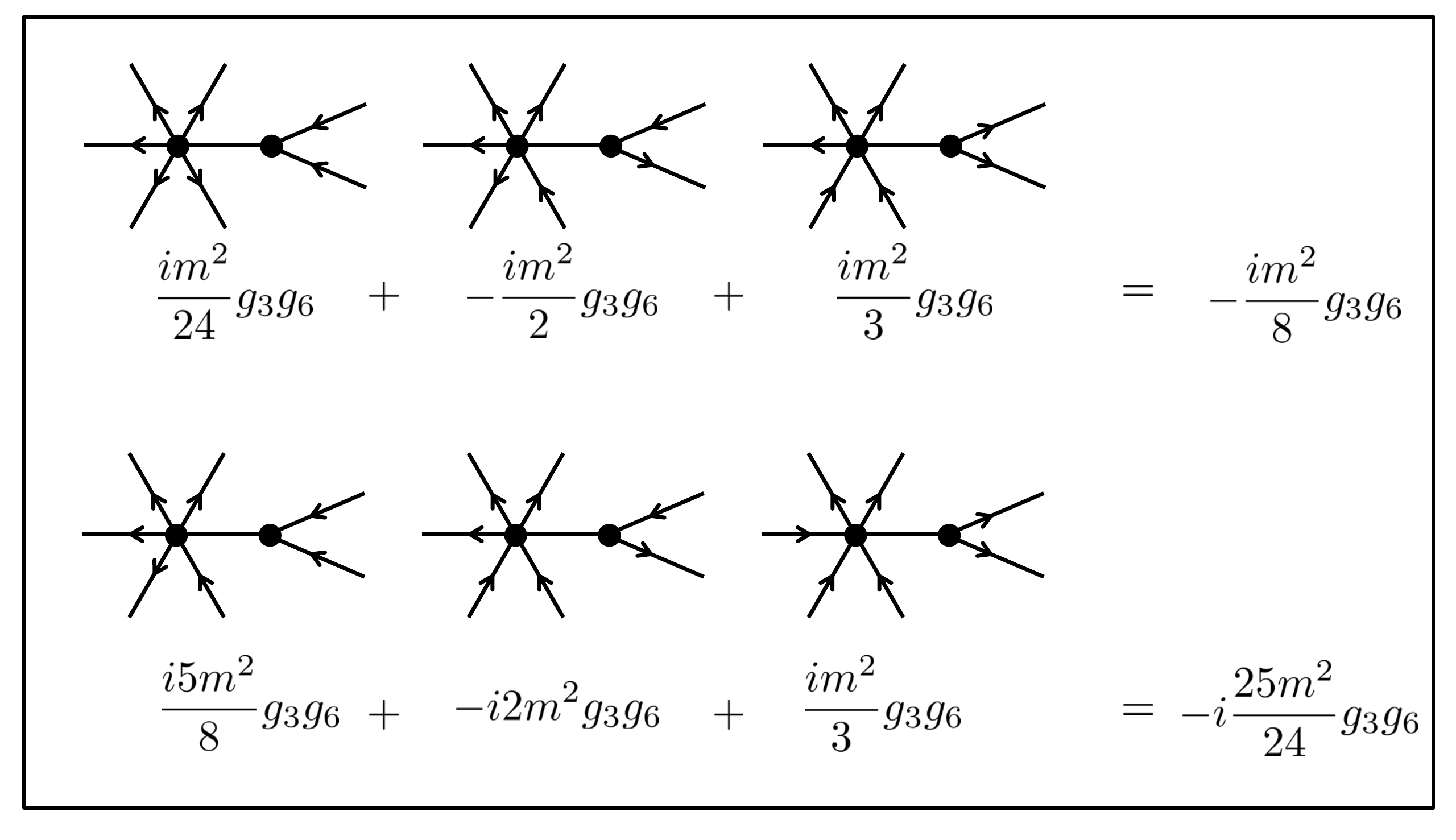}
\caption{{\em Feynman graphs relative the tree level graph (b) of the Figure \ref{manyfeynmans} evaluated on-shell for the process $2 \rightarrow 5$ and 
$3 \rightarrow 4$, where in the latter case the momentum of one of the initial particles was set to zero.}}
\label{graph25graph34}
\end{center}
\end{figure}

\vspace{2mm}
\noindent
{\bf Other integrable models}. 
Repeating the same analysis for a theory made also of only one field but not invariant under a $Z_2$ symmetry, one finds that in order to 
have vanishing production processes it is necessary to have an infinite number of coupling constants and, moreover, their values 
must be properly tuned. For instance, the sum of the diagrams shown in Figure \ref{manyfeynmans} vanishes either if we consider these 
graphs as entering the amplitude $\Gamma_2^5$ or the amplitude $\Gamma_3^4$: this means that, for each graph, we can take for instance 2 external 
legs as in-coming particles (and 5 external legs as out-going particles) or 3 external legs as in-coming particles (and 4 as out-going particles).  
As discussed in more detail below, these two ways of putting differently on-shell the external legs of the Feynman diagrams produces 
two different sums (functions of the same couplings): take for instance the diagram (b) of Figure \ref{manyfeynmans} and evaluate 
explicitly the various terms coming from it in the case of $2 \rightarrow 5$ or in the case $3 \rightarrow 4$.  As seen in Figure \ref{graph25graph34}, given that the overall coupling dependence of these graph is $m^2g_3 \, g_6$, the coefficient of each diagram is however different, according to the process considered. The different sums originating from the different ways of putting on-shell the external particles have to be simultaneously zero if the couplings are properly tuned.  The resulting Lagrangian in this case can be written  
\begin{equation}
{\cal L} \,=\, \frac{1}{2} (\partial_{\mu}\phi)^2 -\frac{m^2}{6 g^2} \left(e^{2 g \phi} + 2 \, e^{-g \phi} -3 \right) \,\,\,,
\label{BDLagrangian}
\end{equation}
which is known as the Bullough-Dodd model and it also belongs to the set of Toda Field Theories. 

In a {\em tour de force} computation, Khastgir has extended the previous analysis to the class of QFT made of two fields, $\phi_1$ and $\phi_2$ \cite{Khastgir}. To carry on the classification of possible integrable Lagrangian made of these two fields, it was useful to distinguish a-priori several cases, for instance if the fields are degenerate in mass and conjugated each other, or self-conjugate fields with different masses. In all cases analysed by Khastgir \cite{Khastgir}, as before to ensure 
the absence of no production processes it was necessary to have infinitely many and fine tuned couplings. Moreover, there were strong indications that the various Lagrangians identified in this way fall into the class of Toda Field Theories, in particular those based on the algebras $a^{(1)}_2, a^{(2)}_4, c^{(1)}_2, d^{(2)}_3, d^{(3)}_4$ and $g^{(1)}_2$ of rank equal to 2 (the notation will be clear in the next Section). 

\subsection{Comments}
Let's pause to comment on what we have presented so far and to draw, for the moment, some important indications: 

\vspace{2mm}
\noindent
{\bf Non-integrability of the LG  Lagrangians}. The simple request to have no production processes at the tree level has led us to an important result, namely to exclude from the class of integrable Lagrangians those with purely polynomial interactions, alias Landau-Ginzburg (LG) theories.  For instance, the $\Phi^4$ Landau-Ginzburg theory is excluded from the list of integrable models just by the tree level non-zero amplitude for the process $2 \rightarrow 4$, similarly the $\Phi^6$ theory is also excluded to be an integrable theory by the non-zero amplitude of the process $2 \rightarrow 6$, etc. Hence, for any QFT with finite number of 
polynomial interaction it is relatively easy to identify the non-zero production processes which tag them as non-integrable models. 

\vspace{2mm}
\noindent
{\bf The integrable sieve}. Chosen a class of models -- specified by the number of fields present in the Lagrangian and its symmetry -- the systematical analysis of all production processes acts,  as a matter of fact, as a {\em sieve}, in the sense that at the end of this procedure we are left out with very special Lagrangians as possible candidates of integrable QFT. This has been the case, for instance, for the Lagrangians given in eqs.\,(\ref{SinhGordonLagrangian}) and (\ref{BDLagrangian}), which are the only two theories left after considering, as a sieve, the production channels $2 \rightarrow n$. As discussed in more detail below, even in presence of a larger number of particle species, the basic production channels of the form $2 \rightarrow n$ can be used to fix relations between the couplings since different channels, e.g. $m \rightarrow n$ with $m > 2$, can be obtained from the basic one using the crossing relation. However, in principle these other channels will give other constraints for the couplings. Consistency of these equations requires a peculiar fine-tuning between the particle masses.

At the end of the day, those Lagrangian which pass all these screenings are integrable -- by construction --  at the classical level, and therefore what remains to be done is to see whether they are also integrable at the quantum level.  A sufficient condition to establish the quantum integrability of the Lagrangians selected by the sieve is that they are stable under the renormalization procedure. This means that all relations found at the tree-level must remain also valid by keeping into account the finite and infinite order corrections of higher order loops. To better appreciate this point, it is worth to reflect how astonishingly well-tuned is the set of values of the masses and couplings  -- from a probabilistic point of view --  which ensures at the tree level the absence of any production process. This property emerges from the considerations that follows. 

\begin{figure}
\begin{center}
\includegraphics[scale=0.4]{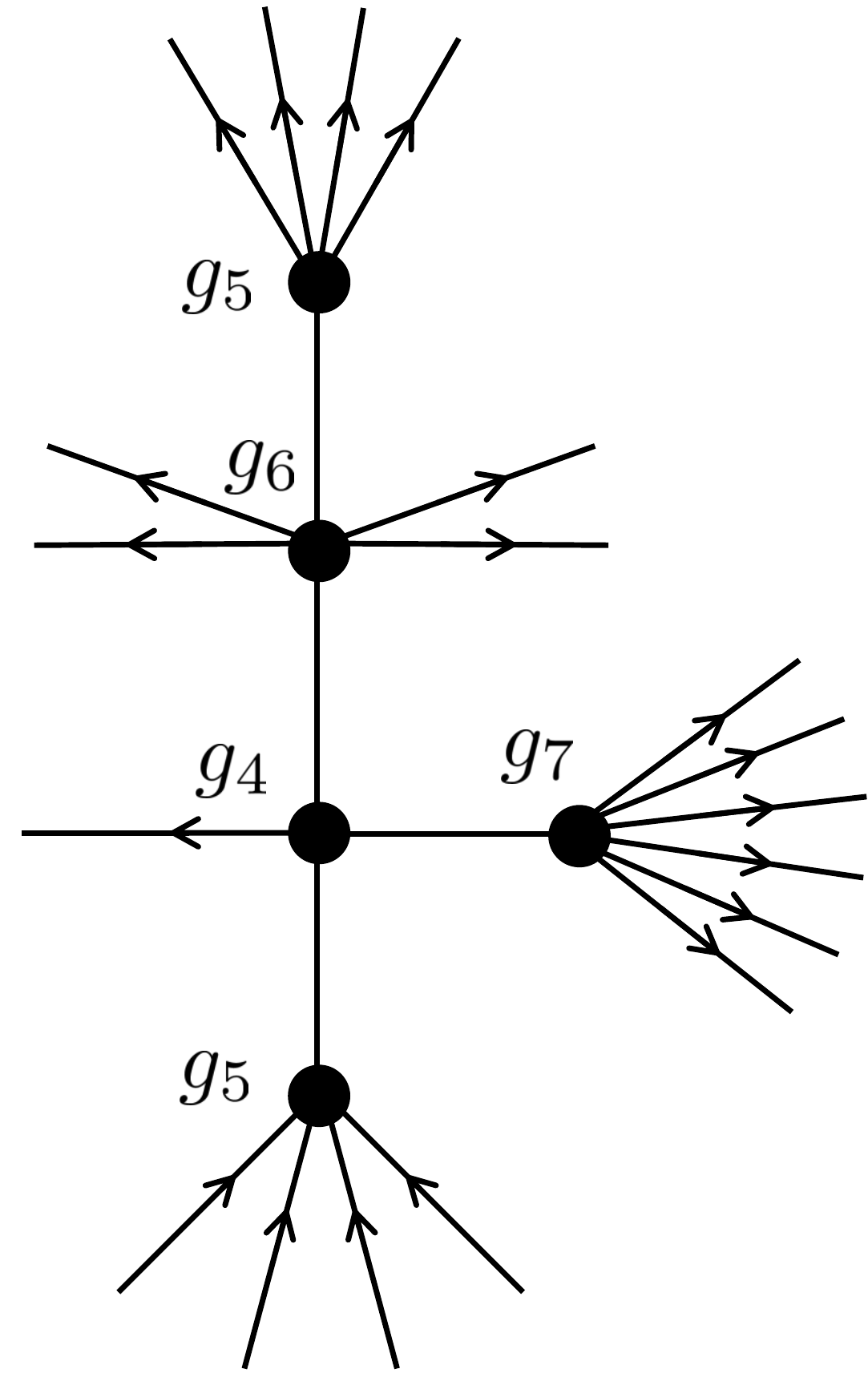}
\caption{{\em One of the tree level diagrams entering the vertex function $\Gamma_4^{15}$ relevant for the production process $4 \rightarrow 15$.}}
\label{prod4_15graph}
\end{center}
\end{figure}

\vspace{2mm}
\noindent \ \\
{\bf Properties of the tree level Feynman diagrams}.
At the tree level, a generic diagram that contributes to the amplitude $n \rightarrow m$ is given in Figure \ref{prod4_15graph} (where, for simplicity, we have 
avoided to put extra indices on the lines to distinguish the various particles). For all tree level diagrams entering the scattering processes, there are a series of relations that link together the number $E = n+m $ of external lines (given by the sum of {\em in} and {\em out} particles), the number $I$ of propagators and the numbers $d_k$ of vertices with $k$ legs: these relations are simply given by 
\begin{eqnarray}
E &\,=\, & \sum_k d_k (k - 2) + 2 \,\,\,\,
\label{diofanto1}\\
I & \,=\,& \sum_k d_k -1 \label{diofanto2}\,\,\,.
\end{eqnarray}
For instance, for the graph of Figure \ref{prod4_15graph} we have $E = 19$, $d_4 = d_6 =d_7 =1, d_5 =2$, $I = 4$, and one can easily check that the relations given above are indeed verified.  

It is now easy to see that all tree level diagrams have dimension $[m]^2$, in terms of the mass scale $m$:  
given that at each vertex the momentum is conserved and the external particles are on-shell (with the total energy of the $n$ particles of the {\em in-state} set to be at the threshold energy $E_{th} = \sum_{k=1}^m m_k$ for the production of the $m$ particles of the {\em out-state}), in light of eqs.\,(\ref{diofanto1}) and (\ref{diofanto2}) the general expression $T$ of any tree-level diagram is given by   
\begin{equation}
T \,=\, m^2 \, \frac{{\cal N}(\{ g_n\})}{{\cal D}(\tilde p_i,\{\mu_k\})} \,\,\,,
\end{equation}
where the numerator ${\cal N}$ is simply the product of all the dimensionless coupling constants entering the graph
$${\cal N} \,=\, \prod_k g_k^{d_k}$$
while the denominator ${\cal D}$ is the product of the propagators of the internal lines 
$$ 
{\cal D} \,=\,\prod_i \frac{1}{\tilde p_i^2 - \hat\mu^2_i}
$$
but evaluated at the values of the internal momenta $\tilde p_i^2$ fixed by the conservation law at each vertex. Notice that for the processes $2 \rightarrow n$, 
with all external particles put on-shell, the denominator ultimately depends only on the masses of all the particles involved in the graph, both those entering the propagators and the external lines. 

\vspace{2mm}
\noindent
{\bf Integrable models are statistically extremely rare}. 
It is important to notice that the number of tree level Feynman diagrams of theories with arbitrarily high power interactions grows {\em exponentially} 
with the number of external lines: first of all, fixed $E$, the number of solutions for $d_k$ of the diophantine equation  (\ref{diofanto1}) grows as the number of ways $P(E)$ of writing the integer $(E-2)$ as a sum of positive integers. The order of magnitude of these numbers is evident from the table 1 below, 
and they scale asymptotically as \cite{ramanujan}
\begin{equation}
P(E) \simeq \frac{1}{4 (E-2) \sqrt{3}} \exp\left(\pi \sqrt{\frac{2 (E-2)}{3}}\right) 
\,\,\,\,
,
\,\,\,\,\,
E \rightarrow \infty \,\,\,. 
\end{equation}
\begin{table}[b]\label{pifre}
\begin{center}
\begin{tabular}{|l|l|} \hline 
$E$ & \,\,\,\,\,$P(E)$ \\
\hline \hline
$5 $ &  \,\,\,\,\,\,\,\,$ 3$ \\
$ 10$ & \,\,\,\,\,\,\,\,$ 22$ \\
$ 15$ & \,\,\,\,\,\,\,\,$101 $ \\
$ 20$ & \,\,\,\,\,\,\,\,$385$ \\
$ 30$ & \,\,\,\,\,\,\,\,$3718$ \\
$ 40$ & \,\,\,\,\,\,\,\,$26\, 015$ \\
$ 50$ & \,\,\,\,\,\,\,\,$147\, 273$ \\
$ 60$ & \,\,\,\,\,\,\,\,$ 715\, 220$ \\
$ 80$ & \,\,\,\,\,\,\,\,$ 12\,132\,164$ \\
$ 100$ & \,\,\,\,\,\,\,\,$ 150\,198\,136$ \\
\hline 
\end{tabular} 
\end{center}
Table 1. Numbers of tree level Feynman diagrams with a given number $E$ of external lines.  
\end{table}
However the number of tree level Feynman graphs for a given amplitude $\Gamma_n^m$ is even larger than $P(E)$ because, for given numbers $\{d_k\}$ of the various vertices with $k$-legs, we can connect in different way the external and the internal particles. In any case, this exponential growth of the Feynman diagrams provides already the best perspective on how rare could be, from a purely statistical or probabilistic point of view, to come across an integrable QFT. 

Imagine, in fact, we have found some particular values of $m_k^*$ and $g_n^*$ that ensure the vanishing of all the on-shell vertex functions , say in the channel $2 \rightarrow n$, i.e. solutions of the infinite number of equations 
\begin{equation}
\Gamma_2^n|_{on-shell} = 0 \,\,\,.
\label{2ngamma}
\end{equation}
One could naively imagine that these infinitely many conditions could be satisfied in terms of infinitely many unknowns (the couplings $g_n$, together with the finite numbers $r$ of the masses $m_k$), in other words a one-to-one correspondence between the number of equations and the number of unknowns. Strictly speaking, however, this is already a deceptive way of presenting the real complexity of the mathematical problem, first of all because for large $n$ also the $\Gamma_2^n$ are made of sums with an exponentially large number of terms, and secondly because these equations are not linear in each coupling 
constant. Hence, the fact alone that all these exponentially large combinations of masses and couplings vanish for some particular values of $m_k^*$ and $ g_n^*$ is already remarkable. But even more remarkable is that the same values $m_k^*$ and $g_n^*$ that ensure the vanishing of the amplitudes 
$\Gamma_2^n|_{on-shell}$ must ensure the vanishing of infinitely many other non-linear sums in terms of the coupling, in turns also made of exponential 
number of terms! We are referring to all the sums that correspond to the various other vertex functions $\Gamma_3^k, \Gamma_4^k, \ldots
\Gamma_l^k$ for arbitrarily large value of $l$ and $k$, computed on-shell. As discussed above, posing $k+l = E$, all these amplitudes can be obtained by $\Gamma_2^p$ (with $p+2 = E$) in terms of crossing transformations but, implementing these crossing transformation alter in an essential way the structure of the sums, which involve in fact  completely different combinations of the masses and the couplings with respect to those entering the sums of $\Gamma_2^n$ 
(see, for instance, the previous example shown in Figure \ref{graph25graph34}). So, $m_k^*$ and $g_n^*$ must ensure the validity of the infinite set of non-linear equations 
\begin{eqnarray}
\Gamma_2^n|_{on-shell} &\,=\,& \sum_{i=1}^{P(n)} T_i^{2 \rightarrow n}(m_k^*,g_n^*) \,=\, 0 \nonumber \\
\Gamma_3^n|_{on-shell} &\,=\,& \sum_{i=1}^{P(n+1)} T_i^{3 \rightarrow n}(m_k^*,g_n^*) \,=\, 0 \nonumber \\
\Gamma_4^n|_{on-shell} &\,=\,& \sum_{i=1}^{P(n+2)} T_i^{4 \rightarrow n}(m_k^*,g_n^*) \,=\, 0 \label{NPproblem}\\
\cdots &  & \,\,\,\,\ \cdots \nonumber \\
\Gamma_l^k|_{on-shell}&\,=\,& \sum_{i=1}^{P(l+k-2)} T_i^{l \rightarrow k}(m_k^*,g_n^*) \,=\, \nonumber  0 \nonumber \\
\cdots &  & \,\,\,\,\ \cdots \nonumber 
\end{eqnarray}
In light of these considerations, only particularly very well-tuned values of the masses and the couplings can ensure the simultaneous vanishing of all production processes. The intrinsic mathematical difficulty of this problem rules out this could happen for generic values of the masses and couplings. Moreover, 
in order to survive quantum fluctuations, under the renormalization of the QFT the non-trivial solutions $m_k^*$ and $g_n^*$ found at the tree-level must still remain solutions -- a condition that can be fulfilled if, for instance, the renormalization simply consists in a rescaling of the overall mass scale $\mu_0$. Apart this simple rescaling, it is conceivable that the non-trivial solutions $m_k^*$ and $g_n^*$ of the infinite number of equations (\ref{NPproblem}) form a {\em discrete} set. This means that would be impossible to move with continuity their values and still getting new solutions of the equations (\ref{NPproblem}). 

\vspace{2mm}
\noindent
{\bf Toda Field Theories as peculiar integrable QFT}. 
The previous considerations were aimed to clarify why the problem of the classification of the integrable QFT is particularly complex, due to the presence of an infinite number of equations, each of them involving an exponentially large set of terms. Nonetheless, verifying that a certain set of values $m_k^*$ and $g_n^*$ gives rise to a solution is  a much simpler task. It is suggestive to have in mind an analogy with the computational problem related to the NP class:
NP problems have the distinguished peculiarity that any given solution can be verified quickly while there is no known efficient way to locate a solution, or all solutions, in the first place. With this respect, the Toda Field Theories represent a non-trivial example of solution, and probably the only ones. For such theories, all masses and couplings come in discrete set of values, all given in terms of the roots of the Lie algebras, which is the key point why these theories fulfil at once the infinite number of identities (\ref{NPproblem}). Of course the magic with Lie algebras, on which the Toda Field Theories are based, is that they are themselves very well-tuned mathematical structures: given a random set of $p$ vectors in $R^r$, the probability that such a set is closed under the all reflections with respect their orthogonal planes is essentially zero and the only way to enforce such a property is to fine-tune the set of these vectors, by choosing proper values of both the relative angles and their lengths: in this way one arrives to the Dynkin classification of the simple roots of the Lie algebras.   

At least for those Toda Theories based on the simply-laced algebras, the roots of the Lie algebras are also responsible of the multiplicative infinite and finite renormalization of the overall mass scale $\mu_0$ which ensures that they are not only integrable at the classical level but also to the quantum level \cite{Zamo1989,AFZ1983,BCDS1990,CM1989,ChM1990,M1992,Mbook,D1997,Oota,Delius}. In the next Section we will briefly comment also on the Toda Field Theories based on the non-simply laced algebras. 

\vspace{2mm}
\noindent
{\bf Elastic $S$-matrices}. 
There is an additional reason for suspecting that, for theories with Lagrangian density as in eq.\,(\ref{LagrangianGeneral}), most probably there no other integrable QFT than the Toda Field Theories. This piece of information comes from the analysis of the closure of the bootstrap equations for factorizable and elastic 
$S$-matrices, a program initially considered in \cite{BootKM,M1992}. To expose in what consists this program, we have firstly to recall some main properties of the elastic scattering $S$-matrices \cite{Zamo1989,AFZ1983,BCDS1990,CM1989,ChM1990,M1992,Mbook,D1997}.  

\begin{figure}
\begin{center}
\includegraphics[scale=0.27]{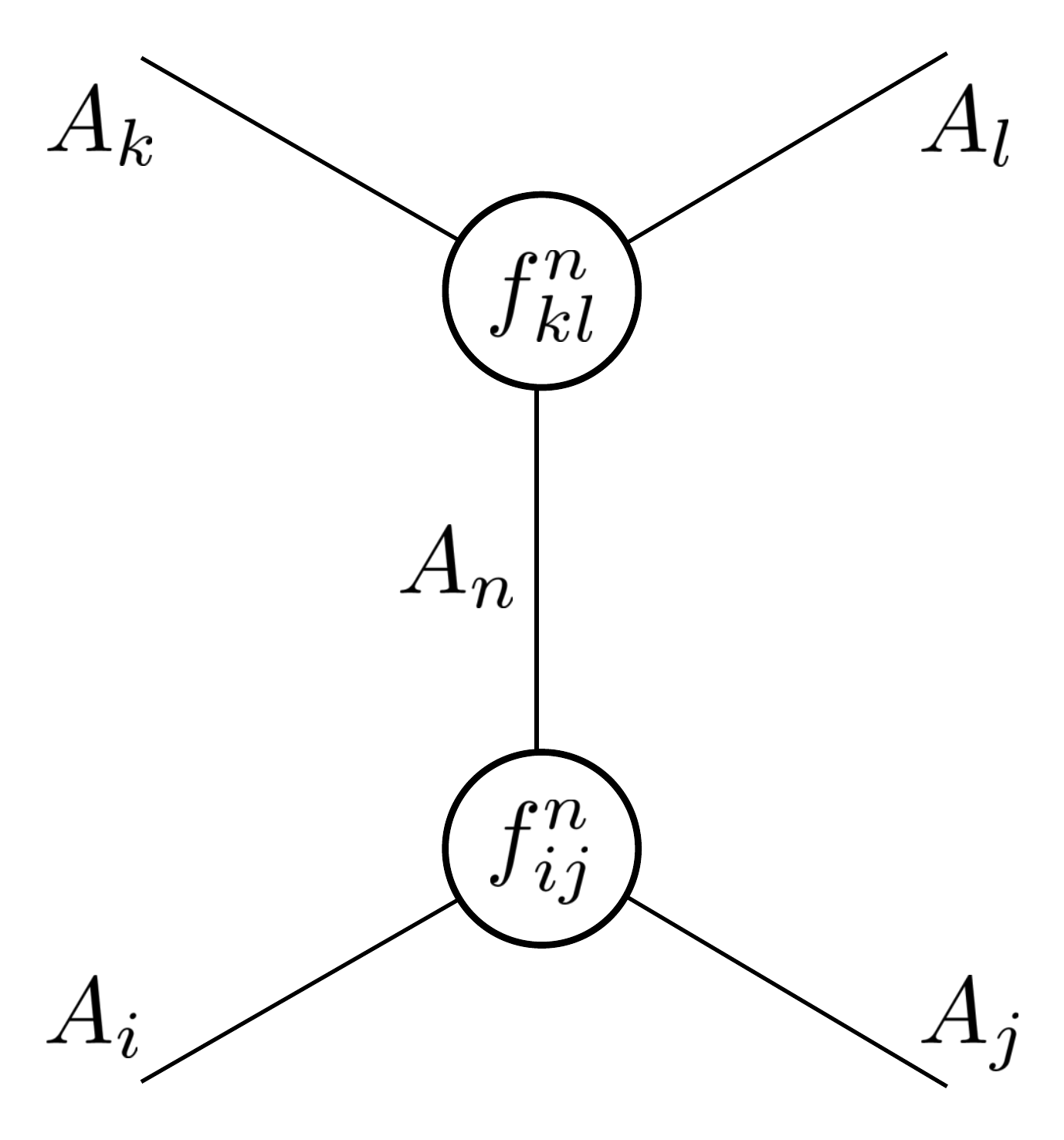}
\caption{{\em Residue of the pole and its expression in terms of the {\em on-shell} coupling constants. }}
\label{couplingresidue}
\end{center}
\end{figure}

First of all, for integrable Lagrangian theories as the ones in (\ref{LagrangianGeneral}), it is possible to prove that one can distinguish all particles in terms of their different eigenvalues with respect to the conserved charges ${\cal Q}_n$ (this essentially comes from the absence of conserved topological charges of spin 
$0$ among the infinite set of conserved charges \cite{BCDS1990}). Consequently, any two-body scattering $S_{ab}(\theta_a-\theta_b)$ relative to the particles $A_a$ and $A_b$ does not have the reflection channel and therefore it is a purely transmissive amplitude \cite{AFZ1983}. The rapidity $\theta$ is the parameter enters the relativistic dispersion relation of these particles as $E_a = m_a \cosh\theta_a$, $ p_a = m_a \sinh\theta_a$. $S$-matrix amplitudes which are purely transmissive can be expressed in terms of product of the functions $s_x(\theta)$, given by
\begin{equation}
s_x(\theta) \,=\,\frac{\sinh\left(\frac{\theta}{2} + i \pi \frac{x}{2}\right)}{\sinh\left(\frac{\theta}{2} - i \pi \frac{x}{2}\right)} \,\,\,,
\end{equation}
as 
\begin{equation}
S_{ab}(\theta) \,=\, \prod_{x \in {\mathcal A}_{ab}} s_x(\theta) \,\,\,. 
\label{product}
\end{equation}
The functions $s_x(\theta)$ satisfy the unitarity equation 
\be
s_x(\theta) \, s_x(-\theta) \,=\, 1\,\,\,, 
\ee
and therefore the same is true for the amplitudes $S_{ab}(\theta)$. Under the crossing transformation $\theta \rightarrow i \pi - \theta$, the functions $s_x(\theta)$ 
go to 
\be 
s_x(\theta) \rightarrow s_{1-x}(\theta) \,\,\,, 
\ee
and therefore the two-body $S$-matrix amplitudes of a self-conjugate particle can be expressed in terms of product of the crossing symmetric 
functions $f_x(\theta)$ given 
by 
\begin{equation}
f_x(\theta) \,\equiv \,s_x(\theta) s_{1 -x}(\theta) \,=\,   \frac{\tanh\frac{1}{2}\left(\theta + i \pi x \right)}{\tanh\frac{1}{2}\left(\theta - i \pi x \right)} 
\,=\,\frac{\sinh \theta + i \sin\pi x}{\sinh\theta - i \sin\pi x}
\,\,\,.
\label{fxfunction}
\end{equation}
The parameters $x$ in both $s_x(\theta)$ and $f_x(\theta)$ are related to the location of the poles of the $S$-matrix amplitudes $S_{ab}$. The bound states correspond to the simple poles with positive residue along the imaginary segment $(0,i \pi)$ of the $\theta$ variable. Consider a $S$-matrix with incoming particles $A_i$ and $A_j$ that has a simple pole in the $s$-channel at $\theta = i \, u_{ij}^n$: in correspondence of this pole, the amplitude can be expressed as   
\EQ
S_{ij}^{kl} \,\simeq \, i \,\frac{R^{(n)}}{\theta - i u_{ij}^n} 
\,\,\,, 
\EN 
and the residue $R^{(n)}$ is related to the {\em on-shell} three-particle vertex functions of the incoming particles and the bound state $A_n$, as shown in Figure  \ref{couplingresidue}
\EQ
R^{(n)} \,=\, f_{ij}^n \, f_{kl}^n \,\,\,.
\label{3couplresidue}
\EN 

\begin{figure}
\begin{center}
\includegraphics[scale=0.5]{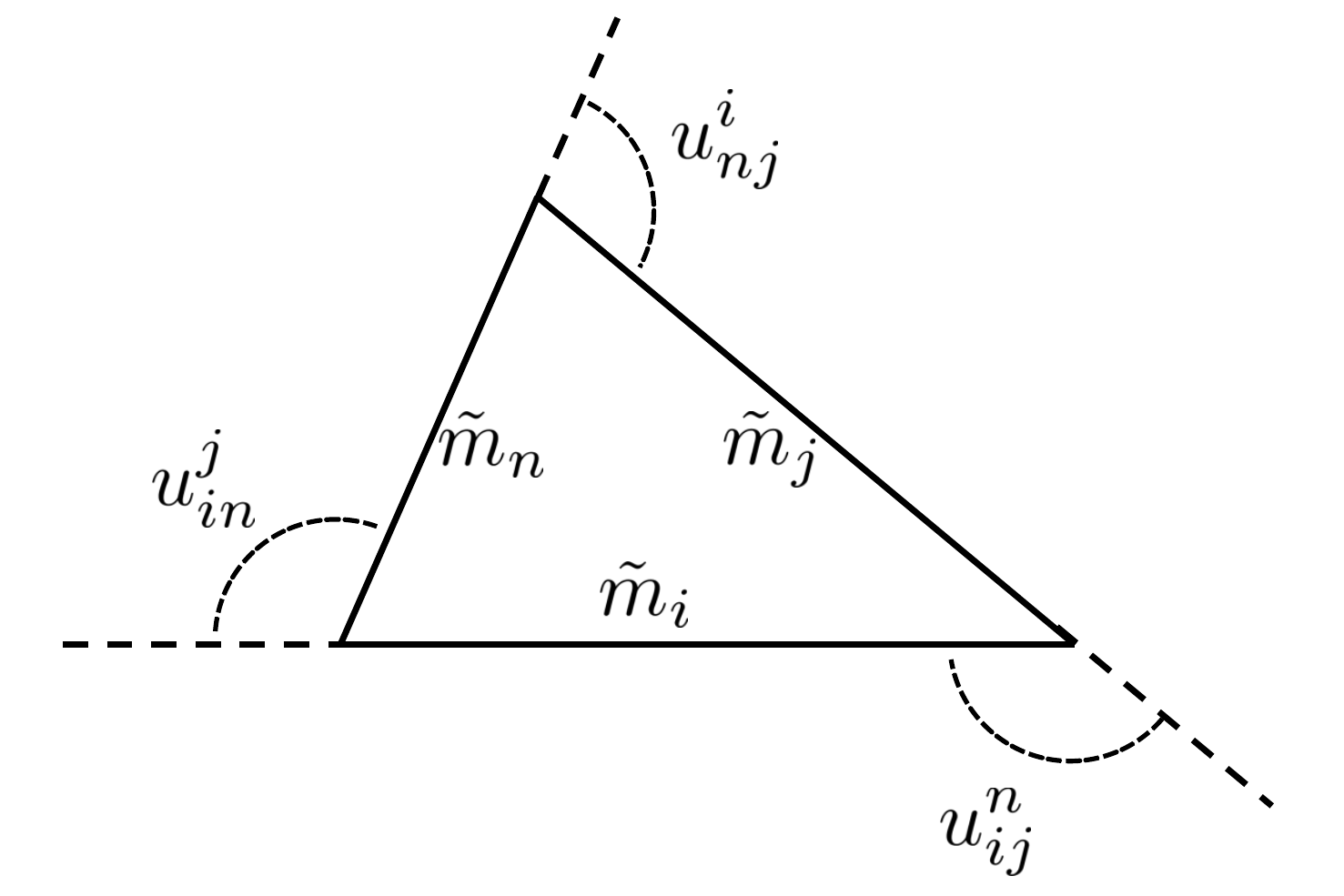}
\caption{{\em Mass triangle.}}
\label{triangolo}
\end{center}
\end{figure}

A non-zero value of $f_{ij}^n$ obviously implies a pole singularity in the other two amplitudes $S_{in}$ and $S_{jn}$ as well, where the poles are now due to the bound states $A_j$ and $A_i$. Since in the bootstrap approach the bound states are on the same footing than the asymptotic states, there is an important relation among the masses of the system: if $\theta = i u_{ij}^n$ is the position of the pole in the scattering of the particles $A_i$ and $A_j$, the mass of the bound state is given by 
\EQ
\tilde{m}^2_n \,=\,\tilde{m}_i^2 + \tilde{m}_j^2 + 2 \tilde{m}_i \tilde{m}_j \, \cos u_{ij}^n \,\,\,.
\label{triangolodellemasse}
\EN 
Hence, the mass of the bound state is the side of a triangle made of also with the masses of the other two particles, where $u_{ij}^n$ is one of the external angle as shown in Figure \ref{triangolo}. This figure clearly highlights the symmetric role played by the three particles.

Moreover, it is easy to show that the positions of the poles in the three channels satisfy  
\EQ
u_{ij}^n + u_{in}^j + u_{jn}^i \,=\, 2 \pi \,\,\,, 
\EN 
a relation that expresses a well-known properties of the external angles of a triangle.

\begin{figure}[b]
\begin{center}
\includegraphics[scale=0.35]{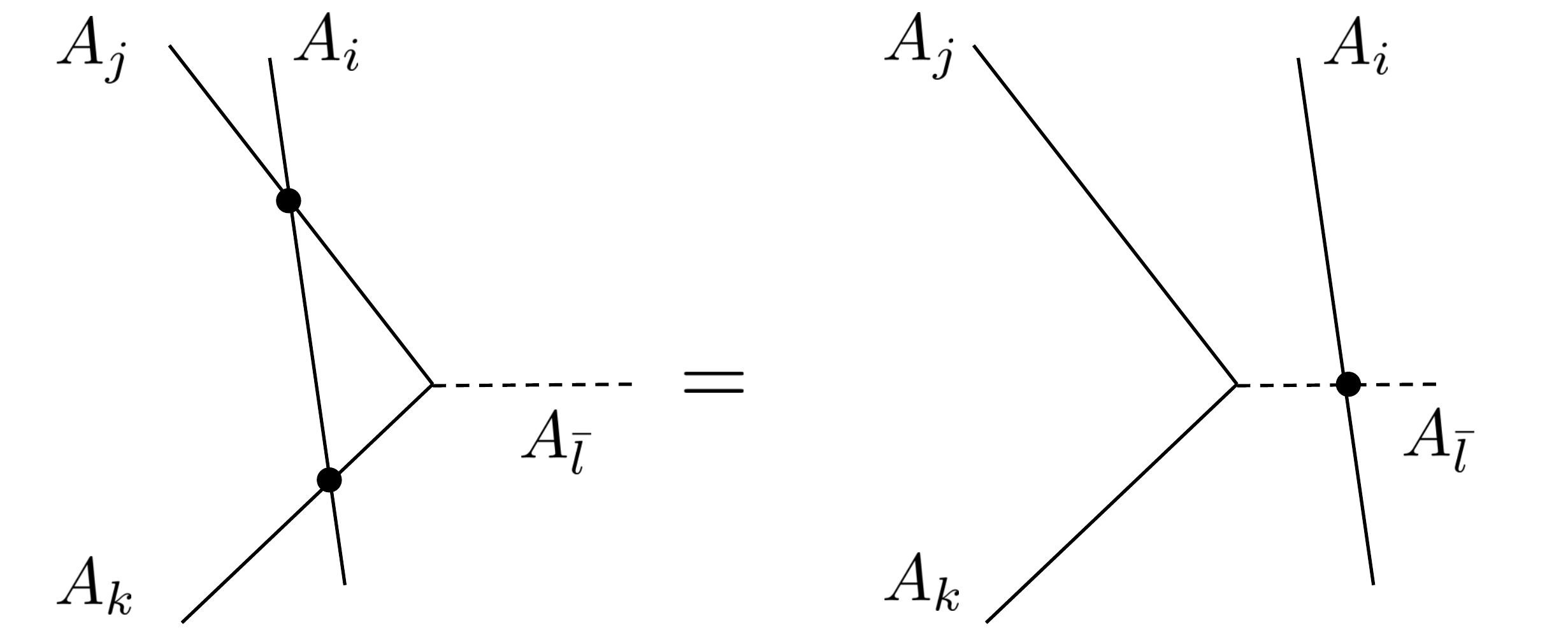}
\caption{{\em Bootstrap equation that links the $S$-matrix amplitudes, where $A_{\bar l}$ is the bound state in the scattering process of the particles 
$A_j$ and $A_k$.}}
\label{boot}
\end{center}
\end{figure}

\vspace{1mm}
\noindent
{\bf Bootstrap principle}. 
The unitarity and crossing symmetry equations alone are not able to fix the position of the poles of these amplitudes, namely to determine the sets 
 ${\mathcal A}_{ab}$ in the amplitude (\ref{product}). To achieve this goal it is necessary to make use of a dynamical condition. This is provided by the bootstrap principle \cite{Zamo1989} that posits that the bound states are on the same footing of the asymptotic states. As a consequence, the amplitudes relative to the bound states can be obtained in terms of the amplitudes of the external particles and viceversa. This translates in an additional non-linear equation that must be satisfied by the scattering amplitudes    
\EQ
S_{i \bar l}(\theta) \,=\,
S_{ij}(\theta + i {\bar u}_{jl}^k) \, 
S_{ik}(\theta - i {\bar u}_{lk}^j) \,\,\,,
\label{bootstrapboundstate}
\EN 
where  
\EQ
{\bar u}_{ab}^c \,\equiv \,\pi - u_{ab}^c \,\,\,.
\EN 
This equation comes from the commutativity of the two processes shown in Figure \ref{boot}, obtained one from the other by the translation of the world-line of the asymptotic particle $A_i$.  

\vspace{1mm}
\noindent
{\bf Classification in terms of the $S$-matrices}. One can think then to classify the integrable QFT in terms of the set of amplitudes $S_{ab}$ which close the bootstrap procedure, alias those which have a set of poles compatible with the bootstrap equation (\ref{bootstrapboundstate}) and that can be interpreted in terms of bound states or multi-particle scattering processes of the asymptotic particles themselves.  The masses of the particles are determined by the relation (\ref{triangolodellemasse}) while the on-shell three-coupling vertices by the residues (\ref{3couplresidue}). In practice, to implement this program, one needs to start  from the amplitude that involves the lighest particle, therefore with the simplest pole structure, and then iteratively applying the bootstrap equations   (\ref{bootstrapboundstate}) and see whether this procedure consistently closes after a finite number of steps.  

For the nature of this procedure, it is clear that the bootstrap will not close with a generic location of the poles: these have to be properly fine tuned in order 
that the iterative application of the bootstrap equations (\ref{bootstrapboundstate}) produce meaningful new amplitudes. For instance, if there is no partial cancellation between poles and zeros of the amplitudes $S_{ij}(\theta + i {\bar u}_{jl}^k)$ and $S_{ik}(\theta - i {\bar u}_{lk}^j)$ involved in the bootstrap equation (\ref{bootstrapboundstate}), the $S$-matrix amplitudes $S_{i \bar l}$ will double the number of poles at each step of the iteration, so that the overall number of the poles of the $S$-matrix which grows multiplicatively, finally leading to inconsistencies with respect to basic requirement of QFT, such as the impossibility to have accumulation of poles in the amplitudes. 

Up to now there has never been found a consistent set of purely transmissive amplitudes than do not coincide with those of the Toda Field Theories or to theories easily related to them.

\section{The Toda Field Theories}\label{TODAFT}
In this Section we are briefly recalling some basic properties and formulas of the Toda Field Theories: an exhaustive review of these theories is beyond our present program and therefore, in the due course, we refer the reader to the well developed literature on this subject for all relevant details. 

The Toda Field Theories associated to a Lie algebra $G$ of rank $r$ are lagrangian models, as the ones discussed in Section 2, involving $r$ bosonic fields which is convenient to collect altogether in a vector $\phi=(\phi_1,\ldots,\phi_r)$. The Lagrangian density of these models are given by 
\EQ                                                             
{\mathcal L}\,=\,\frac{1}{2} \, (\partial_{\mu}\phi) \cdot (\partial^{\mu}\phi) -
\frac{c^2m^2}{g^2} \sum_{i=1}^{r+1} \, q_i [\exp(g
\alpha_i \cdot \phi)-1]\,\,\,,
\label{toda2}                                                   
\EN
where $m^2$ and $g$ are real parameters while $\{\alpha_i\}_{i=1}^r$ is the set of the simple roots of $G$. In the perspective of considering the NR limit, we make explicit the speed of light, thus $\partial_0=c^{-1}\partial_t$. 
At the classical level the equations of motion of these theories admit a Lax pair formulation and therefore there is an infinite hierarchy of conserved charges responsible for the classical integrability of the models for all Lie algebras. In the following we mainly focus our attention on the simply-laced algebras $A_n$, $D_n$ and $E_n$, i.e. those with simple roots of the same length, here taken to be $\sqrt{2}$. The Toda Field Theories based on the non-simply laced algebras can be defined, at least at the classical level, by the so-called {\em folding} procedure, shown in Table A.a and Table A.b in Appendix \ref{DynkinToda}, which consists of an identification of the roots using the symmetry properties of the Dynkin diagrams. The quantum properties of the non-simply laced models requires however a separate discussion and, for this reason, we refer the reader to the Refs.\,\cite{Delius} for all details on this particular issue. From now on, as said above, we focus only on the simply-laced models. 

The set of the integer numbers $\{q_i\}$ presented in eq.\,(\ref{toda2}) is different for each algebra and it is related to the definition of the {\em maximal root} of the algebra, given by
\EQ
\alpha_{r+1}\,=\,-\sum_{i=1}^r q_i\, \alpha_i\,\,\,.
\label{linearmaxima}
\EN
The extended set of roots, obtained by adding the maximal root, form the Dynkin diagram of the {\em affine Lie algebras}. For these systems, posing 
$q_{r+1} = 1$, we have 
\EQ
\sum_{i=1}^{r+1} \, q_i\alpha_i=0, \hspace{1cm} \sum_{i=1}^{r+1} \, q_i = h \,\,\,, 
\label{charge}
\EN
where, for the simply-laced algebras, $h$ coincides with the Coxeter number of $G$. 
At the level of perturbation theory, the Taylor expansion of the Lagrangian provides a quadratic term plus an infinite number of interaction vertexes:
\begin{equation}
\mathcal{L}=\frac{1}{2}\sum_{a=1}^r\partial_\mu\phi^a\partial^\mu\phi^a-\sum_{j=2}^\infty\frac{mc^2g^{j-2}}{j!}\sum_{a_1,...,a_j}C^{(j)}_{a_1,...,a_j}\phi^{a_1}...\phi^{a_j}\label{todaexp}
\end{equation}

\vspace{2mm}
\noindent
{\bf Mass spectrum}. 
It is easy to see that the massive nature of these field theories comes from the last exponential term in (\ref{toda2}), the one with the maximal root. 
Indeed, expanding in series of $g$ the Lagrangian density, it is for the presence of this term that we have a well-defined minimum of the Lagrangian, with the classical values of the masses of the various particles $A_a$ obtained diagonalising the quadratic terms of the Lagrangian  
\EQ
M^2_{ab} \,=m^2C^{(2)}_{ab}=\, m^2\sum_{i=1}^{r+1} \, q_i\alpha_i^a\alpha_i^b \,\,\,.
\label{mmm}
\EN
In the basis of the eigenvectors, $C^{(2)}_{ab}=\mu_a^2 \delta_{ab}$. It is simple to see that the mass spectrum is degenerate if the group of the automorphisms of the Dynkin diagram is non trivial.

For the simply-laced algebra, there is a remarkable result concerning the mass spectrum \cite{BCDS1990}: collecting the mass values into a $r$-dimensional vector 
\[
{\bf m}\,=\,(m_1,m_2,\cdots m_r) \,\,\,,  
\]
it can be proven that this vector is the Perron-Frobenius eigenvectors of the $(r \times r)$ incidence matrix ${\mathcal I}$ of the algebra ${\mathcal G}$, defined by 
\begin{equation}
{\mathcal I} \,=\,  2 \,{\bf 1} - \mathcal{C} \,\,\,,
\label{incidencematrix}
\end{equation} 
where $\mathcal{C}_{ij} = 2 \alpha_{i} \cdot \alpha_j/\alpha_j^2$ is the Cartan matrix of the algebra ${\mathcal G}$. The classical mass spectrum of the various Toda Field Theories are summarized in the Tables A.c and A.d in the Appendix \ref{DynkinToda}.   

\vspace{3mm}
\noindent
{\bf Coupling constants}

\noindent
After the quadratic term - responsible for the mass spectrum of the theories -- the next perturbative term of the Lagrangian theories (\ref{toda2}) consists of the three-particle coupling constants (in the adimensional units)
\EQ                                                        
C^{(3)}_{abc} \,= \, \sum_i \, q_i\alpha_i^a\alpha_i^b\alpha_i^c \,\,\,. 
\label{tre}
\EN    
As discussed in \cite{BCDS1990,CM1989,ChM1990}, on the basis in which the mass term is diagonal $C^{(2)}_{ab}=\mu_a^2\delta_{ab}$, these three-coupling constants $C^{(3)}_{abc}$ have an interesting geometrical interpretation, since it is possible to prove that they vanish if it is impossible to draw a triangle with sides made of $\mu_a$, $\mu_b$, and $\mu_c$ whose internal angles are rational fraction of $\pi$ -- a natural consequence of the fact that the masses are expressed by algebraic numbers. Moreover, when they are different from zero, the quantities $C^{(3)}_{abc}$ are proportional to the area ${\mathcal{A}}^{abc}$ of the aforementioned mass triangle and, for the simply-laced algebras, we have  
\EQ 
\left| C^{(3)}_{abc} \right| = \frac{4}{\sqrt{h}} \, {\mathcal{A}}^{abc}\,\,\,. 
\label{strong}                                                         
\EN                                                                    
Obviously the non-vanishing values of $C^{(3)}_{abc}$  indicate the possible scattering processes in which enter the particles and their bound states, see Figure \ref{couplingresidue}. The three-coupling constants are very important quantities because, for the renormalization properties of the simply-laced models, their exact $S$-matrix can be written down requiring tree-level consistency with the classical mass spectra and the non-zero $C^{(3)}_{abc}$ responsible for the 
one-particle exchanges.

\vspace{2mm}
\noindent
{\bf Renormalization}. 
All the perturbative divergences of Lagrangian theories as (\ref{toda2}) come from the tadpole diagrams, that can be eliminated by defining the normal ordering. 
This appears, though, to introduce into the theories an infinite set of counterterms which do not necessarily preserve the form of the Lagrangian 
(\ref{toda2}) which only depends on two parameters, $m$ and $g$. However, for exponential terms, their normal ordering only induces a 
multiplicative change in them, and therefore for the Lagrangian (\ref{toda2}) only a renormalization of the mass scale $m^2$  
\EQ
m^2 \,\rightarrow \,m^2\, 
\left(\frac{\Lambda^2}{m^2}\right)^{\frac{g^2}{4\pi}} \,\,\,. 
\label{ren}
\EN
Hence, for the Toda Field Theories the infinities are renormalised only in terms of the mass scale, while the coupling constant $g$ does not renormalised. In $(1+1)$ dimensions there are however also the {\em finite} corrections induced by the loop diagrams. For the Toda Field Theories based on the simply-laced algebras, it is pretty remarkable that they do not alter the mass ratios \cite{BCDS1990,CM1989,ChM1990} as well as the inherent integrable structure of the theory that emerges from the tree level diagrams. It is for this reason that one can write down the $S$-matrix of these simply-laced Toda Field Theories which have their physical poles fixed by the fusion angles given by the classical value of the masses and the non-zero three-coupling constants $C^{(3)}_{abc}$. 

\vspace{2mm}
\noindent
{\bf Exact $S$-matrix amplitudes}. The exact two-body $S$-matrix of various simply-laced Toda Field Theories has been computed and compared with the 
lowest order perturbation theory in several papers \cite{AFZ1983,BCDS1990,CM1989,ChM1990,M1992,Mbook,D1997,Oota}. The reader can consult these 
reference for spelling out all the precious details of these computation: an exemplary representative is given by the $S$-matrix of the Toda Field Theories based on the simply-laced algebras $A_n$, discussed below. There is, however, a remarkable integral representation formula that encodes, at once, all $S$-matrix amplitudes relative to the scattering of the particles $A_a$ and $A_b$ of a generic simply-laced Toda Field Theory: the formula reads \cite{Oota} 
\begin{equation}
S_{ab}(\theta)\,=\,\exp\left[-i\int_0^\infty\frac{dt}{t}\left(8\sinh\left(\frac{B t}{2}\right)\sinh\left(\frac{\pi t}{h}-\frac{B t}{2}\right)\left(2\cosh\frac{\pi t}{h}- {\mathcal I}\right)^{-1}_{ab}-2\delta_{ab}\right)\sin(\theta t)\right]
\label{generalSmatrixamplitudes}
\end{equation}
and employs the incidence matrix ${\mathcal I}$ of the algebra ${\mathcal A}$, see eq.\,(\ref{incidencematrix}), the Coxeter exponent $h$ and the function $B$ of the coupling constant $g$ given by 
\begin{equation}
B\,=\,\frac{1}{4h}\frac{c g^2}{1+c g^2/8\pi}\,\,\,, 
\label{Bfunction}
\end{equation}
where in this formula we have already conveniently inserted the speed of light $c$.   

\section{Lieb-Liniger models} \label{LLSect}
In this Section we will briefly recall the main definitions and properties of the main actor of this paper, alias the Lieb-Liniger model. We first address  
the simplest version of this model, based on a single species of undistinguishable bosonic particles, and then we pass to study its multi-species generalisation, commonly known in literature as Yang-Gaudin model. 
While the former version of the model is always integrable, the latter version is instead generically non-integrable. Below we discuss, in particular, the 
conditions which lead to the integrability of the multi-species Lieb-Liniger model, which is the model that will be recovered by the non-relativistic
limit of the $O(N)$ sigma model, later considered in this paper. 

It is worth to remind that there are two different but equivalent formulations of the Lieb-Liniger model, associated to the second or the first quantization formalisms respectively. Each formulation has its own advantage, in particular the one based on the second quantization formalism privileges those aspects of quantum field theories which are particularly useful for our later considerations.

\vspace{1mm}
\noindent
\subsection{Single species Lieb-Liniger model} Let's start our discussion from the simplest version of the Lieb-Liniger model, the one in which there is 
only one species of undistinguishable bosonic particles of mass $m$. We discuss in order its second and first quantization formulations.  

\vspace{1mm}
\noindent
{\bf Non-linear Schr\"{o}dinger model}. In the second quantization formalism, the formulation of the model consists of a local non-relativistic field theory (also known as the non-linear Schr\"{o}dinger model) for a complex Bose field field $\psi(t,x)$ that satisfies the canonical commutation relations
\begin{equation}
[\psi(t,x),\psi^\dagger(t,x')] \,=\,\delta(x-x') \,\,\,\,\, \,\,,
\,\,\,\,\,\, 
[\psi(t,x), \psi(t,x')] \,=\,0 \,\,\,.
\label{CRPSI}
\end{equation}
The Hamiltonian for the field $\psi(t,x)$ is given by 
\be
 H\,=\,\int\,\left(\frac{\partial_x \psi^\dagger\partial_x \psi }{2m}+ \lambda\,\psi^\dagger\psi^\dagger\psi\psi\right)\,
\mathrm{d}x\,\,\,  ,
\label{HLL}
\ee
and the associated Lagrangian density of the model can be expressed as 
\be
{\cal L}\,=\, i\,\frac{1}{2}\left(\psi^\dagger\partial_t\psi
- \partial_t\psi^\dagger
\psi\right) -
\frac{1}{2m} \partial_x\psi^\dagger\partial_x\psi - \lambda\, \psi^\dagger\psi^\dagger\psi\psi\,.
\label{LagrangianNLS}
\ee
Notice that the Hamiltonian (\ref{HLL}) is invariant under the continuos $U(1)$ transformation 
$$\psi \rightarrow e^{i\alpha} \, \psi \,\,\,\,\,,\,\,\,\,\, 
\psi^\dagger \rightarrow e^{-i \alpha} \, \psi^\dagger$$ 
and the corresponding conserved quantity ${\cal Q}$, coming from the Noether's theorem 
\be
{\cal Q}  \,=\, \int  \,\psi^\dagger\,\psi\, \mathrm{d}x \,\,\,, \label{totalnumer}
\ee
expresses the conservation of the total number of particles. The Hamiltonian also commutes with the momentum ${\cal P}$ of the field theory 
given by 
\be
{\cal P}  \,=\,-\frac{i}{2}\,\int  \, \left(\psi^\dagger \, \partial_x\psi - \partial_x\psi^\dagger \, \psi\right) 
\, \mathrm{d}x \,\,\,.
\label{momentum}
\ee
In addition to ${\cal Q}$ and ${\cal P}$, the non-linear Schr\"{o}dinger model is also supported by an infinite number of higher order conserved charges that ensures 
its quantum integrability (see for instance \cite{Korepin}).   

\vspace{1mm}
\noindent
{\bf Lieb-Liniger Hamiltonian}. The conservation laws alone of ${\cal P}$ and ${\cal Q}$ allow us to express the model in terms of the first quantization formalism \cite{Korepin}: this is done by initially defining a translation invariant Fock vacuum $\mid 0 \rangle$ in terms of the condition 
\be 
\psi(x) \,\mid 0 \rangle \,=\, 0 \,\,\,, 
\ee
and then looking for common eigenfunctions $\psi_N$ of the operators $ H$, ${\cal P}$ and ${\cal Q}$ 
\be
\mid \chi_N (\lambda_1,\ldots, \lambda_N \rangle \,=\, \int d^Nz \, \chi_N(z_1,\ldots,z_N | \lambda_1,\ldots,\lambda_N) 
\psi^\dagger(z_1) \ldots \psi^\dagger(z_N) \mid 0 \rangle \,\,\,.
\ee
For the bosonic statistic of the particles, $\chi_N(z_1,\ldots,z_N | \lambda_1,\ldots,\lambda_N)$ are symmetric functions of all $z_i$. Restricting to the subspace of the Hilbert space where the number of particles $N$ is fixed, the eigenvalue 
equations 
\begin{eqnarray}
&& H \, \mid \chi_N \rangle \,=\, E_N \, \mid \chi_N \rangle \nonumber \,\,\,, \\
&& {\cal P} \mid \chi_N \rangle \,=\, P_N \, \mid \chi_N \rangle \,\,\, \\
&& {\cal Q} \mid \chi_N \rangle \,=\, N \, \mid \chi_N \rangle \nonumber \,\,\, 
\end{eqnarray}
translates into eigenvalue equations for the function $\chi_N$, that has to be an eigenfunction of the quantum Hamiltonian $H_N$ and 
the quantum operator $P_N$ 
\begin{eqnarray}
&& H_N  \,=\, -\frac{1}{2m}\sum_{i=1}^N\frac{\p^2}{\p x_i^2}+2\lambda\,
\sum_{i<j}\delta(x_i-x_j)\,\,\,, \label{LLHAMILTONIAN}\\
&& P_N  \,=\,- i\, \sum_{i=1}^N\frac{\p}{\p x_i}
\end{eqnarray}
In the first quantization formalism, the model is thus defined by the Hamiltonian $H_N$ -- which is indeed the one generally known as the Lieb-Liniger Hamiltonian \cite{LL} --  and consists of a gas of $N$ non-relativistic bosons of mass $m$ interacting by means of local two-body repulsive delta-functions.  

\vspace{1mm}
\noindent
{\bf Elastic S-matrix} For infinite number of conserved charges of the non-linear Schr\"{o}dinger model, in the time evolution of a state made of a given number $N$ of excitations, not only this number will be conserved but the same also holds for the individual momenta of all its excitations. As shown by Lieb and Liniger in their original paper \cite{LL}, the eigenvalue problem relative to the Hamiltonian $H_N$ of eq. (\ref{LLHAMILTONIAN}) can be solved in terms of a Bethe Ansatz which employs the two-body elastic $S$-matrix. As shown in Appendix A, for the scattering of two undistinguishable particles interacting through a delta-function 
potential the scattering can be considered purely transmissive (since it is impossible to distinguish between transmission and reflection channels) 
and the two-body $S$-matrix is given by 
\be
S_{\text{LL}}(p,\lambda)\,=\,\frac{p- 2 i m \,\lambda}{p +2 i m \,\lambda}\,\, \equiv e^{i \delta(p,\lambda)}
\label{SmatrixLLM}
\ee
where $p = p_1 - p_2$ is the momentum difference of the two scattering particles and $\delta(p,\lambda)$ is the phase-shift. 
This quantity can be expressed as 
\begin{equation}
\delta(p,c)  \,=\,  - i \log \frac{p - 2 i m \lambda}{p + 2 i m \lambda} \,=\, 2 \arctan \frac{p}{2 m \lambda}  \,=\, 2 \int_0^\infty \frac{dt}{t} 
\,e^{-2 m \lambda t } \, \sin(p t) \,\,\,.
\label{integralrepresentationdelta}
\end{equation}
We will show that this is the expression in term of which we can write down all $S$-matrix amplitudes of the non-relativistic limit of the Toda field theories.

\subsection{Multi-species Lieb-Liniger model} 
\label{secmultiLL}

As in the previous case, we can have two equivalent formulations of the model. Let's start also in this case by its formulation in terms of non-relativistic field theory.  

\vspace{3mm}
\noindent
{\bf Non-linear Schr\"{o}dinger model}. In the second quantization formalism, we consider a local non-relativistic field theory for $r$ complex Bose field fields $\psi_i(t,x)$ with masses $m_i$ ($i = 1,\ldots, r$) that satisfy the canonical commutation relations
\begin{equation}
[\psi_i(t,x),\psi^\dagger_j(t,x')] \,=\,\delta_{ij}\, \delta(x-x') \,\,\,\,\, \,\,,
\,\,\,\,\,\, 
[\psi_i(t,x), \psi_j(t,x')] \,=\,0 \,\,\,.
\label{CRPSI2}
\end{equation} 
As the most general expression for the Hamiltonian of such a multi-species system we can take  
\be
 H\,=\,\int\mathrm{d}x\,\left(\sum_{i=1}^r \frac{\partial_x\psi_i^\dagger
\partial_x\psi_i}{2m_i} + \sum_{i,j} \lambda_{ij} \,\psi^\dagger_i\psi^\dagger_j\psi_i\psi_j\right)\,. 
\label{eq:mHLL}
\ee
The corresponding Lagrangian density is given by 
\be
{\cal L}\,=\,\sum_{k=1}^r \left[ i\,\frac{1}{2}\left(\psi^\dagger_k\partial_t\psi_k
- \partial_t\psi^\dagger_k\psi_k\right) - 
\frac{1}{2m_i} \partial_x\psi^\dagger_k 
\partial_x\psi_k\right] - \sum_{k,l}^r \lambda_{k,l}\, \psi^\dagger_k\psi^\dagger_l\psi_k\psi_l\,.
\label{LagrangianmNLS}
\ee
Notice that in this model the Hamiltonian (\ref{eq:mHLL}) preserves individually the number of particles ${\cal Q}_k$ of each species and, of course, 
the total momentum ${\cal P}$
\begin{eqnarray}
&{\cal P} & \,=\,-\frac{i}{2}\,\int dx \, \sum_{k=1}^r \left(\psi^\dagger_k \, \partial_x\psi_k - \partial_x \psi^\dagger_k \, \psi_k\right) \,\,\,,
\label{momentumt}\\
&{\cal Q}_k & \,=\, \int dx \,\psi^\dagger_k\,\psi_k\,\,\,\,\, , \,\,\,\,\, k=1,\ldots r\,\,\,.. \label{totalnumerr}
\end{eqnarray}

\vspace{1mm}
\noindent
{\bf Lieb-Liniger Hamiltonian}. Using, as the previous case, the conservation of the number of particles of each species, we can go to the first quantization formalism and express the quantum Hamiltonian of the model as a coupled set of Lieb-Liniger models 
\be
H_N  \,=\, -\sum_{k=1}^r \frac{1}{2m_k}\sum_{i_k=1}^{N_k}\frac{\p^2}{\p x_{i_k}^2} + 2 \sum_{k,l}^r \lambda_{k,l}\,
\sum_{i_k<j_l}\delta(x_{i_k} -x_{j_l})\,\,\,, \label{mLLHAMILTONIAN}\\
\ee

\vspace{1mm}
\noindent
{\bf Integrability and exact $S$-matrix}. The integrability of the Hamiltonian (\ref{mLLHAMILTONIAN}) is obtained 
by choosing equal value of all the masses and equal value of all couplings \cite{yang,suth} (this case is usually called Yang Gaudin model)
\begin{eqnarray} 
&& m_k \,=\, m \label{equalmass} \,\,\,,\,\,\, k=1,\ldots r \\
&& \lambda_{k,l} \,=\,\lambda \,\,\,\,\,, \,\,\,\, \, k,l =1, \ldots r 
\end{eqnarray}

As discussed in Appendix \ref{AppendixA}, the contact potential is never purely transmissive, thus integrability forces the masses to be equal; moreover integrability is attained only if the couplings are degenerated $\lambda_{k,l}=\lambda$.
At its integrable point, the model is characterised by the two-body $S$-matrix. For the multi-species Lieb-Liniger  model we have to consider separately two cases: (a) the scattering of two undistinguishable particles of the same species; (b) the scattering of two distinguishable particles of two different species. In the case (a), the $S$-matrix coincides with the expression (\ref{SmatrixLLM}) given above and there is no reflection amplitude. In the case (b), as shown in Appendix \ref{AppendixA}, we have simultaneously a reflection and a transmission amplitudes given by 
\begin{equation}
R(v) \,=\,\frac{-i2\lambda}{v + 2 i \lambda} \,\,\,,\,\,\,\,T(v)\,=\,\frac{v}{v + 2 i \lambda}
\end{equation}
where $v$ is the relative velocity of the two particles. 

\section{Non-relativistic limit of the Sinh-Gordon model}\label{SinhGordon}
In this Section we analyse the non-relativistic limit of the Sinh-Gordon model, mainly following the steps of refs. \cite{KMT}. This analysis plays the role of 
a pedagogical exercise for addressing richer QFT later. 

The ShG model is described in terms of a relativistic Lagrangian of a real bosonic field $\phi(t,x)$ 
\be
\mc{L}= \frac12\partial_\mu\phi\partial^\mu\phi-
\frac{m^2c^2}{g^2 }\left(\cosh(g \phi)-1\right)\,, 
\label{LL6}
\ee
where $m$ is a mass scale and $c$ is the speed of light and, as before, $\partial_0=c^{-1}\partial_t$. The parameter $m$ is related to
the physical (renormalized) mass $\tilde{m}$ of the particle by \cite{babkar}
\begin{equation}
m^2\,=\,\tilde{m}^2\frac{B}{\sin B}\,\,\,, 
\label{mu_m}
\end{equation}
where $B$ is the function (\ref{Bfunction}) of the coupling constant $g$, specialised to the Coxeter number $h=2$ relative to the Sinh-Gordon model
\begin{equation}
B \,=\,\frac{1}{8} \,\frac{c g^2}{1 + c g^2/8\pi} \,\,\,.
\end{equation}
As mentioned in Section \ref{integrability}, the ShG model has been recognized to be integrable both at a classical and quantum level \cite{AFZ1983,FMS1993,AM} and it is the simplest example of Toda Field Theories. Its integrability implies the absence of particle production process and its $n$-particle scattering amplitudes are purely elastic and factorized. The energy $E$ and the momentum $P$ of a particle can be written as $E=\tilde{m} c^2 \cosh\th$, $P=\tilde{m} c \sinh\th$, where $\th$ is the rapidity, and in terms of the particle rapidity difference $\th$, the two-body $S$-matrix is given by \cite{AFZ1983}
\be
\mathcal{S}_{\text{ShG}}(\th,\alpha)=\frac{\sinh\th-i\,\sin(B)}{\sinh\th+i\,\sin(B)}\,\,\,.
\ee

\vspace{2mm}
\noindent
{\bf Lieb-Liniger model as double limit of the Sinh Gordon model}. We want to show that, taking the non-relativistic limit of the Sinh-Gordon model, the result 
is nothing else but the simplest version of the Lieb-Liniger model. The easiest way to see this, is to show that the $S$-matrix of the Sinh-Gordon model reduces in this limit to the one of the Lieb-Liniger. Naively one would expect that the non-relativistic limit is simply given by sending $c \rightarrow \infty$ 
but instead a non trivial result requires a fine-tuning of the coupling constant $g$. In other words, the proper non-relativistic limit of the 
Sinh-Gordon model is realised in terms of the double limit where $c\to\infty$ while $g \rightarrow 0$ in such a way the product $g c$ remains constant \cite{KMT}. If we perform such a limit keeping constant the momenta of the particles, the relative rapidity must go to zero as $\theta\simeq k/(mc)$, where we used that $\tilde{m}\to m$ in this limit, as it is clear from (\ref{mu_m}). Taking this combined double limit we immediately see that 
\begin{equation}
\lim_{\text{NR}}\mathcal{S}_{\text{ShG}}(\theta)=\frac{k-imc^2 g^2/8}{k+imc^2 g^2/8}\,=\,
\frac{k-2im\lambda}{k+2im\lambda}\hspace{2pc} ,\hspace{2pc}\lambda=\frac{g^2 c^2}{16}\,\,\,.
\end{equation}
Above, with $\lim_{\text{NR}}$ we mean the combined double limit. Of course, in order to establish the mapping between the two models, it is highly 
desirable to perform the limit at the level of fields and Hamiltonian (or Lagrangian). We firstly present the quickest way of setting up this mapping (which 
works well for the Sinh-Gordon model) and secondly the more refined way, which passes through the equations of motion, which instead holds for any other Toda Field Theory. 

\vspace{2mm}
\noindent
{\bf Mapping of the field theories: quickest way}. Consider the action of the ShG model obtained from the Lagrangian density (\ref{LL6}). Since in the NR limit $g\sim c^{-1}$, we can consider the power expansion of the interaction and neglect all terms beyond $\phi^4$ that will finally disappear 
taking $c \rightarrow \infty$. Defining the coupling $\beta= g c$, which is constant in the NR limit, expanding we have ($\partial_0 \equiv 
\frac{\partial}{c \partial t}$) 
\begin{equation}
S_{\text{ShG}}=\int dxdt\;\mathcal{L}_{\text{ShG}}=\int dxdt\;\frac{1}{2}:\partial_{\mu}\phi\partial^{\mu}\phi:-\frac{m^2c^2}{2}:\phi^2:-\frac{1}{4!}m^2 
\beta^2:\phi^4:+\mathcal{O}(c^{-2})\label{LL14}
\end{equation}
where $: :$ means that the expression must be normal ordered with respect the free modes of the field.
The power expansion of the action generates a $c-$divergent mass term and a finite quartic interaction, plus further interactions negligible in the NR limit.
In order to extract from $\phi$ the proper NR fields we split it in two parts associated with positive and negative frequencies:
\begin{equation}
\phi(t,x)=\frac{1}{\sqrt{2m}}\left(e^{imc^2t}\psi^{\dagger}(t,x)+e^{-imc^2t}\psi(t,x)\right)\label{LL11}\,\,\,.
\end{equation}
The exponential factors take care of the fast oscillating behavior induced by the $m^2c^2$ term of the Lagrangian (\ref{LL14}), while in the NR limit the smooth dynamics of the theory is encoded into the $\psi$ fields. From the Lagrangian we get the field $\pi$ conjugated to $\phi$ 
\begin{equation}
\pi(t,x)\,=\,\frac{1}{c^2}\partial_{t}\phi(t,x)\,=\,i\sqrt{\frac{m}{2}}\left(e^{imc^2t}\psi^{\dagger}(t,x)-e^{-imc^2t}\psi(t,x)\right)+\mathcal{O}(c^{-2})
\end{equation}
where we used that $\psi$ and all its derivatives are supposed to be non singular in the NR limit. Imposing the canonical equal time commutation rules $[\phi(t,x),\pi(t,y)]=i\delta(x-y)$, it is easy to show that in the NR limit the $\psi$ fields behave as canonical non relativistic bosonic fields:
\begin{equation}
[\psi(t,x),\psi(t,y)]=0,\hspace{3pc}[\psi(t,x),\psi^{\dagger}(t,y)]\,=\,\delta(x-y) \,\,\,. 
\end{equation}
Notice that with this definition of the relativistic fields, the normal ordering of the action with respect the free modes of $\phi$ is completely equivalent to the normal ordering with respect the $\psi^\dagger$ fields, that must appear always on the left of the $\psi$ fields.

Substituting now (\ref{LL11}) in (\ref{LL14}) we obtain a cancellation of the mass term:
\begin{eqnarray}
\nonumber&&S_{\text{ShG}}=\int dxdt\;\frac{i}{2}:\left(e^{imc^2t}\partial_t\psi^\dagger+e^{-imc^2t}\partial_t\psi\right)\left(e^{imc^2t}\psi^\dagger-e^{-imc^2t}\psi\right):+\\
&&-\frac{1}{4m}:\left(e^{imc^2t}\partial_x\psi^\dagger+e^{-imc^2t}\partial_x\psi\right)^2:-\frac{1}{4!}\frac{\beta^2}{4}:\left(e^{imc^2t}\psi^\dagger+e^{-imc^2t}\psi\right)^4:+\mathcal{O}(c^{-2})\label{LL15}
\end{eqnarray}
However the limiting procedure has not been completed yet: indeed we have to expand the above expression and retain only the terms in which the oscillating phases are absent. This last passage is the one that needs extra care, because it will be problematic in all other Toda Field Theories. 
Expanding the expression in (\ref{LL15}), we find a sum of terms as
\begin{equation}
\int dxdt\; e^{imc^2 nt}\mathcal{O}(t,x) \,\,\,, 
\end{equation}
where $\mathcal{O}(t,x)$ are fields made of products of $\psi$ and $\psi^\dagger$ operators (therefore smooth for $c\to\infty$) and $n$ an integer. It is not correct to state that $e^{imc^2nt}\mathcal{O}(t,x)\xrightarrow{c\to\infty} 0$ for $n\ne0$, but this is true once we consider time integrations over an infinitesimal (but not zero) time interval $\Delta$:
\begin{equation}
\int_{t_0}^{t_0+\Delta}dt \;e^{imnc^2t}\mathcal{O}(t,x)\simeq \mathcal{O}(t_0,x)\int_{t_0}^{t_0+\Delta}dt\; e^{imnc^2t}\,=\, \mathcal{O}(t_0,x) e^{imnc^2t_0}\frac{e^{imc^2n\Delta}-1}{imc^2}\sim c^{-2}
\end{equation}
Above, $n\ne 0$ and $\Delta$ is kept fixed in the $c\to\infty$ limit, but it is considered to be small enough to approximate $\mathcal{O}(t,x)\simeq\mathcal{O}(t_0,x)$ in the whole integration domain. With this caveat, the NR limit of (\ref{LL15}) can be extracted neglecting all the fast oscillating phases and 
the resulting Lagrangian density is 
\begin{equation}
\lim_{\text{NR}}{\cal L}_{\text{ShG}}\,=\, \frac{i}{2}\left(\partial_t\psi^\dagger\psi-\psi^\dagger\partial_t\psi\right)-\frac{1}{2m}\partial_x\psi^\dagger\partial_x\psi-\frac{\beta^2}{16}\psi^\dagger\psi^\dagger\psi\psi\label{LL18} \,\,\,. 
\end{equation}
The final expression of the Lagrangian we got in the non-relativistic limit can be readily recognized as the one associated with the Lieb-Liniger 
model (\ref{LagrangianNLS}), with $\lambda=\beta^2/16$, where this value of the coupling constant is consistent with the previous analysis of the scattering matrix.

\vspace{2mm}
\noindent
{\bf Mapping of the field theories: a more refined way}. Let's now present a more refined way to compute the NR limit of the dynamics, directly based on the Heisenberg equations of motion: this way of proceeding will be used to study other Toda Field Theories, while taking the limit directly from the action would instead fail. It is worth stressing that the discrepancy between the two approaches is due to the fact that the NR limit of the action does not take in account the proper limit of the path integral: in this perspective we should rather consider all the Feynman graphs and take the NR limit of those, instead than directly of the action. Such an approach would be completely equivalent to proceed through the equation of motion, although more complicated at a technical level.
For this reason let's then consider the Heisenberg equation of motion
\begin{equation}
\frac{1}{c^2}\partial_{t}^2\phi \,=\,\partial_x^2\phi-\frac{m^2c^2}{g}:\sinh(g\phi):\label{LL19}
\end{equation}
To remove the divergences of the theories we need to normal order the expression with respect to the eigenmodes of the free theory or, equivalently, to the $\psi$ fields of (\ref{LL11}). In principle, to extract the non relativistic limit of the dynamics, we should solve the equation of motion given above and then take the $c\to\infty$ limit of the solution. Of course this is rather impossible and we should instead aim to obtain a non relativistic equation of motion whose solution is exactly the non relativistic limit of the solution of (\ref{LL19}). Similarly to what done before, let's use (\ref{LL11}) and retain only the non vanishing terms in the NR limit, although keeping all the oscillating phases 
\begin{equation}
i\left(e^{imc^2t}\partial_t\psi^\dagger-e^{-imc^2t}\partial_t\psi\right)=\frac{1}{2m}\partial_{x}^2\left(e^{imc^2t}\psi^\dagger+e^{-imc^2t}\psi\right)-\frac{\beta^2}{4!}:\left(e^{imc^2t}\psi^\dagger+e^{-imc^2t}\psi\right)^3:+\mathcal{O}(c^{-2})
\end{equation}
As one could expect, the fast oscillating phases will give vanishing contributions, but to show it properly we need to consider time integrations. 
Hence, let's write the previous differential equation as an equivalent integral equation
\begin{eqnarray}
\nonumber&&i\psi^\dagger(t_0+\Delta)\,=\, i\psi^{\dagger}(t_0)+\epsilon\int_{t_0}^{t_0+\Delta}dt\;e^{-i2mc^2(t-t_0)}i\partial_t\psi+\\
\nonumber&&+\frac{1}{2m}\partial_{x}^2\left(\psi^\dagger+e^{-i2mc^2(t-t_0)}\psi\right)-\frac{\beta^2 e^{-imc^2(t-t_0)}}{4!}:\left(e^{imc^2(t-t_0)}\psi^\dagger+e^{-imc^2(t-t_0)}\psi\right)^3:\\ \label{LL21}
\end{eqnarray}
This equation must holds for any $t_0$ and any $\Delta$, but in the end we will be interested in $\Delta\to 0$ after have sent $c\to\infty$. The $\epsilon$ term is an additional parameter introduced for power counting and at the end of the calculation will be posed $\epsilon=1$. We can now imagine to recursively solve (\ref{LL21}): in the solution, each time integral over an oscillating phase gives a $\mathcal{O}(c^{-2})$ contribution, instead the integration of terms in which the fast oscillating phase is absent contributes as $\Delta$. This means that a recursive solution of (\ref{LL21}) will be organized as:
\begin{equation}
i\psi^{\dagger}(t_0+\Delta)=i\psi^{\dagger}(t_0)+\sum_{n,j}\epsilon^n\Delta^{n-j} c^{-2j}\mathcal{O}_{n,j}(t_0)\label{LL22}\,\,\,.
\end{equation}
Where the operators $\mathcal{O}_{n,j}(t_0)$ are normal ordered functions of $\psi(t_0)$ and $\psi^{\dagger}(t_0)$, the $\epsilon$ parameter efficiently keeps track of the recursive order of the solution. After having ideally solved recursively the equation of motion, we can take $c\to\infty$ and obtain
\begin{equation}
i\psi^{\dagger}(t_0+\Delta)\,=\,i\psi^{\dagger}(t_0)+\sum_{n,j}\epsilon^n\Delta^n\mathcal{O}_{n,0}(t_0)\label{LL23}
\end{equation}
At this stage, after we took the NR limit, we take a derivative with respect to the $\Delta$ parameter and compute the resulting expression at $\Delta=0$
\begin{equation}
i\partial_t\psi^{\dagger}(t_0)\,=\,\epsilon\mathcal{O}_{1,0}(t_0) \,\,\,. 
\end{equation}
In this way, we obtain the desired equation of motion, since the expression must be valid for any $t_0$. So, in conclusion, we get that in the NR equation of motion only the first order recursive solution of (\ref{LL21}) matters and, among the various contributions, only the non oscillating terms. Therefore we simply have:
\begin{equation}
i\partial_t\psi^\dagger\,=\,\frac{1}{2m}\partial_x^2\psi^\dagger-\frac{\beta^2}{8} \psi^\dagger\psi^\dagger\psi\label{LL25} \,\,\,. 
\end{equation}
Based on this result, we can now easily reconstruct the Hamiltonian that corresponds to this equation of motion and its Hermitian conjugate, using $i\partial_t\psi^\dagger=[\psi^\dagger,H]$ and the commutation rules of $\psi,\psi^\dagger$ we find (\ref{HLL}) with $\lambda=\beta^2/16$ as it should be. The Hamiltonian is completely fixed apart an inessential additive constant.

\section{Non-relativistic limit of the Bullough-Dodd model} \label{BD}

The easiest integrable theory after the Sinh-Gordon model is the Bullogh-Dodd (BD) model \cite{AFZ1983,DB1977,ZS1979,FMSBD,FLZZ}.  
This integrable model is described by a relativistic Lagrangian with only one real field
\begin{equation}
\mathcal{L}_{\text{BD}}\,=\,\frac{1}{2}\partial_\mu\phi\partial^\mu\phi-\frac{m^2c^2}{6 g^2}\left(:e^{2 g \phi}:+:2e^{-g \phi}:-3\right) \,\,\,. 
\end{equation}
In particular, as discussed in Section \ref{integrability}, the ShG and BD models are the only integrable QFT whose action is written in terms of a single bosonic field \cite{D1997}. This integrable field theory is described by only one stable particle whose scattering matrix is \cite{AFZ1983} 
\begin{equation}
\mathcal{S}_{\text{BD}}\,=\,f_{\frac{2}{3}}(\theta)f_{-\frac{{\mathcal B}}{3}}(\theta)f_{\frac{{\mathcal B}-2}{3}}(\theta)\,\,\,,
\end{equation}
where
\begin{equation}
{\mathcal B}\,=\,\frac{c g^2}{2\pi}\frac{1}{1+\frac{c g^2}{4\pi}}\,\,\,,
\end{equation}
and the functions $f_x(\theta)$ were defined in eq.\,(\ref{fxfunction}). With the experience gained on the NR of the ShG model, let's see what happens taking the double limit $c\to\infty$ and $g \rightarrow 0$, with $g c=\beta$ kept constant in the BD model. We anticipate that the result is going to be once again the LL model, as it can be easily guessed by the NR limit of the scattering matrix
\begin{equation}
\lim_{\text{NR}}\mathcal{S}_{\text{BD}}\,=\,\frac{k-i\frac{\beta^2}{6}m}{k+i\frac{\beta^2}{6}m}\label{BD29}
\end{equation}
that matches the expression given in eq.\, (\ref{SmatrixLLM}) provided we choose $\lambda = \beta^2/12$. This result strongly suggests the LL model is also the NR limit of the BD model, but in order to prove it, we must consider also the NR limit of the dynamics. Differently from the ShG case, the NR limit of the dynamics obtained from the action gives an incorrect result and the proper limit must be taken from the equations of motion, as we are going to show explicitly. 

\vspace{2mm}
\noindent
{\bf Incorrect NR limit got from the action}. If we blindly follow the steps that brought us from the relativistic ShG action to the non relativistic LL action, we will see  that this gives rise to an inconsistent result with the limit of the scattering matrix obtained above. Indeed, after we split the field once again in its modes as in 
(\ref{LL11}), following the previous strategy we expand the BD action and retain only the non vanishing terms in the $c\to\infty$ limit:
\begin{equation}
S_{\text{BD}}\,=\,\int dxdt\; :\frac{1}{2}\partial_\mu\phi\partial^\mu\phi:-\frac{m^2c^2}{2}:\phi^2:-cm^2\frac{\beta}{6}:\phi^3:- m^2\frac{\beta^2}{8}:\phi^4:+\mathcal{O}(c^{-1})\label{BD30}
\end{equation}
where $\beta=g c$. Notice that a new cubic term is present in this expression, in contrast to the ShG case (\ref{LL14}). Moreover this term is divergent 
when $c\rightarrow \infty$. On the other hand, though, if we plug the expansion (\ref{LL11}) of the field in the above action, the cubic term seems to not contribute at all, since always gives oscillating terms 
\begin{equation}
c\int_{t_0}^{t_0+\Delta}dt :\left(e^{imc^2t}\psi^\dagger+e^{-imc^2t}\psi\right)^3:\sim c^{-1}
\end{equation}
Therefore, it is tempting to neglect the cubic term in (\ref{BD30}). Proceeding in this way, the NR limit of the BD action gives 
\begin{equation}
\lim_{\text{NR}}S_{\text{BD}}=\int dxdt\; \frac{i}{2}\left(\partial_t\psi^\dagger\psi-\psi^\dagger\partial_t\psi\right)-\frac{1}{2m}\partial_x\psi^\dagger\partial_x\psi-\frac{3}{16} \beta^2\psi^\dagger\psi^\dagger\psi\psi \,\,\,. 
\end{equation}
Even though we can recognize in the above a NR action associated with a LL model, the value of the coupling is not the correct one to match with the NR limit of the scattering matrix (\ref{BD29}), where the coupling appearing in the LL model (\ref{HLL}) turns out to be $\lambda=\beta^2/12$. This discrepancy is due to a subtle role of the cubic term of the action (\ref{BD30}): its effect cannot be neglected and renormalizes the density-density interaction of the LL model, as explained below. 

\vspace{2mm}
\noindent
{\bf Correct NR limit got from the equation of motion}
In light of the discrepancy found between the NR limit of the scattering matrix and the action, let's proceed more carefully and consider the NR limit employing directly the equation of motion. We will see that, in this picture, the role of the cubic term in (\ref{BD30}) is correctly taken in account and the NR limit of the dynamics matches with that of the scattering matrix. Consider then the Heisenberg equation of motion for the BD model
\begin{equation}
\frac{1}{c^2}\partial_t^2\phi\,=\,\partial_x^2\phi-\frac{m^2c^2}{3 g}:\left(e^{2 g\phi}-e^{-g \phi}\right):
\end{equation}
As in the ShG case, we power expand the equation of motion keeping only the non vanishing terms
\begin{equation}
\frac{1}{c^2}\partial_t^2\phi=\partial_x^2\phi-m^2c^2\phi-cm^2\frac{\beta}{2}:\phi^2:-m^2\frac{\beta^2}{2}:\phi^3:\label{BD34}
\end{equation}
Using (\ref{LL11}) and recasting the above in an integral equation we find
\begin{eqnarray}
\nonumber&&i\psi^\dagger(t_0+\Delta)=i\psi^{\dagger}(t_0)+\epsilon\int_{t_0}^{t_0+\Delta}dt\;\left[e^{-i2mc^2(t-t_0)}i\partial_t\psi+\frac{1}{2m}\partial_{x}^2\left(\psi^\dagger+e^{-i2mc^2(t-t_0)}\psi\right)+ \right.\\
\nonumber&& -\frac{c \beta\sqrt{m}e^{-imc^2(t-t_0)}}{4\sqrt{2}}:\left(e^{imc^2(t-t_0)}\psi^\dagger+e^{-imc^2(t-t_0)}\psi\right)^2:+\\
&&\left.-\frac{\beta^2e^{-imc^2(t-t_0)}}{8}  :\left(e^{imc^2(t-t_0)}\psi^\dagger+e^{-imc^2(t-t_0)}\psi\right)^3:\right]\label{BD35}
\end{eqnarray}
We can now recursively solve the above and obtain an expansion similar to (\ref{LL22}), but the presence of a $c$ divergent factor in the equation of motion changes the $c$ power counting
\begin{equation}
i\psi^{\dagger}(t_0+\Delta)=i\psi^{\dagger}(t_0)+\sum_{n,j}\epsilon^n\Delta^{n-j} \sum_{a=0}^{n}c^{a-2j}\mathcal{O}_{n,j,a}(t_0)\label{BD36} \,\,\,, 
\end{equation}
where now the index $a$ counts how many times the $c-$divergent cubic term is used in the iterative solution up to order $n$. 
As the scattering matrix admits a well-defined NR limit, we expect that also this expression will be well-behaved. Based on the power counting, one must have then that $\mathcal{O}_{n,j,a}=0$ whenever $a-2j>0$. In this case, we can focus on the terms proportional to $ \Delta$ in (\ref{BD36}). As a matter of fact, through the same steps that led us from (\ref{LL22}) to (\ref{LL25}),
one can see that these are the relevant ones for the NR equations of motion. Inspecting Eq. (\ref{BD36}), one immediately realizes that terms linear in $\Delta$ can appear both for $n=1$ and $n=2$: therefore, unlike the ShG case, there could be important terms hidden in the second order iterative solution of (\ref{BD35}). A rather lengthy, but straightforward, computation of (\ref{BD35}) up to the second iteration shows that there are not $c$-divergent terms (as it should be) and the presence of a new term linear in $\Delta$ at the second iterative solution. In particular, the new term comes from the cubic order in the Taylor expansion of the BD potential and it is associated with $\mathcal{O}_{2,1,2}(t_0)$. We leave the technical computations to Appendix \ref{A} where we derive the final NR Hamiltonian from the equation of motion 
\begin{equation}
H=\int dx \; \frac{\partial_x\psi^\dagger\partial_x\psi}{2m}+\frac{\beta^2}{12}\psi^\dagger\psi^\dagger\psi\psi\label{BD38} \,\,\,. 
\end{equation}
This Hamiltonian is indeed the one of the Lieb Liniger model (\ref{HLL}) with the choice $\lambda=\beta^2/12$, consistently to what we got from the NR limit of the scattering matrix in the BD model. 

Therefore, we can conclude that the NR limit of the BD theory is once again the LL model, as for the ShG case. In the next Section we are going to show that the same happens for all other Toda Field Theories which, in the NR limit, end up to be  a set of decoupled LL models.

\section{Non-relativistic limit of the Toda Field Theories} \label{Todanon}

The NR limit of the dynamics can be worked out for all the Toda theories at once through a simple generalization of what has been done in Section \ref{BD} for the Bullogh Dodd model, i.e. passing through the equations of motion. Here we present only the main steps of the procedure. 
First of all, we remove the original coupling $g$ through the definition $\beta=g c$, and then we consider the Heisenberg equation of motion for 
the fields $\phi^a$, in the basis in which the mass term is diagonal 
\begin{equation}
\frac{1}{c^2}\partial_{t}^2\phi^a=\partial_x^2\phi^a-m^2c^2\mu_a^2\phi^a-m^2\sum_{j=2}^{\infty}\frac{\beta^{(j-1)}c^{3-j}}{j!}C^{(j+1)}_{a,a_1,...,a_j}:\phi^{a_1}...\phi^{a_j}:\label{TD41}
\,\,\,
\end{equation}
The equations of motion are readily derived from the expansion of the Lagrangian (\ref{todaexp}), where we renormalized the mass parameter (\ref{ren}) through the normal ordering.
Next, we split all the fields in modes
\begin{equation}
\phi^a(t,x)\,=\,\frac{1}{\sqrt{2m_a}}\left(e^{im_ac^2t}\psi_a^{\dagger}(t,x)+e^{-im_ac^2t}\psi_a(t,x)\right)\,\,\,,
\end{equation}
where $m_a=m\mu_a$. By mean of canonical commutation rules for the fields $\phi^a$ and their conjugate momenta, we find that in the NR limit the fields $\psi_a$ are independent NR bosonic species 
\begin{equation}
[\psi_a(x),\psi_b(y)]\,=\, 0
\,\,\,\,\,\, 
, 
\,\,\,\,\,\,\,
[\psi_a(x),\psi^\dagger_b(y)]\,=\,\delta_{a,b}\delta(x-y)
\,\,\,.
\label{CRpsia}
\end{equation}
Proceeding now as in Section \ref{BD}, first we neglect all the terms that will be suppressed in the limit $c\rightarrow \infty$, arriving in this way to the simplified equation of motion 
\begin{equation}
\frac{1}{c^2}\partial_{t}^2\phi^a=\partial_x^2\phi^a-m^2c^2\mu_a^2\phi^a-m^2\frac{c \beta }{2}C^{(3)}_{a,a_1,a_2}:\phi^{a_1}\phi^{a_2}:-m^2\frac{\beta^{2}}{3!}C^{(4)}_{a,a_1,...,a_j}:\phi^{a_1}\phi^{a_2}\phi^{a_3}:\label{TD43}
\end{equation}
Notice the presence, as it was already the case for the BD model, of a $c$ divergent term for the three-couplings. Inserting in eq.\,(\ref{TD43}) the NR fields $\psi_a$ and converting this equation in an integral equation analogue to eq.\,(\ref{BD35}), we can essentially repeat the same analysis with the same conclusion, namely that the three-couplings $C^{(3)}$ contributes to the NR limit through a two steps process. 

Of course, for the case of general Toda Field Theories, the presence of different masses implies extra technicalities: for example the absence of a $c$ divergent term in the iterative solution of (\ref{TD43}) is not completely obvious. Consider the first order iterative solution of (\ref{TD43}): the possible troublesome term comes from the time integration related to $C^{(3)}$:
\begin{equation}
c\int_{t_0}^{t_0+\Delta}dt\; e^{i(-m_a\pm_1 m_{a_1}\pm_2 m_{a_2})c^2(t-t_0)}\;C^{(3)}_{a,a_1,a_2}\psi^{\pm_1}_{a_1}(t_0)\psi^{\pm_2}_{a_2}(t_0)\label{TD44}
\end{equation}
where we adopted the convention $\psi^+=\psi^\dagger$ and $\psi^-=\psi$. Such a term is $c$ divergent if and only if we have mass degeneracy $m_a=\pm_1 m_{a_1}\pm_2 m_{a_2}$, but this -- we know -- is impossible in the Toda theories. As we already mentioned, the three body coupling is proportional to the area of a triangle whose sides are the masses of the involved particles \cite{BCDS1990,CM1989,ChM1990}. The mass degeneracy is incompatible with any of such triangles, therefore if by chance the kinematics condition $m_a=\pm_1 m_{a_1}\pm_2 m_{a_2}$ is satisfied, the relative coupling is zero and (\ref{TD44}) is never divergent
in the limit $c \rightarrow \infty$.

After long but straightforward calculations we can compute the second step in the iterative solution of the equations of motion and then extract the $\sim \Delta$ term, exactly as in the BD model. The resulting NR equation of motion can be written as:
\begin{equation}
i\partial_t\psi^\dagger_a=\frac{1}{2m_a}\partial_x^2\psi^\dagger_a-\frac{\beta^2}{24}\sum_{a_1,a_2,a_3,\pm_1,\pm_2,\pm_3}\frac{1}{\sqrt{\mu_a \mu_{a_1}\mu_{a_2}\mu_{a_3}}}\left[\mathcal{M}_{a,a_1,a_2,a_3}^{\pm_1,\pm_2\pm_3}\; \psi^{\pm_1}_{a_1}\psi^{\pm_2}_{a_2}\psi^{\pm_3}_{a_3}\right]_{\mu_a\pm_1 \mu_{a_1}\pm_2 \mu_{a_2}\pm_3 \mu_{a_3}=0}\label{TD45}
\end{equation}
Above $[...]_{...}$ means we are summing only those terms that satisfy the zero mass condition and the coupling $\mathcal{M}$ is a pure geometric quantity:
\begin{figure}
\begin{center}
\includegraphics[scale=0.5]{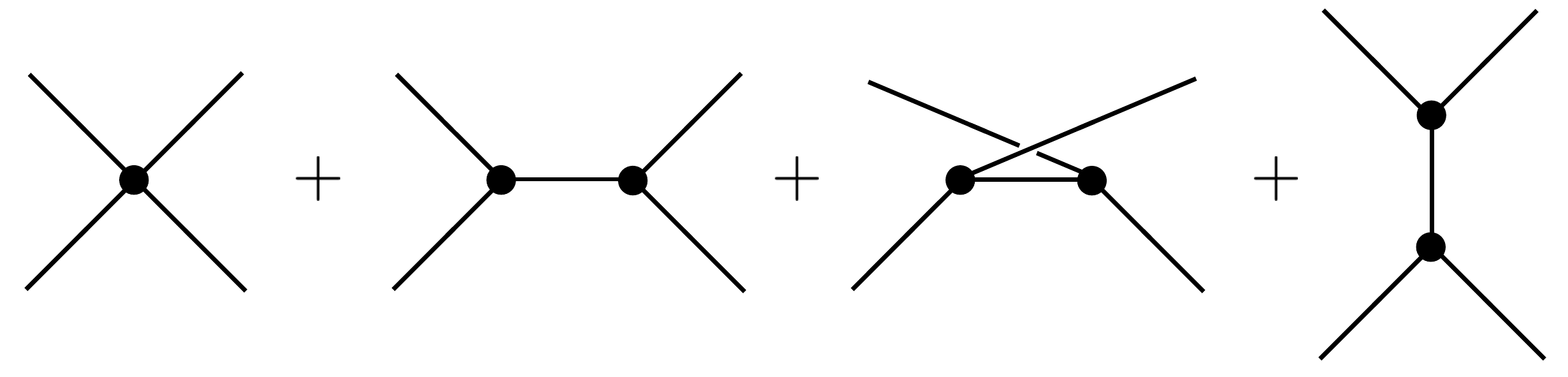}
\caption{\emph{Representation of the coupling $\mathcal{M}$ of (\ref{TD46}) through Feynman graphs.}}\label{scattering}
\end{center}
\end{figure}
\begin{equation}
\mathcal{M}^{\pm_1,\pm_2,\pm_3}_{a,a_1,a_2,a_3}=C^{(4)}_{a,a_1,a_2,a_3}+\sum_{b}\left[\frac{C^{(3)}_{a,a_{3},b}C^{(3)}_{b,a_{1},a_{2}}}{(\mp_{1}\mu_{a_1}\mp_{2}\mu_{a_2})^2-\mu_{b}^2}+\frac{C^{(3)}_{a,a_{1},b}C^{(3)}_{b,a_{2},a_{3}}}{(\mp_{2}\mu_{a_2}\mp_{3}\mu_{a_3})^2-\mu_{b}^2}+\frac{C^{(3)}_{a,a_{2},b}C^{(3)}_{b,a_{3},a_{1}}}{(\mp_{3}\mu_{a_3}\mp_{1}\mu_{a_1})^2-\mu_{b}^2}\right]\label{TD46}
\end{equation}
Equation (\ref{TD45}) is still quite involved, but it can be immediately simplified if $\mathcal{M}$ is recognised to be a scattering amplitude. Notice, indeed, 
that (\ref{TD46}) coincides with the sum of the tree level scattering amplitudes of Figure \ref{scattering}, these graphs computed at zero rapidity and in 
adimensional units, where the particle $a$ is outgoing and the particles $a_j$ are outgoing/ingoing depending on the sign $\pm_j$, as the zero mass condition in (\ref{TD46}) suggests. This feature is already present in the Bullogh Dodd model, as discussed in Appendix \ref{A}. For the integrability of the original Toda Field Theories, scattering events do not change the ingoing and outgoing particles and therefore most of the $\mathcal{M}$ amplitudes are indeed zero. In this way, we arrive to the final NR equations of motion relative to a density-density interaction:
\begin{equation}
i\partial_t\psi^\dagger_a=\frac{1}{2m_a}\partial_x^2\psi^\dagger_a-\frac{\beta^2}{8}\sum_{a,a'}\frac{\Lambda_{a,a'}}{\mu_a \mu_{a'}}\;\psi_{a}^\dagger\psi^\dagger_{a'}\psi_{a'}\label{TD47}
\end{equation}
\begin{equation}
\Lambda_{a,a'}=C^{(4)}_{a,a,a',a'}+\sum_{b}\left[\frac{[C^{(3)}_{a,a',b}]^2}{(\mu_{a'}-\mu_{a})^2-\mu_{b}^2}+\frac{[C^{(3)}_{a,a',b}]^2}{(\mu_{a}+\mu_{a'})^2-\mu_{b}^2}-\frac{C^{(3)}_{a,a,b}C^{(3)}_{b,a',a'}}{\mu_{b}^2}\right]\label{TD48}
\end{equation}
The NR Hamiltonian associated to these equations of motion, from which (\ref{TD47}) can be derived, is readily written as 
\begin{equation}
H\,=\,\int dx\;\left[\sum_{a}\frac{\partial_x\psi^\dagger_a\partial_x\psi_a}{2m_a}+\frac{\beta^2}{16}\sum_{a,a'}\frac{\Lambda_{a,a'}}{\mu_a \mu_{a'}}\psi^\dagger_a\psi^\dagger_{a'}\psi_a\psi_{a'} \right] \,\,\,. 
\label{TD49}
\end{equation}
Hence, it seems that the NR limit of the Toda Field Theories consists of a set of bosons of different species, coupled together through a density-density interaction. It remains though to check whether some of the couplings $\Lambda_{a,a'}$ vanish. Since in the Toda Field Theories the scattering is purely transmissive, such a feature must be also true in the NR limit. However, as shown in Appendix \ref{AppendixA}, an inter-species density-density interaction such as in the Hamiltonian (\ref{TD49}) is never purely transmissive and therefore, to be consistent with the scattering of the Toda theories, different species must be necessarily decoupled. This can be explicitly checked in the $A_n$ and $D_n$ series, since in \cite{SZ1992} the tree level scattering amplitude has been calculated: specialising the result of this paper to the zero rapidity case, one can check that the coupling constants in (\ref{TD49}) are indeed diagonal $\Lambda_{a,a'}\propto \delta_{a,a'}$. 
In \cite{ChM1990} the tree level diagonal scattering amplitude at zero rapidity has been computed for all simply-laced Toda Field Theories and this leads to the expression
\begin{equation}
\Lambda_{a,a'}\,=\,\frac{2\mu_a^2}{h}\delta_{a,a'}\,\,\,. 
\end{equation}
With this extra piece of information, we finally arrive to the conclusion that the Hamiltonian coming from the NR limit of the Toda Field Theories consists of 
a set of $r$ decoupled Lieb-Liniger models, all with the same interaction but different masses 
\begin{equation}
H\,=\,\int dx\;\sum_{a}\left[\frac{\partial_x\psi^\dagger_a\partial_x\psi_a}{2m_a}+\frac{\beta^2}{8h}\psi^\dagger_a\psi^\dagger_a\psi_a\psi_a\right]\label{TD52}\,\,\,.
\end{equation}

\subsection{NR limit of the scattering matrix of the $A_n$ series}
\label{TDS}
It is interesting to explicitly check that the Toda Field Theories reduce to decoupled Lieb-Liniger models by considering the NR limit of the scattering matrix. 
Let's present in detail this computation for the $A_n$ series, starting from the expression of their scattering amplitudes \cite{AFZ1983}:
\begin{equation}
\mathcal{S}_{\text{TD}}^{aa'}(\theta)=\prod_{j=0}^{a-1}\prod_{j'=0}^{a'-1}\mathcal{S}_{\text{TD}}^{11}\left(\theta+\frac{i\pi(a-2j)}{n+1}-\frac{i\pi(a'-2j')}{n+1}\right)\label{TD53}
\end{equation}
where
\begin{equation}
\mathcal{S}_{\text{TD}}^{11}(\theta)\,=\,\frac{\sinh\left(\frac{\theta}{2}+\frac{i\pi}{n+1}\right)\sinh\left(\frac{\theta}{2}-\frac{i\pi}{n+1}+\frac{i B}{2}\right)\sinh\left(\frac{\theta}{2}-\frac{i B}{2}\right)}{\sinh\left(\frac{\theta}{2}-\frac{i\pi}{n+1}\right)\sinh\left(\frac{\theta}{2}+\frac{i\pi}{n+1}-\frac{i B}{2}\right)\sinh\left(\frac{\theta}{2}+\frac{i B}{2}\right)}\label{TD54}
\end{equation}
$\theta$ is the relative rapidity of the two particles and $B$ is the function given in eq.\,(\ref{Bfunction}), with $h = n+1$. The particle momentum is 
\begin{equation}
k_{a}\,=\,\mu_{a}\tilde{m}c\sinh(\theta_{a})\,\,\,,
\end{equation}
where $\tilde{m}$ is a coupling dependent mass scale such that $\lim_{\text{NR}}\tilde{m}= m$. Therefore in the NR limit we have
\begin{equation}
\theta_{a} - \theta_{a'} \simeq \frac{1}{c}\left(\frac{k_{a}}{m_{a}}-\frac{k_{a'}}{m_{a'}}\right)\,=\,\frac{v}{c}\,\,\,,
\end{equation}
where $v$ is the relative velocity of the particles. The NR limit of the relativistic scattering matrix acquires Galilean invariance. 

Plugging now (\ref{TD54}) in (\ref{TD53}), we see that the scattering matrix becomes a product/ratio of terms $\sinh(\theta/2+i\chi_{j,j'})$ and $\sinh(\theta/2\pm i B/2+i\chi_{jj'}')$ with $\chi_{jj'}$, $\chi_{jj'}'$ numbers independent from $B$ and $\theta$. In the non relativistic limit, both $\theta$ and $B$ go to zero, therefore in order that $\mathcal{S}_{\text{TD}}^{aa'}$ is not trivially constant we should have that at least one of the constants $\chi_{jj'}$, $\chi'_{jj'}$ is not zero. This means that it must be fulfilled at least one of these requirements 
\begin{equation}
a-a'=2(j-j')\,\,\, ,\hspace{2pc}a-a'=2(j-j')+2\\,\,\, , \hspace{2pc}a-a'=2(j-j')-2\label{TDS58}
\,\,\,
\end{equation}
otherwise $\mathcal{S}_{\text{TD}}^{jj'}$ in the NR limit becomes simply a constant. In particular, it is easy to see that if $a-a'$ is an odd number, the NR limit of the scattering matrix is a constant. Consider then the case in which $a-a'$ is even and, without loss of generality, consider the case $a\ge a'$. The condition (\ref{TDS58}) permits to find the contributions that in the NR limit retain a non trivial dependence on the relative velocity 
\begin{equation}
\mathcal{S}_{\text{TD}}^{aa'}\to \text{const.}\prod_{j\ge 0,\frac{a-a'}{2}-1}^{j\le a-1,\frac{a+a'}{2}-2}\frac{v+i\frac{\beta^2}{4(n+1)}}{v}\prod_{j\ge 0, \frac{a-a'}{2}+1}^{j\le a-1,\frac{a+a'}{2}}\frac{v}{v-i\frac{\beta^2}{4(n+1)}}\prod_{j\ge 0,\frac{a-a'}{2}}^{j\le a-1,\frac{a+a'}{2}-1}\frac{v-i\frac{\beta^2}{4(n+1)}}{v+i\frac{\beta^2}{4(n+1)}}
\end{equation}
It is evident that there will be many cancellations. With a little thought, it is easy to see that if $a'\le a-2$ all products have the same number of terms. Therefore they cancel each others and the NR limit of the scattering matrix is just a constant. We already said that if $a-a'$ is odd, then the NR limit is again a constant: therefore, the only non trivial scattering is the diagonal $a=a'$. In this case we have 
\begin{equation}
\lim_{\text{NR}}\mathcal{S}_{\text{TD}}^{aa}\propto\frac{v-i\frac{\beta^2}{4(n+1)}}{v+i\frac{\beta^2}{4(n+1)}}\,\,\,.
\end{equation}
The proportionality constant in the diagonal scattering, as well the constant value of the off diagonal case, is fixed from the unitarity condition to be $\pm 1$: in principle we could fix the sign through a more detailed computation of the NR limit, but we simply notice that for zero interaction we must have free bosons. This fixes the constant to be $1$. To summarize, in the NR limit the only non trivial matrix elements are the diagonal one and their reduce to those of a LL model:
\begin{equation}
\lim_{\text{NR}}\mathcal{S}_{\text{TD}}^{aa}=\frac{k-i\frac{m_{a}\beta^2}{4(n+1)}}{k+i\frac{m_{a}\beta^2}{4(n+1)}} \,\,\,. 
\end{equation}
Above we have parametrized the scattering matrix in terms of the relative momentum $k$ for an easier comparison with the LL scattering matrix. Such a scattering matrix is perfectly consistent with the Hamiltonian (\ref{TD52}), since in the $A_n$ series the Coxeter number is simply $h=n+1$.

\subsection{General case} 
We have presented the $A_n$ case in order to show in detail all the effects that contribute at the end to have a NR scattering which is only diagonal. 
This conclusion can be reached in full generality, for all simply-laced Toda Field Theory, simply employing the general expression (\ref{generalSmatrixamplitudes}) 
of the $S$-matrix amplitudes of these theories and taking its NR limit. For $c\to\infty$ the relative rapidity is approximatively $\theta\sim \frac{v}{c}$ with $v$ the relative velocity of the two particles: in order to make manifest the NR limit, in the integral of eq.\,(\ref{generalSmatrixamplitudes}) we can make 
the change of variable $\tau=c^{-1}t$, so that 
\begin{equation}
S_{ab}(\theta)\,=\,\exp\left[-i\int_0^\infty\frac{d\tau}{\tau}\left(8\sinh\left(\frac{B c\tau}{2}\right)\sinh\left(\frac{\pi c\tau}{h}-\frac{B c\tau}{2}\right)\left(2\cosh\frac{\pi c \tau}{h} - {\mathcal I}\right)^{-1}_{ab}-2\delta_{ab}\right)\sin(v\tau)\right]
\end{equation}
In the double limit  $c \rightarrow \infty$ and $g \rightarrow 0$, we have that $B c\to \frac{\beta^2}{4h}$. On the other hand, in the limit 
$c\rightarrow \infty$ the matrix $T_{ab} \equiv \left(2\cosh\frac{\pi c \tau}{h}  - {\mathcal I}\right)^{-1}_{ab}$ becomes purely diagonal 
\begin{equation}
T_{ab} \simeq \delta_{ab} \,e^{-\frac{\pi c \tau}{h}}
\,\,\,\,\,\,
,
\,\,\,\,\,\,
c\rightarrow \infty
\end{equation}
and thus the NR limit of the scattering amplitudes becomes purely diagonal 
\begin{equation}
S_{ab}(\theta)\,\to\,\exp\left[i2\delta_{ab}\int_0^\infty\frac{d\tau}{\tau}e^{-\frac{\beta^2}{4h}\tau}\sin(v\tau)\right]\,=\,\exp\left[\delta_{ab}\ln\left(\frac{\frac{\beta^2}{4h}+iv}{\frac{\beta^2}{4h}-iv}\right)\right]\label{stoda4}\,\,\,,
\end{equation}
where the position of the branch cut of the logarithm is fixed along the negative real semiaxis. Thus different particles are decoupled in the NR limit (their scattering matrix is simply $1$) and only the scattering of identical particles survives
\begin{equation}
S_{aa}\to\frac{k - i 2 m_a\frac{\beta^2}{8h}}{k + i 2 m_a\frac{\beta^2}{8h}}\,\,\,. 
\end{equation}
The above parametrization of the NR scattering matrix is written in term of the relative momenta for an easier comparison with the LL scattering matrix: the above scattering describes a collision of two particles of mass $m_a$ and coupling $\beta^2/8h$, consistently with the result from the equations of motion.

\section{$O(N)$ non-linear sigma model and coupled Lieb-Liniger models}
\label{secsigma}

As we showed in the previous section, the remarkable richness of Toda field theories is essentially lost in the NR limit, which consists in a trivial set of decoupled Lieb Liniger models. As anticipated, such a conclusion has indeed a deep reason: it results from a conspiracy between the purely transmissive scattering of Toda theories and the density-density form of the interaction term, which seems to be ubiquitous in all the NR limit. This kind of interaction will always induce a non-vanishing reflection coefficient. Moreover, it is generically true that whenever two species of particles can be distinguished using the conserved quantities of the theory, integrability imposes a purely transmissive scattering. Therefore, the only consistent possibility in such case is that different species end up decoupled in the NR limit.

The most promising way-out to this dungeon is to have an additional symmetry, so that different species have the same eigenvalues under the conserved charges of non zero spin. This in particular requires mass degeneracy among the particles. The simplest candidate that fulfills all these requirements is the $O(N)$ non-linear sigma model, that is to say a integrable relativistic field theory \cite{ZZ} based on $N$ constrained bosonic fields
\begin{equation}
S=\int dxdt\;\sum_j^N \frac{1}{2}\partial_{\mu}\phi_j\partial^\mu\phi_j\hspace{1pc} , \hspace{1pc} 
\sum_{j}^N\phi_j^2=\omega N \label{sgm1} \,\,\,.
\end{equation}
The presence of the constraint makes the model interacting and massive, the mass of the particles (identical for all the species) depends on the parameter $\omega$ in eq.\,(\ref{sgm1}), which has to be properly renormalized to absorb UV divergences. Equivalent formulations of the model rescale the fields in such a way the constraint is normalized to one $\sum_{j}\phi^2_j=1$, then $\omega N$ appears as a coupling constant in front of the action.
The scattering matrix of the model is \cite{Zamsigma} 
\begin{equation}
A_{i}(\theta_1)A_{j}(\theta_2)=\delta_{i,j}\mathcal{S}^1(\theta_1-\theta_2)\sum_{k=1}^{N}A_{k}(\theta_2)A_{k}(\theta_1)+\mathcal{S}^2(\theta_1-\theta_2)A_{j}(\theta_2)A_{i}(\theta_1)+\mathcal{S}^{3}(\theta_1-\theta_2)A_{i}(\theta_2)A_{j}(\theta_1)\label{sgm2}
\end{equation}
where
\begin{equation}
\mathcal{S}^{3}(\theta)=-\frac{1}{N-2}\frac{i2\pi}{\theta}\mathcal{S}^{2}(\theta),\hspace{3pc}\mathcal{S}^1(\theta)=-\frac{1}{N-2}\frac{i2\pi}{i\pi-\theta}\mathcal{S}^{2}(\theta)\label{sgm3}
\end{equation}
\begin{equation}
\mathcal{S}^{2}(\theta)=U(\theta)U(i\pi-\theta),\hspace{3pc}U(\theta)=\frac{\Gamma\left(\frac{1}{N-2}-i\frac{\theta}{2\pi}\right)\Gamma\left(\frac{1}{2}-i\frac{\theta}{2\pi}\right)}{\Gamma\left(\frac{1}{2}+\frac{1}{N-2}-i\frac{\theta}{2\pi}\right)\Gamma\left(-i\frac{\theta}{2\pi}\right)}\label{sgm4}
\end{equation}
The presence of the constraint (\ref{sgm1}) makes the NR limit of this theory much more difficult to be performed when compared to the previous cases, in particular we cannot proceed through the equation of motion any longer, rather we must rely on the path integral. For the seek of clarity, let's first present and comment the results, leaving the technicalities for later. 

From the scattering matrix analysis (see Section \ref{scatmat} below), it is simple to understand that a sensible NR limit can only be obtained in the combined limit $c\to\infty$ and $N\to\infty$, but keeping $Nc^{-1}$ constant. The NR limit of the scattering matrix is compatible with that of multiple species NR bosonic Hamiltonian (\ref{eq:mHLL}) at the integrable point (Yang Gaudin model)
\begin{equation}
H=\sum_j^N \frac{\partial_x\psi^\dagger_j\partial_x\psi_j}{2m}+\lambda\sum_{jj'}^N\psi^\dagger_j\psi^\dagger_{j'}\psi_j\psi_{j'}\label{sgm5}\,\,\,. 
\end{equation}
Matching the scattering matrices requires $\lambda=\pi cN^{-1}$, a quantity that must be finite in the NR limit. The mass of the NR model can be written as a complicated function of the renormalized $\omega$ coupling of (\ref{sgm1}), but more conveniently we can use $m$ as the free parameter and rather consider $\omega$ to be determined from $m$. 

We will indeed confirm that the Hamiltonian (\ref{sgm5}) is the Hamiltonian emerging from the NR of the $O(N)$ model in Section \ref{NRsgmdynamics}, where the NR limit will be studied at the level of correlation functions in the Fourier space, with the ground state of the non linear sigma model sent to the vacuum state of the NR model (thus, the state with zero particles). The correlators of the relativistic theory are mapped in those of the NR model provided we identify the NR fields splitting the relativistic ones in positive and negative frequencies
\begin{equation}
\psi_{j}(k^0,k^1)=\sqrt{2m}\;\phi_{j}(k^0-mc^2,k^1)\;\Theta(mc^2-k^0),\hspace{2pc}
\psi_{j}^{\dagger}(-k^0,-k^1)=\sqrt{2m}\;\phi_{j}(k^0+mc^2,k^1)\;\Theta(mc^2+k^0)\label{sgmbis6}\,\,\,.
\end{equation}
Thus, at the level of correlation functions we have
\begin{equation}
\braket{\prod_i\frac{\psi_{j_i}(k^0_i,k^1_i)}{\sqrt{2m}}\prod_{i'}\frac{\psi_{j_{i'}}^\dagger(q^0_{i'},q^1_{i'})}{\sqrt{2m}}}=\lim_{\text{NR}}\left[\braket{\prod_i \phi_{j_i}(k^0_i-mc^2,k^1_i)\prod_{i'}\phi_{j_{i'}}(-q^0_{i'}+mc^2,-q^1_{i'})}\right]\label{sgmbis7}\,\,\, ,
\end{equation}
where $\Theta$ is the Heaviside Theta function.

The correspondence between the correlation functions will be found through an analysis of the Feynman graphs in the Fourier space: the NR limit of the Feynman graphs of the sigma model reduces to the Feynman diagrams of (\ref{sgm5}). It must be said that the relation in Fourier transform between the $\phi_j$ and $\psi_j$ fields is in perfect agreement with the usual splitting of the relativistic field (\ref{LL11}). In order to see this we can write $\phi_j$ in the coordinate space:
\begin{eqnarray}
\nonumber&&\phi_j(t,x)=\int \frac{d^2k}{(2\pi)^2}e^{ik^0t-ik^1x}\phi_{j}(k^0,k^1)=\int_{k^0>0} \frac{d^2k}{(2\pi)^2}e^{ik^0t-ik^1x}\phi_{j}(k^0,k^1)+\int_{k^0<0} \frac{d^2k}{(2\pi)^2}e^{ik^0t-ik^1x}\phi_{j}(k^0,k^1)=\\
&&=\frac{1}{\sqrt{2m}}\left[e^{imc^2t}\psi_j^{\dagger}(t,x)+e^{-imc^2t}\psi_{j}(t,x)\right]  \,\,\,. \label{new125}
\end{eqnarray}

These results will be presented in more detail in the next subsections. For the time being, let's notice an interesting property of the Hamiltonian (\ref{sgm5}) got in the NR limit. For the $O(N)$ sigma model, its NR limit requires $Nc^{-1}$ to be constant, therefore necessarily $N\to \infty$ and then the Hamiltonian (\ref{sgm5}) in principle describes an \emph{infinite} set of different bosonic species. The requirement of dealing with infinite species is  of course rather unphysical, but we can easily circumvent this issue with the simple observation that the Hamiltonian (\ref{sgm5}) separately conserves the number of particles of each different species. Therefore, we can simply project on states with a \emph{finite} number of different particle species. If we pose a bound to the possible indexes we plug in the correlation functions, for example we impose all the indexes $j_{i},j_{i'}$ that appear in (\ref{sgmbis7}) to be comprised from $1$ to $n$, then we are describing the NR model with $n$ different species, since all the particles associated with fields $\psi_{j>n}$ are absent by construction. In this way from the non linear sigma model we can extract, in the NR limit, Yang Gaudin models with a finite and arbitrary number of different species: in particular we can even get the single species LL model, just posing to zero the density of all other particles.

\subsection{Non relativistic limit of the scattering matrix}
\label{scatmat}
The NR limit of the scattering matrix of the non linear sigma model can be easily studied. When $c \rightarrow \infty$ the rapidities vanish $\theta\sim k/(cm)$ with $m$ the mass of the particle. Considering the relativistic scattering amplitudes (\ref{sgm3}-\ref{sgm4}) it is immediate to understand that the only sensible way to take the NR limit is keeping $Nc^{-1}$ constant while $c\to\infty$. In this case we easily obtain
\begin{equation}
\mathcal{S}^{1}\xrightarrow{\text{NR}}0,\hspace{2pc}\mathcal{S}^{2}\xrightarrow{\text{NR}} \frac{k}{k+i2\pi mcN^{-1}},\hspace{3pc}\mathcal{S}^{3}\xrightarrow{\text{NR}} -\frac{i2\pi mcN^{-1}}{k+i2\pi mcN^{-1}}\label{6}
\end{equation}
Therefore we obtain the following non relativistic scattering
\begin{equation}
A_{j'}(k_1)A_j(k_2)=\frac{k_1-k_2}{k_1-k_2+i2\pi mcN^{-1}}A_j(k_2)A_{j'}(k_1)+\frac{-i2\pi mcN^{-1}}{k_1-k_2+i2\pi mcN^{-1}}A_{j'}(k_2)A_j(k_1)
\end{equation}
This NR scattering is exactly the same of the Yang Gaudin model (\ref{sgm5}) (see Section \ref{secmultiLL}) and strongly suggests the latter as the correct NR limit of the non linear sigma model. However, the mapping of the scattering matrix alone is not enough to safely reach this conclusion, in particular the scattering matrices cannot fix the mapping between the correlation functions of the two models: in order to do this, in the next section we analyse the NR limit of the dynamics.

\subsection{NR limit of the dynamics}
\label{NRsgmdynamics}

As we have already anticipated, the NR limit at the level of the dynamics will be performed at the level of Feynman diagrams.
The Feynman diagrams of the Yang Gaudin model provide a power expansion in terms of the coupling $\lambda=\pi N^{-1}c$, thus it is natural to compare such an expansion with the large $N$ expansion of the non linear sigma model that produces a power series in $1/N$, then of course we must study the NR limit of the latter.
In this perspective, we firstly proceed in reviewing the Feynman rules for the NR bosons (Section \ref{secNRFeynman}), then the large $N$ expansion  of the non linear sigma model (Section \ref{secsigmaFeynman}). Finally in Section \ref{sgmNR} we see how, in the NR limit, the graphs of the sigma model become those of the Yang Gaudin model.
However, the NR mapping is complicated because of the subtle issue of the renormalization that the sigma model must undergo before of taking the proper limit.
In particular, the model requires mass and field strength renormalization \cite{polyakov}: since any order in $1/N$ introduces new divergent counterterms, we are in principle forced to verify the mapping order by order. Because of this issue we verify the NR mapping in the two cases, i.e. at the three level where no renormalization is needed and at $\mathcal{O}(N^{-1})$, where a non trivial renormalization must be introduced.
However we feel confident that this non trivial check, joint with the mapping of the exact (non perturbative) scattering matrix, is enough to identify the bosonic Yang Gaudin model as the NR limit of the non linear sigma model.

\subsubsection{Feynman diagrams of the NR model}
\label{secNRFeynman}

\begin{figure}
\begin{center}
\includegraphics[scale=0.35]{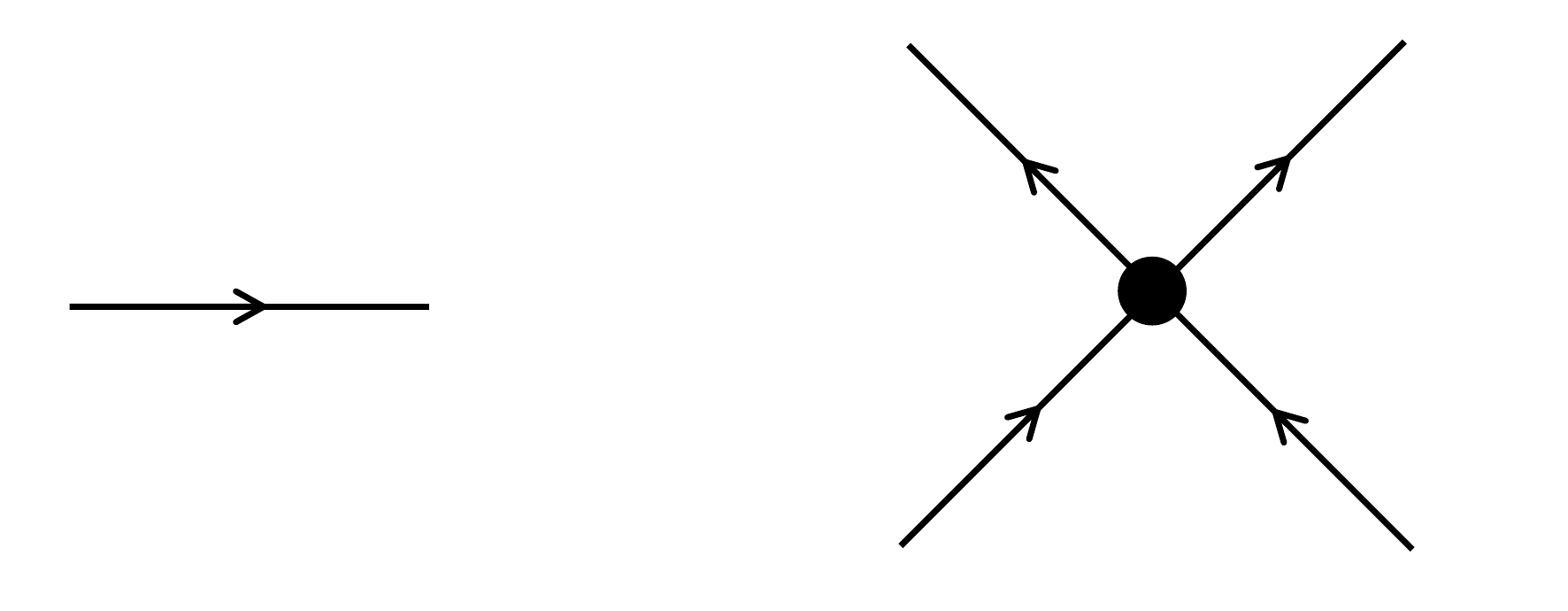}
\caption{\emph{Diagrammatic representation of the propagator and interaction vertex of the Yang Gaudin model.}}\label{figNRprop}
\end{center}
\end{figure}

The Feynman graphs of the NR model (\ref{sgm5}) are easily constructed in terms of a free (retarded) NR propagator $G_{\text{NR}}$ and an interaction vertex involving four fields.
Since $G_{\text{NR}}$ connects a creation field $\psi^\dagger_j$ with an annihilation one $\psi_j$, we must distinguish graphically the two sides of the propagator. Therefore, we represent it with an arrow that flows from $\psi^\dagger_j$ towards $\psi_j$ (Figure \ref{figNRprop}). In the Fourier space we denote the NR propagator as $\tilde{G}_{\text{NR}}$ and its value is (the momentum flow is chosen in agreement with the direction of the arrow):
\begin{equation}
\tilde{G}_{\text{NR}}(k^0,k^1)=\frac{-i}{k^0+\frac{(k^1)^2}{2m}-i\epsilon}\label{NRprop}
\end{equation}

Since the interaction vertex conserves the number of particles, it must have an equal number of incoming and outgoing arrows (Figure \ref{figNRprop}) and its value is simply $-i\lambda$. The fact that the fields in the interaction of (\ref{sgm5}) are normal ordered prevents any self interaction of the vertex.
It must be said that in the Feynman diagrams we should also keep track of the different species, thus is principle we should attach a label on each arrow: the propagators can connect only $\psi^\dagger_j$ with the same species $\psi_j$. The interaction conserves separately the number of each species, thus if the incoming arrows are associated with $(j,j')$ species, even the outgoing arrows are labeled $(j,j')$.

\subsubsection{Large N expansion of the non linear sigma model}
\label{secsigmaFeynman}

We now review the large $N$ expansion for the non linear sigma model \cite{abdalla,polyakov}. The first step is to represent the constraint that appears in the action (\ref{sgm1}) by mean of an auxiliary field $\Lambda$ and an integral representation of the functional Dirac delta:

\be
\delta\left(\sum_j\phi_j^2-\omega N\right)=\int \mathcal{D}\Lambda \;e^{-i\int dxdt\;\frac{\Lambda}{2}\left(\sum_j\phi_j^2-\omega N\right)}\,\,\, .
\ee

This representation of the constraint leads naturally to define an \emph{auxiliary action} with unconstrained bosons and a ghost field
\be
\mathcal{S}_\text{aux}=\int dxdt\;\sum_j \frac{1}{2}\partial_\mu\phi_j\partial^\mu\phi_j -\frac{\Lambda}{2}\left(\sum_j \phi^2_j-N\omega\right)\label{N60}\,\,\, .
\ee

Correlation functions in the non linear sigma model can be equivalently computed with the constrained action (\ref{sgm1}) or with the above unconstrained one at the price of introducing an extra field in the path integral, but the second formulation naturally leads to the large $N$ expansion.
Since the action is quadratic in terms of the bosonic fields, in principle we could perform exactly the path integral over $\phi_j$ and reduce ourselves to a path integral for the single field $\Lambda$, however such an operation makes less transparent the construction of the Feynman rules, so we proceed along another direction.
Notice that the ghost field plays the role of an effective space-time dependent mass for the bosonic field: knowing in advance that the bosons become massive because of the constraint, we shift the ghost field $\Lambda\to \Lambda+m^2c^2$. This operation does not affect the correlation functions of the bosonic fields, but has the advantage of making explicit a constant mass term for the bosons (we discard in the action an inessential additive constant):
\be
\mathcal{S}_\text{aux}=\int dxdt\;\sum_j \frac{1}{2}\left(\partial_\mu\phi_j\partial^\mu\phi_j -m^2c^2\phi_j^2\right)-\frac{\Lambda}{2}\left(\sum_j \phi^2_j-N\omega\right)\,\,\, \label{new131}.
\ee

\begin{figure}[t]
\begin{center}
\includegraphics[scale=0.4]{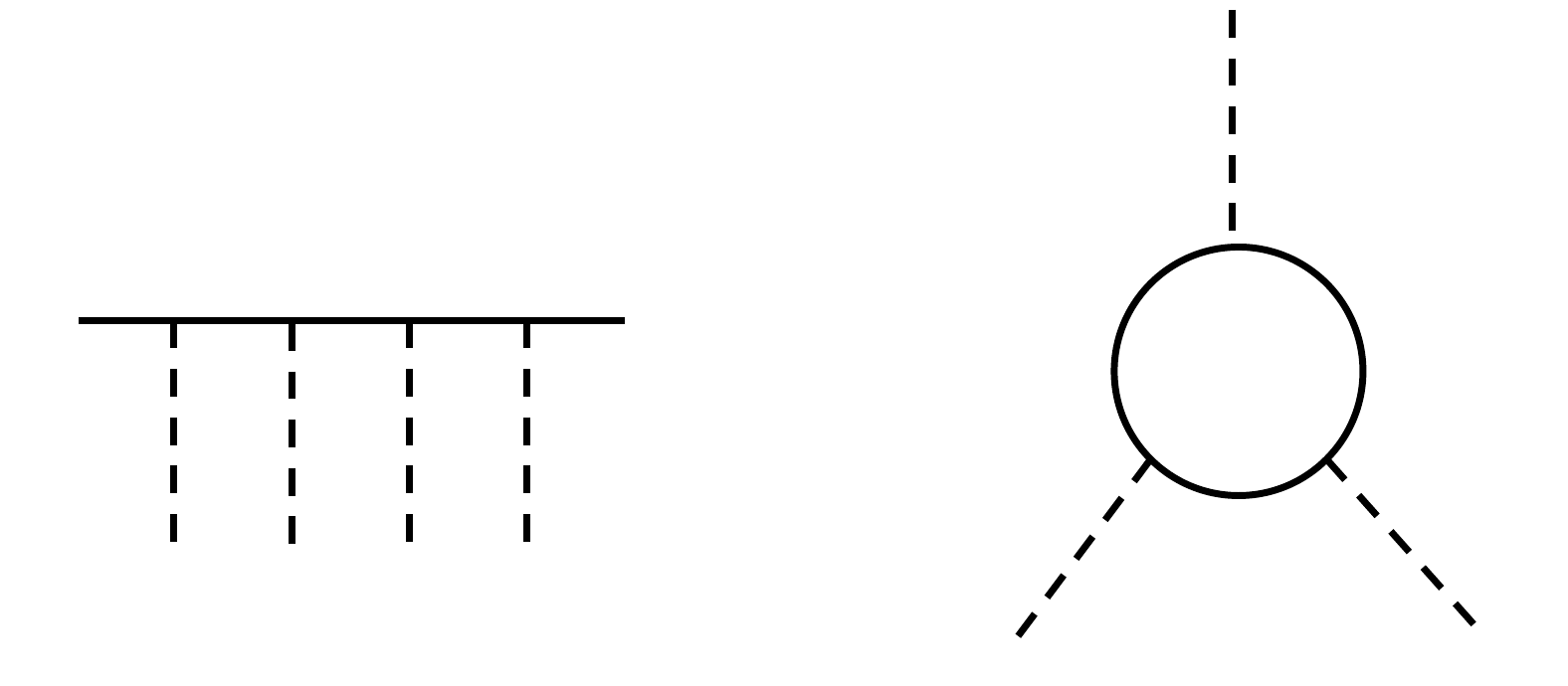}
\caption{\emph{In the Feynman diagrams that describe the path integral in $\phi_j$ at fixed ghost field there are two classes of connected diagrams. The graphs involving external continuous legs (left) and the graphs in which only the external legs of the ghost field appear (right).}}\label{sgmfig1}
\end{center}
\end{figure}
For the time being the mass $m$ is a free parameter that will be fixed soon.
It is convenient to perform the path integral over the $\phi_j$ fields through Feynman graphs, using the massive action to generate the free propagators. The path integral over the ghost field is left for a second moment and at this stage the latter behaves as a classical current coupled with the $\phi_j$ fields. In these Feynman graphs we represent free propagators as continuous lines, the ghost field is a dotted line and, as it is clear from (\ref{new131}), it interacts with two bosonic fields of the same species, but (for the moment) it does not propagate. These Feynman diagrams have two kinds of external legs: continuous external legs represent the $\phi_j$ fields we are considering in the actual correlators to be computed, then  of course we have dotted external legs representing the ``external source" $\Lambda$. Therefore the (connected) Feynman diagrams are naturally organized in two classes: those with external continuous legs and those that have only dotted external legs, as described in Figure \ref{sgmfig1}.
With these Feynman diagrams we describe the integration over the $\phi_j$ fields, but in order to compute the correlators we should also integrate over the ghost field $\Lambda$
\be
\braket{...}=\int \mathcal{D}\Lambda\; e^{i\int dx dt\;\frac{N\omega}{2}\Lambda}\; \; \;\sum\left[\text{Feynman diagrams}\right] \label{N65}\; \; \; ,
\ee
where the Feynman diagrams are of course $\Lambda$ dependent because of the dotted external legs.

\begin{figure}[b]
\begin{center}
\includegraphics[scale=0.5]{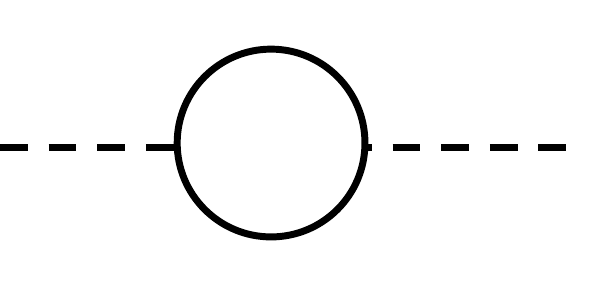}
\caption{\emph{The connected diagram that contributes to the gaussian part of the effective action in the large $N$ limit.}}\label{nfig1}
\end{center}
\end{figure}

Clearly it is impossible to compute exactly the path integral and we must rely on perturbation theory, but in order to do so we need, as a starting point, a gaussian action for the ghost field to generate a $\Lambda$ propagator: we are going to extract it from the sum of the Feynman diagrams.
Among the various connected Feynman diagrams with only dotted external legs (thus associated with only the auxiliary field as ``external current"), we are interested in the graphs with two dotted legs drawn in Figure \ref{nfig1}. Since this graph has only two dotted external lines, it is quadratic in the ghost field: we label as $\mathcal{G}$ both this graph and its value.
We now use the fact that the sum of the Feynman diagrams can be riexponentiated in terms of the connected diagrams \cite{weinberg} to write the sum in (\ref{N65}) as
\be
\sum\left[\text{Feynman graphs}\right]=\sum\left[\text{Feynman graphs without }\mathcal{G}\right]e^{\mathcal{G}}\,\,\, . \label{GFeynman}
\ee
Thus (\ref{N65}) can be written as
\be
\braket{...}=\int \mathcal{D}\Lambda\; e^{i\int dx dt\;\frac{N\omega}{2}\Lambda+\mathcal{G}}\; \; \;\sum\left[\text{Feynman graphs without }\mathcal{G}\right]\label{N67}\,\,\, .
\ee

We can now use the graph $\mathcal{G}$, quadratic in the ghost field, to define a propagator for $\Lambda$
\begin{equation}
\braket{\Lambda(k^\mu)\Lambda(q^\mu)}_0=(2\pi)^2\delta\left(k^\mu+q^\mu\right)\Gamma\left(k^\mu k_\mu\right)\,\,\, ,
\label{LambdaProp}
\end{equation}

where $\Gamma^{-1}$ is simply the value of the loop diagram in Figure \ref{nfig1}:
\begin{equation}
\Gamma^{-1}\left(k^\mu k_\mu\right)=-\frac{N}{2}\int \frac{d^2 q}{(2\pi)^2}\frac{1}{q^\mu q_\mu-m^2c^2+i\epsilon_1}\frac{1}{\left(k^\mu-q^\mu\right)\left(k_\mu-q_\mu\right)-m^2c^2+i\epsilon_2}\label{N72}\,\,\, .
\end{equation}

The factor $N$ comes from the summation over all the species of particles in the internal loop and the contractions are done with the Minkowski metric with restored $c$:
\be
\eta_{\mu\nu}=\begin{pmatrix} \frac{1}{c^2} && 0 \\ 0 &&-1 \end{pmatrix}\,\,\, .
\ee

This integral can be done exactly and it is computed in Appendix \ref{gammaprop}, where we get
\be
\Gamma\left(k^\mu k_\mu\right)=i4\pi c N^{-1} m^2\sqrt{\left(\frac{k_\mu k^\mu}{m^2c^2}-4\right)\frac{k_\mu k^\mu}{m^2c^2}}\left(\log\left[\frac{\sqrt{4-\frac{k_\mu k^\mu}{m^2c^2}}+\sqrt{-\frac{k_\mu k^\mu}{m^2c^2}}}{\sqrt{4-\frac{k_\mu k^\mu}{m^2c^2}}-\sqrt{-\frac{k_\mu k^\mu}{m^2c^2}}}\right]\right)^{-1}\,\,\, .\label{new138}
\ee

Thanks to this propagator we can now perturbatively compute the path integral in the ghost field and find the final Feynman rules to compute the correlators of the bosonic fields $\phi_j$.
In particular, the new Feynman diagrams are constructed from the old ones simply letting the auxiliary field (dotted line) to propagate, thus the final Feynman diagrams can be constructed out of these different objects:

\begin{enumerate}
\item Massive bosonic standard propagators (continuous lines).
\item The propagators of the ghost field (dotted lines).
\item The interaction vertexes that can be read from the the auxiliary action (\ref{new131}). One vertex couples a dotted line with two continuous lines, instead the interaction proportional to $\omega$ involves a single dotted line and no continuous lines.
\item Since we are interested in the correlators of the physical fields, the dotted lines appear only as internal propagators in the Feynman diagrams.
\item An extra recipe is needed: since we used the graphs of Figure \ref{nfig1} to construct the propagator of the ghost field, these should not enter in the final Feynman graphs (their contribution has already been resummed in the dotted propagator), thus we should remove them by hand from the Feynman diagrams.
\end{enumerate}

\begin{figure}[t]
\begin{center}
\includegraphics[scale=0.35]{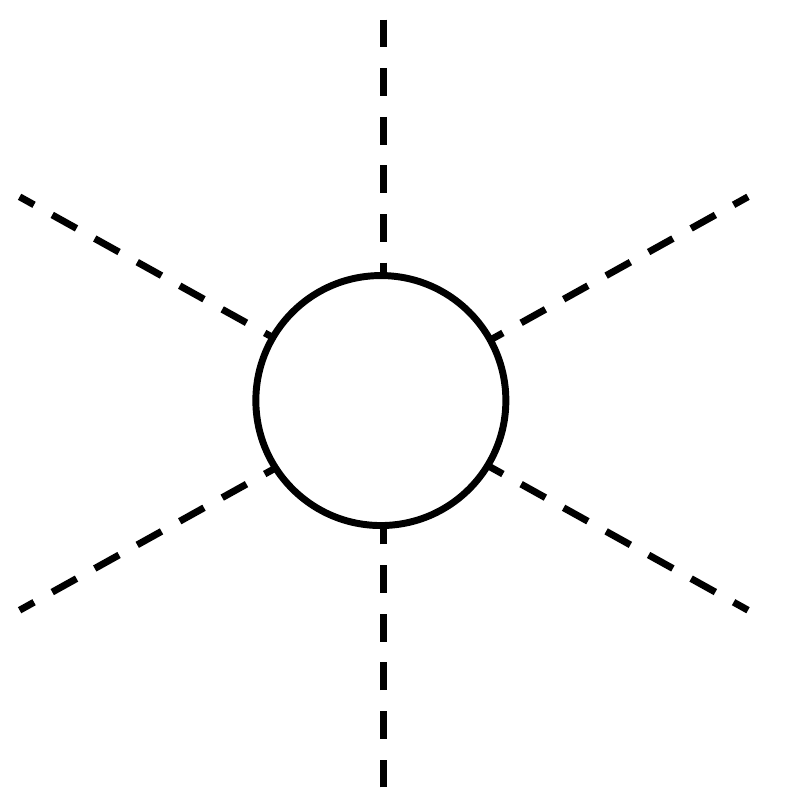}
\caption{\emph{The integration over the internal loops of continuous lines gives a $N$ factor.}}\label{nfig2}
\end{center}
\end{figure}
\begin{figure}[t]
\begin{center}
\includegraphics[scale=0.25]{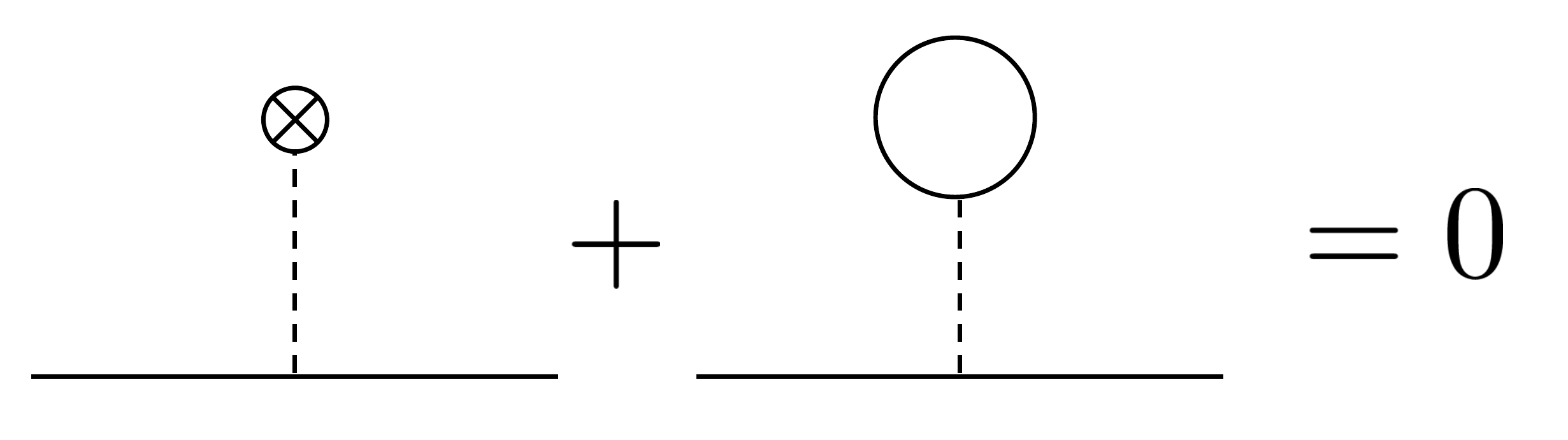}
\caption{\emph{Cancellation of $N$ divergent graphs.}}\label{nfig3}
\end{center}
\end{figure}

In terms of these Feynman diagrams we can readily exploit the large $N$ limit as follows. 
$N$ divergent contributions to the Feynman graphs come from closed loops of continuous lines (namely the internal loops of Figure \ref{nfig2} with an arbitrary number of departing dotted legs). Each closed loop gives an $N$ factor coming from the sum of the internal continuous propagators over the different indexes of $\phi_j$.
On the other hand the $\Gamma$ (\ref{new138}) propagator carries a factor $1/N$, therefore any dotted line counts as $1/N$. Actually it is useful to split the factor $\frac{1}{N}=\frac{1}{\sqrt{N}}\frac{1}{\sqrt{N}}$ and attach each of these factors $\frac{1}{\sqrt{N}}$ at the edges of the dotted lines. With this convention for the power counting, an internal loop with $n$ departing dotted lines contributes as $N^{1-n/2}$, therefore the only loops that are not suppressed in $N$ are those with $n=1$ and $n=2$. However, notice that the loops with two legs ($n=2$) are excluded by the point 5 of the Feynman rules.
We are left with the more dangerous loops with only a departing dotted line, but now we take advantage of the fact that the mass $m$ has not been fixed yet: tuning $m$ properly we can ask the interaction proportional to $\omega$ to cancel the one loop diagrams, as shown in Figure \ref{nfig3}.

This requirement is immediately translated in a mass equation that ties together the coupling $\omega$ with the mass $m$
\be
\omega-\int\frac{d^2k}{(2\pi)^2}\frac{i}{k_\mu k^\mu-m^2c^2+i\epsilon}=0\, \, \, .\label{new139}
\ee

After we impose this mass equation, $\omega$ explicitly disappears from all the Feynman diagrams and it contributes only through the mass $m$.

Notice we can read this equation backward: once we gave the value of $m$ (that is the physical parameter), from the equation we read the positive (infinite) value $\omega$ must attain. 
Now the Feynman diagrams describe a proper large $N$ expansion through the insertion of a progressive number of internal dotted propagators: at the zeroth order of the expansion we get free bosons of mass $m$.

\begin{figure}[b]
\begin{center}
\includegraphics[scale=0.35]{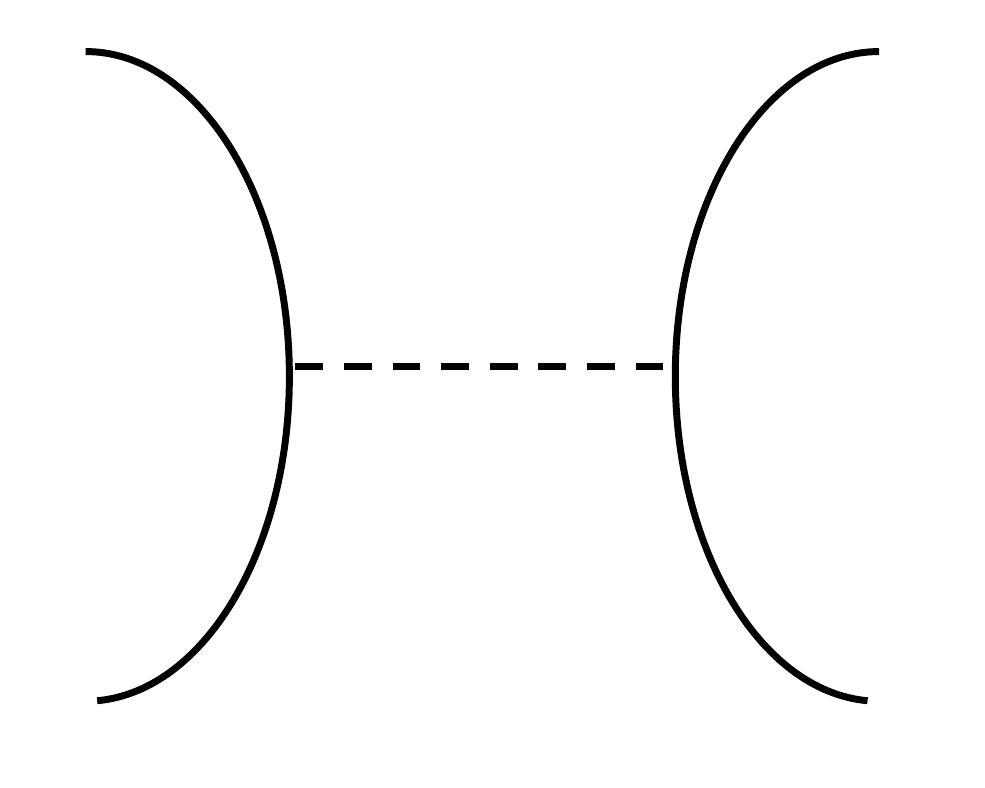}
\caption{\emph{Four field vertex mediated by the ghost field.}}\label{nfig4}
\end{center}
\end{figure}

In particular, we can conveniently think the propagators of the ghost field to define an effective vertex for the $\phi_j$ fields, as depicted in Figure \ref{nfig4}. As a matter of fact, since the dotted lines are never external legs, they can only appear coupled to the continuous lines as in Figure \ref{nfig4}: we can regard such a graph as an effective (non local) vertex of four fields mediated by the ghost field.
Pursuing this interpretation we can write the following effective action (below $\Gamma(x^\mu)$ is the $\Gamma$ propagator written in the coordinate space by mean of a simple Fourier transform):
\be
\mathcal{S}_\text{eff}=\mathcal{S}_\text{free}+\int d^2x d^2y\;\frac{i}{8} \sum_{jj'}\phi_j^2(x^\mu)\phi_{j'}^2(y^\mu)\Gamma(x^\mu-y^\mu)\label{new104}
\ee

Where $\mathcal{S}_\text{free}$ is the massive free action for the bosonic fields $\phi_j$. 
However, this expression is not enough to reproduce  the Feynman diagrams of the large N expansion; this is due to the selection rules on the Feynman diagrams. 
The correct expansion, matching the original one in \eqref{N65}, is recovered adding the following extra rules for the Feynman diagrams extracted from (\ref{new104}):
\begin{enumerate}
\item The mass equation prevents the appearance of closed loops as those of Figure \ref{nfig3}. This exclusion is not implemented in the above effective action, however this is a minor issue, since it is sufficient to add a counterterm proportional to $\sum_j\phi_j^2$ to cancel these graphs (i.e. we add a counterterm equal to the crossed circle in Figure \ref{nfig3}).
\item The large N expansion prevents the appearance of closed loops which would correspond to Figure \ref{nfig1}. Indeed, this diagram was already taken into account in \eqref{GFeynman} to give the propagator \eqref{LambdaProp} to the field $\Lambda$: this selection rule cannot be easily implemented as the previous one and it should be kept in mind in the computation of the Feynman diagrams.
\end{enumerate}

Until now we dealt with the bare action, but the large $N$ expansion of the non linear sigma model presents UV divergences that must be properly renormalized \cite{polyakov}. The renormalization can be attained with a mass renormalization (we can equivalently renormalize the coupling $\omega$ due to the mass transmutation) and the insertion of a field strength counterterm. In particular, the latter amounts to the addition in the effective action (\ref{new104}) of a term proportional to $\sum_j \partial_\mu\phi_j\partial^\mu \phi_j$.

\subsubsection{NR limit of the Feynman diagrams}
\label{sgmNR}

Armed with the large $N$ expansion of the non linear sigma model we can finally study the NR mapping of the dynamics through the correlation functions.
The first step is to show that the free propagator of the putative NR fields defined in (\ref{sgmbis6}) reduces to the NR propagators, as it should be.
The check is immediate: applying the definition (\ref{sgmbis6}) we have
\be
\braket{\psi_{j}(k^0,k^1)\psi_{j'}(q^0,q^1)}_0=2m\Theta(mc^2-k^0)\Theta(mc^2-q^0)\braket{\phi_{j}(k^0-mc^2,k^1)\phi_{j'}(q^0-mc^2,q^1)}_0\,\,\, .
\ee
Then noticing that the relativistic correlator gives an energy conservation law $\delta(k^0+q^0-2mc^2)$ we discover that the above is identically zero, because of the product of the $\Theta$ functions with the $\delta$ over the energies
\be
\Theta(mc^2-k^0)\Theta(mc^2-q^0)\delta(k^0+q^0-2mc^2)=\Theta(mc^2-k^0)\Theta(k^0-mc^2)\delta(k^0+q^0-2mc^2)=0\,\,\, .
\ee

We now consider $\braket{\psi^\dagger\psi}_0$, with the same passages and making explicit the propagator of the relativistic field we get
\be
\braket{\psi_j^\dagger (k^0,k^1)\psi_{j'}(q^0,q^1)}_0=(2\pi)^2\delta_{jj'}\delta\left(k^\mu-q^\mu\right)\Theta(mc^2-k^0)\frac{i2m}{\frac{1}{c^2}(k^0-mc^2)^2-(k^1)^2-m^2c^2+i\epsilon}\,\,\, .\label{new142}
\ee

Then we use the following splitting of the propagator (the identity holds apart from an inessential rescaling of the vanishing regulator $\epsilon$)
\be
\frac{i}{k_\mu k^\mu-m^2c^2+i\epsilon}=\frac{c}{2\sqrt{(k^1)^2+m^2c^2}}\left[\frac{-i}{k^0+c \sqrt{(k^1)^2+m^2c^2}-i\epsilon}+\frac{-i}{-k^0+c \sqrt{(k^1)^2+m^2c^2}-i\epsilon}\right]\,\,\, ,
\ee
then the NR limit of (\ref{new142}) easily follows:
\begin{equation}
\lim_\text{NR}\braket{\psi_j^\dagger (k^0,k^1)\psi_{j'}(q^0,q^1)}_0=(2\pi)^2\delta_{jj'}\delta\left(k^\mu-q^\mu\right)\frac{-i}{k^0+\frac{(k^1)^2}{2m}-i\epsilon}\,\,\, .
\end{equation}

The above expression is in agreement with the NR propagator (\ref{NRprop}), therefore the free part of the effective action (\ref{new104})  correctly reproduces the NR propagators, after the NR limit has been taken.
We now proceed to test the NR mapping at three level in the effective action (\ref{new104}): since at the three level the Feynman graphs do not contain any free momentum to be integrated, the NR limit is easily taken. 
In particular, the additional rule 2) presented below Eq. \eqref{new104},
can be safely ignored at the tree level, as it only concerns loops.
First of all we insert the mode splitting (\ref{new125}) in the interaction of the effective action (\ref{new104}), in this way we obtain several vertexes for the non relativistic fields with oscillating phases.
Now notice that the NR limit of the propagators forces all the momenta (even those of internal propagators) to attain NR values, therefore the only interaction vertexes relevant at three level are those such that all the NR fields of the vertex can attain NR values at the same time. The oscillating phases in the coordinate space mean momentum shifts in the Fourier space: this fact, joint with the overall energy conservation, leads to the conclusion that only the number conserving vertexes (i.e. those of the form $\psi^\dagger\psi^\dagger\psi\psi$) are relevant for the NR limit.
However, in principle two possible indexes structures for the vertex are allowed, i.e.
\be
\int d^2x d^2y\;\frac{i}{2(2m)^2} \sum_{jj'}\psi_j^\dagger(x^\mu)\psi_j(x^\mu)\psi^\dagger_{j'}(y^\mu)\psi_{j'}(y^\mu)\Gamma(x^\mu-y^\mu)\label{new146}
\ee
and
\be
\int d^2x d^2y\;\frac{i}{4(2m)^2} \sum_{jj'}\psi_j^\dagger(x^\mu)\psi^\dagger_j(x^\mu)\psi_{j'}(y^\mu)\psi_{j'}(y^\mu)\Gamma(x^\mu-y^\mu)e^{i2mc^2(x^0-y^0)}\,\,\, .\label{new147}
\ee

Notice that the indexes structure of the vertex in the NR model is the same as the first of the above (i.e. the number of each particle species is separately conserved), on the other hand the Yang Gaudin model does not have an interaction analogue to the second term, that instead allows for tunneling among different species.
In fact, among the two vertexes only the first matters in the NR limit. In order to see this, we Fourier transform the vertexes. Eq (\ref{new146}) becomes
\be
\int \frac{d^8k}{(2\pi)^8}(2\pi)^2\delta\left(k^\mu_1+k^\mu_3-k^\mu_2-k^\mu_4\right)\;\frac{i}{2(2m)^2} \sum_{jj'}\psi_j^\dagger(k^\mu_1)\psi_j(k_2^\mu)\psi^\dagger_{j'}(k_3^\mu)\psi_{j'}(k_4^\mu)\;\Gamma((k_1-k_2)^\mu(k_1-k_2)_\mu)\,\,\, .\label{new148}
\ee

In the NR limit, as we already said, the momenta in the fields are constrained to attain NR values, thus the NR limit of this interaction vertex simply amounts to take the NR limit of $\Gamma$ at fixed momenta $k^\mu_1-k_2^\mu$. Using the explicit expression (\ref{new138}) we get
\be
\lim_\text{NR}\frac{i}{2(2m)^2}\Gamma((k_1-k_2)^\mu(k_1-k_2)_\mu)=-\pi cN^{-1}\,\,\, ,\label{new149}
\ee

thus the NR limit of (\ref{new148}) exactly fits the NR interaction vertex of Section \ref{secNRFeynman}, apart from an $i$ factor. The latter comes from the fact that in the path integral it enters $i$ times the effective action, thus in the Feynman diagrams (\ref{new149}) acquires the extra $i$ needed to match the NR vertex.
It remains to show that in the NR limit the vertex (\ref{new147}) vanishes: moving in the Fourier space we get an expression similar to (\ref{new148}), but the extra phases factor shift the momentum in $\Gamma$ of $2mc^2$. Thus the relevant vertex in the NR limit is
\be
\lim_\text{NR}\;\Gamma\left(\frac{1}{c^2}(k_1^0+k_2^0-2mc^2)^2-(k_1^1+k_2^1)^2\right)=0\, \, \, ,
\ee
as it can be immediately checked from the explicit expression for $\Gamma$ (\ref{new138}).
This last step concludes the desired mapping: the Feynman diagrams at tree level of the effective action (\ref{new104}) are sent to the tree level Feynman graphs of the Yang Gaudin model.

\begin{figure}[t]
\begin{center}
\includegraphics[scale=0.35]{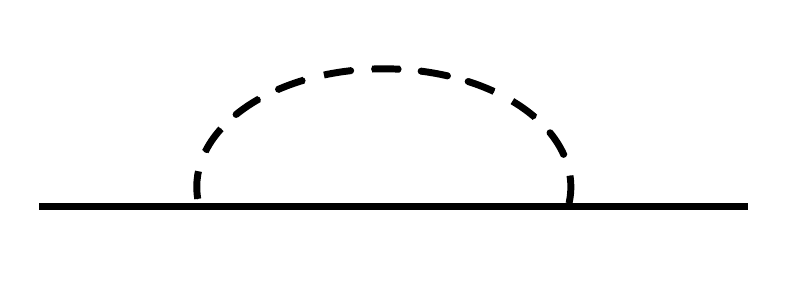}
\caption{\emph{This relativistic graph does not have a NR counterpart, due to the normal ordering in the interaction of the Yang Gaudin model.}}\label{nfignorm}
\end{center}
\end{figure}

This remarkably simple picture get more involved beyond the tree level because of the following reasons:
\begin{enumerate}
\item The relativistic effective vertex is not normal ordered, differently from the NR vertex. 
The unwanted graphs excluded by the normal ordering are obtained making two of the external legs of the effective vertex (Figure \ref{nfig4}) to autointeract. One autointeraction has already been eliminated by the mass equation (Figure \ref{nfig3}), thus we are left with the ``unwanted" graph of Figure \ref{nfignorm} that appears at the $\mathcal{O}(N^{-1})$ expansion of the theory and is of course absent at the tree level. Since Figure \ref{nfignorm} has two legs it is natural to ask ourselves if this graph can be completely removed by a mass (or equivalently the coupling $g$) or of the field strength renormalization. In the relativistic theory an exact cancellation seems to be impossible, because the graph of Figure \ref{nfignorm} has a non trivial momentum dependence. Despite these difficulties it could be that in the NR limit this cancellation occurs: in fact, we are going to explicitly show this fact at $\mathcal{O}(N^{-1})$ of the expansion.

\item The relativistic diagrams obey to an extra ``selection rule" that is irrelevant as long we remain at tree level, but as soon as we allow for loops we must forcefully exclude the loop diagrams of Figure \ref{nfig1}: such a rule is not present in the NR diagrams.
However, such a selection rule is not expected to be relevant in the NR limit: at tree level we showed that only the effective vertex (\ref{new146}) is relevant in the NR limit. Assume this is true beyond tree level and compute the NR counterpart of the internal loop of Figure \ref{nfig1}. Its value is
\be
-N\int \frac{d^2q}{(2\pi)^2}\frac{-i}{q^0+\frac{(q^1)^2}{2m}-i\epsilon_1}\frac{-i}{q^0-k^0+\frac{(q^1-k^1)^2}{2m}-i\epsilon_2}\label{sigma157}\,\,\,,
\ee
where $k$ is the momentum injected in the loop. Notice that the above integral is zero because we are free to deform the integral in $q^0$ in the lower complex plane without embracing any singularity. Therefore, the selection rule we should forcefully impose in the relativistic theory is already naturally implemented in the NR model.

\item The correct procedure to take the NR limit is not on the propagators and vertexes, but on the diagrams themselves. In principle we should consider a whole diagram and take its NR limit, showing it reduces to a diagram of the NR model. 
While at the tree level such a procedure is completely equivalent to take the limit of the building blocks of the diagrams (propagators and interaction vertexes), this is no longer true beyond the tree level because of the presence of UV divergent integrals. In fact, the relativistic theory must be properly renormalized before of taking the NR limit. This is the major obstacle in providing a mapping of the whole set of the Feynman diagrams, since at any order new counterterms are introduced and the theory must be renormalized from scratch.
\end{enumerate}

\begin{figure}[t]
\begin{center}
\includegraphics[scale=0.3]{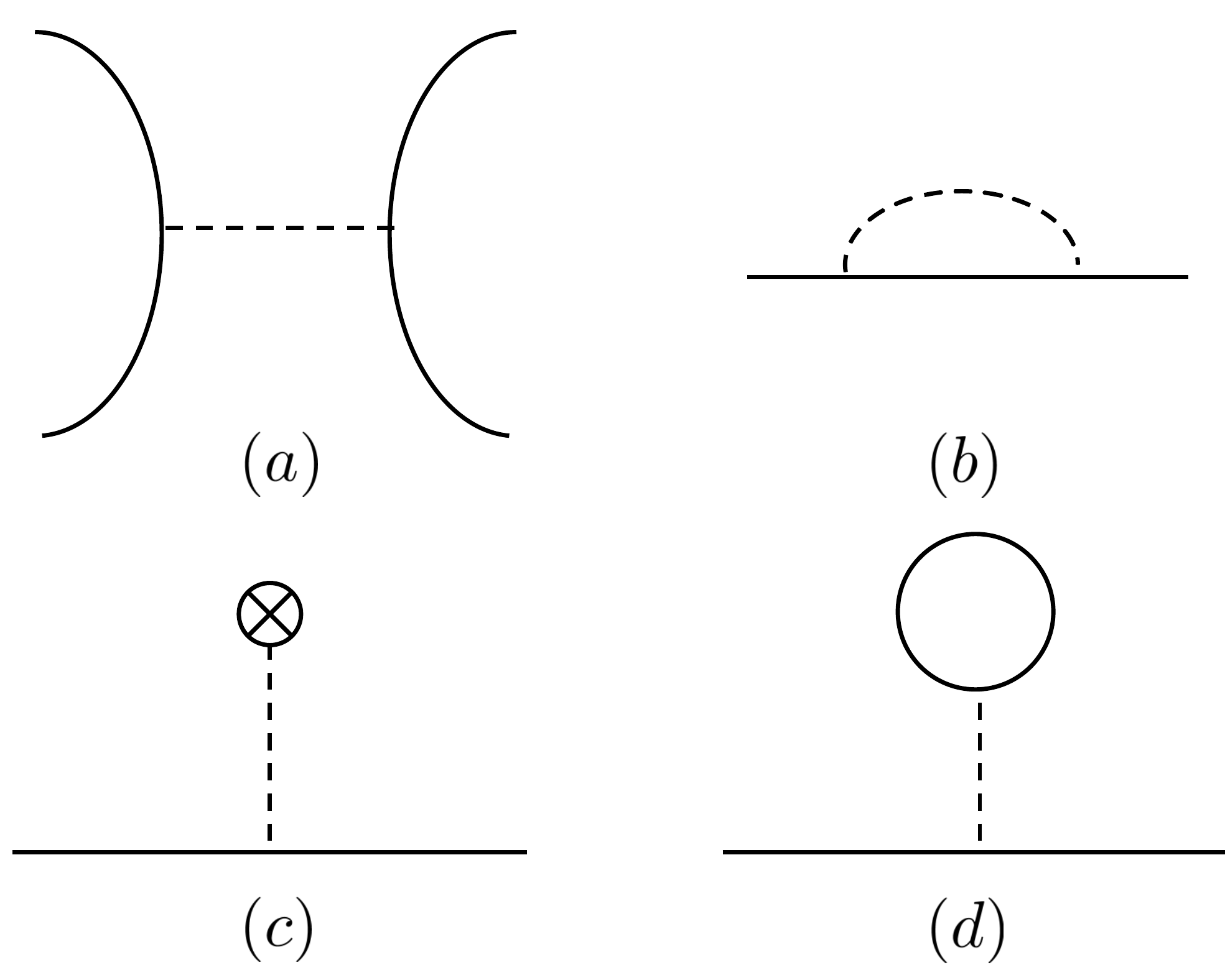}
\caption{\emph{Graphs relevant at $\mathcal{O}(N^{-1})$.}}\label{nfig6}
\end{center}
\end{figure}
To get convinced that these issues do not spoil the NR limit we test the simplest example where the model must be renormalized. Thus we change perspective and instead of considering the Feynman diagrams at tree level, we focus on the $\mathcal{O}(N^{-1})$ diagrams depicted in Figure \ref{nfig6} (here diagrams are not meant to be building blocks for larger diagrams: the external legs are indeed the external legs of the whole Feynman diagram).

At the zeroth order the mass equation makes the $(c)$ and $(d)$ graphs to cancel each other, but the graph $(b)$ (that is $\sim\mathcal{O}(N^{-1})$) is actually divergent and forces us to modify the mass equation, as well as to introduce a field renormalization counterterm.
Of course, the NR limit of the $\mathcal{O}(N^{-1})$ correlators must reduce to the first order expansion in the coupling $\lambda=\pi c N^{-1}$ of the Yang Gaudin model: thus among the graphs of Figure \ref{nfig6} only the four leg vertex $(a)$ must survive in the NR limit. Actually, the fact that the $(a)$ graph reproduces in the NR limit the correct $4$ leg NR graph it is exactly what we showed at tree level (in fact, $(a)$ is a tree-like Feynman diagram). Therefore, it remains to check if the graphs $(b)$, $(c)$ and $(d)$ (or rather their sum) vanish in the NR limit.
These three graphs contribute to the two point correlator $\braket{\phi(k^\mu)\phi(q^\mu)}$. Leaving out the $\delta$ over the momenta, the three graphs contribute as:
\be
(b)=-\int \frac{d^2p}{(2\pi)^2}\frac{i}{(p^\mu-k^\mu)(p_\mu-k_\mu)-m^2c^2+i\epsilon}\Gamma(p_\mu p^\mu)\,\,\, ,
\ee
\be
(c)=\frac{N}{2}\omega\int\frac{d^2 p}{(2\pi)^2}\Gamma(p_\mu p^\mu),\hspace{4pc}
(d)=-\frac{N}{2}\left[\int\frac{d^2 p}{(2\pi)^2}\Gamma(p_\mu p^\mu)\right]\int \frac{d^2k}{(2\pi)^2}\frac{i}{k_\mu k^\mu-m^2c^2+i\epsilon}\,\,\, .
\ee

Since for large momenta the $\Gamma$ propagators diverges as $\sim p^\mu p_\mu$ (with logarithmic corrections), the $(b)$ graph is UV divergent and first of all we should extract its divergent part.
In the graph $(b)$ we perform a Wick rotation $p^\mu\to \bar{p}^\mu=(-ip^0,p^1)$ and $k^\mu\to \bar{k}^\mu=(-ik^0,k^1)$, then we rewrite it as
\begin{eqnarray}
\nonumber&&(b)=\int \frac{d^2p}{(2\pi)^2}\frac{1}{(\bar{p}^\mu \bar{p}_\mu+m^2c^2)}\Gamma(-\bar{p}_\mu \bar{p}^\mu)+\int \frac{d^2p}{(2\pi)^2}\left[\frac{-\bar{k}^\mu \bar{k}_\mu}{(\bar{p}^\mu \bar{p}_\mu+m^2c^2)^2}+\frac{4(\bar{p}_\mu \bar{k}^\mu)^2}{(\bar{p}_\mu \bar{p}^\mu+m^2c^2)^3}\right]\Gamma(-\bar{p}_\mu \bar{p}^\mu)+\\
\nonumber&&+\int \frac{d^2p}{(2\pi)^2}\left[\frac{1}{(\bar{p}^\mu-\bar{k}^\mu)(\bar{p}_\mu-\bar{k}_\mu)+m^2c^2}-\frac{1}{(p^\mu p_\mu+m^2c^2)}+\frac{k^\mu k_\mu}{(\bar{p}^\mu\bar{p}_\mu+m^2c^2)^2}-\frac{4(\bar{p}_\mu \bar{k}^\mu)^2}{(\bar{p}_\mu \bar{p}^\mu+m^2c^2)^3}\right]\Gamma(-\bar{p}_\mu \bar{p}^\mu)\,\,\, ,\\  \label{new154}
\end{eqnarray}
where the contractions are performed through the Euclidean metric
\begin{equation}
\eta^{E}_{\mu\nu}=\begin{pmatrix} \frac{1}{c^2}&& 0 \\ 0 && 1 \end{pmatrix}\,\,\, .
\end{equation}

The second row of (\ref{new154}) is UV finite, differently from the first and second integrals that provide, respectively, a mass and a field strength renormalization. As a matter of fact, thanks to the symmetries of the integration domain, the second integral can be rewritten as
\begin{eqnarray}
\nonumber &&\int \frac{d^2p}{(2\pi)^2}\left[\frac{-\bar{k}^\mu \bar{k}_\mu}{(\bar{p}^\mu \bar{p}_\mu+m^2c^2)^2}+\frac{4(\bar{p}_\mu \bar{k}^\mu)^2}{(\bar{p}_\mu \bar{p}^\mu+m^2c^2)^3}\right]\Gamma(-\bar{p}_\mu \bar{p}^\mu)=\\
&&=k_\mu k^\mu\int \frac{d^2p}{(2\pi)^2}\left[\frac{1}{(\bar{p}^\mu \bar{p}_\mu+m^2c^2)^2}+\frac{2\bar{p}_\mu \bar{p}^\mu}{(\bar{p}_\mu \bar{p}^\mu+m^2c^2)^3}\right]\Gamma(-\bar{p}_\mu \bar{p}^\mu)\,\,\, ,
\end{eqnarray}

i.e. exactly in the form of the field strength counterterm discussed at the end of Section \ref{secsigmaFeynman}. The first UV divergent integral in (\ref{new154}) can be absorbed, rather than in a mass renormalization, in a renormalization of the coupling $\omega$ modifying the mass equation from (\ref{new139}) to
\be
(c)+(d)+\int \frac{d^2p}{(2\pi)^2}\frac{1}{(\bar{p}^\mu \bar{p}_\mu+m^2c^2)}\Gamma(-\bar{p}_\mu \bar{p}^\mu)=0\,\,\, .
\ee

Compared to the mass equation (\ref{new139}), $\omega$ is shifted by a $\mathcal{O}(N^{-1})$ UV divergent term.
Now that the $\mathcal{O}(N^{-1})$ has been made finite by the renormalization, we can take the NR limit of the finite part of the $(b)$ graph, that vanishes in the NR limit.
Thus we conclude that, after the proper renormalization, the NR limit of the $\mathcal{O}(N^{-1})$ order consists in a free non relativistic propagator and a four field correlator that matches the first perturbative order of the Yang Gaudin model, as it should be.

\section{Conclusions}\label{Conclusions}

In this paper we have studied the non-relativistic limit of a large class of massive bosonic integrable QFT for the purpose of identifying possible new non-relativistic 
integrable models. To answer this question we have studied the non-relativistic limit of the simply-laced Toda Field Theories and the $O(N)$ non-linear model, finding that in the first case we end up in a {\em decoupled} set of Lieb-Liniger models of different masses, while in the second case in a symmetrically {\em coupled} set of Lieb-Liniger models, all with the same mass. This analysis seems to indicate that the Lieb-Liniger models may exhaust the list of non-relativistic integrable models that can be obtain as non relativistic limits of integrable QFT. 

A posteriori, in light of the computations shown in this paper, one can put together different hints to see why the Lieb-Liniger may be the only non-relativistic model 
obtained as a non relativistic limit of IQFT.
In general, we found that such a limit provides a non relativistic Hamiltonian of the type (\ref{generalHamiltoniannr1}). In absence of mass degeneracy,
non-relativistic dynamics forces the number of each species to be individually conserved. This implies the presence, for each species, of a $U(1)$ symmetry transformation, acting as $\psi_k \rightarrow e^{i \alpha_k} \,\psi_k$, $\psi^\dagger_k \rightarrow e^{-i \alpha_k}\,\psi^\dagger_k$. Then, the local potential term $V(\{\psi^\dagger_k,\psi_l \})$ -- function of $\psi^\dagger_i$ and $\psi_l$ but not of their derivatives -- must involve only the $U(1)$ invariant expressions $\eta_k \equiv \psi^\dagger_k \psi_k$. The quantity $\eta_k(x)$ is obviously the local density of the $k$-type particles at the position $x$. Even though the $O(N)$ sigma model possesses mass degeneracy, also in this case the interactions has been found to be a functional of the particle densities.
Assuming that such a potential can be expanded in series, we have (sum on the repeated indices) 
\EQ
V(\{\eta_k\}) \,=\,\mu_k \eta_k + \frac{1}{2!} \mu_{kl} \eta_k \eta_l + \frac{1}{3!} \mu_{klm}\,\eta_k \eta_l \eta_m + \frac{1}{4!} \mu_{klmn} \eta_k \eta_l \eta_m \eta_n + \cdots 
\label{expansionV}
\EN 
where the first, linear term, $\mu_k \eta_k$ represents simply a chemical potential and can be disregard in the considerations that follow. So, it remains to 
discuss the presence of quadratic and higher order interactions. Notice that, in non-relativistic models, particle production is forbidden even for virtual processes, therefore an $n$-body local interaction terms (where $n > 2$) in the non relativistic Hamiltonian would represent a genuine higher-body interaction, involving 
$n$-particles simultaneously present at the point $x$. It is hard to imagine how these processes can be compatible at least with Yang-Baxter integrability, since they would spoil the possibility of factorising an $n$-body scattering process into a sequence of $2$-body interactions. This seems to suggest that the interaction in NRIM is restricted to be density-density like, i.e. to the quadratic order only. 

On the other hand, for non-relativistic models presenting a mass degeneracy, there is no physical principle forbidding the tunnelling between different particle species. On the contrary, this kind of processes have a clear physical interest as they permit the description of particles with spin. Moreover, the tunnelling is not incompatible with integrability as testified by few exactly solvable models including these terms \cite{essler}.
However, the explicit non-relativistic limit performed in this paper on the simply-laced Toda Field Theory and the $O(N)$ sigma model has shown that the potentials defined by this 
limit are indeed only quadratic in $\eta_k$ and has excluded the possibility that another type of integrability emerges from this procedure. It remains an intriguing question to understand 
whether tunnelling processes are forbidden by the formal procedure involved in the NR limit, or if on the contrary, there are relativistic models giving rise to this effect. 
To proceed in this direction, one could try to extend the approach presented here to other integrable QFT, for instance other sigma models or those involving fermions. A natural question would be whether the conclusions presented in this work can be established as a firm and general result, beyond each and any individual case that can be analysed.

\vspace{3mm}
{\it Acknowledgements:} We would like to thank Patrick Dorey and Boris Dubrovin for useful discussions and Angela Foerster for pointing us 
reference \cite{Lamacraft}.

\newpage

\appendix
\section{Dynkin diagrams and masses of Toda Field Theories}\label{DynkinToda}
In this Appendix we simply report the relations involving the roots of the Lie algebras entering the Toda Field Theories and the classical values of the masses of these theories.

\begin{center}
\begin{tabular}{|ccc|} \hline
$A_{2r}^{(1)} / Z_2$  & & $A_{2r}^{(2)}$    \\               
\begin{picture}(180,70)
\thicklines            
\multiput(40,20)(15,0){3}{\circle{3}}             
\multiput(41.5,20)(15,0){2}{\line(1,0){12}}
\multiput(40,12)(15,0){3}{\makebox(0,0){{\scriptsize 1}}}
\put(30,20){\makebox(0,0){$\alpha_1$}} 
\put(41.5,20.3){\line(5,1){47}}
\multiput(110,20)(15,0){3}{\circle{3}}             
\multiput(111.5,20)(15,0){2}{\line(1,0){12}}       
\multiput(110,12)(15,0){3}{\makebox(0,0){{\scriptsize 1}}}
\put(160,20){\makebox(0,0){$\alpha_{2r}$}}                   
\put(138.5,20.3){\line(-5,1){47}}
\multiput(73,20)(5,0){7}{.}            
\put(90,30){\circle{3}}                            
\put(85,35){\makebox(0,0)[b]{{\scriptsize 1}}}
\put(95,35){\makebox(0,0)[lb]{$\alpha_{2r+1}$}}
\multiput(90,27)(0,-5){5}{\line(0,-1){3}}          
\multiput(90,33)(0,5){6}{\line(0,1){3}}
\put(90,55){\vector(1,0){10}}
\put(90,55){\vector(-1,0){10}}
\put(105,55){\makebox(0,0)[l]{$Z_2$}}
\end{picture}  
& \begin{picture}(50,70)                       
  \put(25,35){\makebox(0,0){$\Longrightarrow$}}
  \end{picture}                                
&                                              
\begin{picture}(155,70)
\thicklines            
\put(20,35){\circle{3}}                
\put(20,35){\circle{6}}                
\put(23,36){\line(1,0){11.5}}
\put(23,34){\line(1,0){11.5}}
\put(17,20){\makebox(0,0){$\alpha_{r+1}$}}
\put(35,20){\makebox(0,0){$\alpha_{1}$}}
\put(35,35){\circle{3}}                 
\put(36.5,35){\line(1,0){12}}
\put(50,35){\circle{3}}
\put(20,40){\makebox(0,0)[b]{{\scriptsize 1}}}
\put(35,40){\makebox(0,0)[b]{{\scriptsize 2}}}
\put(50,40){\makebox(0,0)[b]{{\scriptsize 2}}}
\multiput(53,35)(5,0){7}{.}             
\put(90,35){\circle{3}}                 
\put(103.5,35){\line(-1,0){12}}                     
\put(105,35){\circle{3}}                
\put(120,35){\circle*{4}}                           
\put(120,36){\line(-1,0){14}}                       
\put(120,34){\line(-1,0){14}}                       
\put(90,40){\makebox(0,0)[b]{{\scriptsize 2}}}      
\put(105,40){\makebox(0,0)[b]{{\scriptsize 2}}}     
\put(120,40){\makebox(0,0)[b]{{\scriptsize 2}}}     
\put(120,20){\makebox(0,0){$\alpha_{r}$}}           
\end{picture}
\\ \hline
$D_{r+1}^{(1)} / \sigma$    & & $B_r^{(1)}$ \\    
\begin{picture}(180,70)
\thicklines               
\put(55,35){\circle{3}}                
\put(56.5,35){\line(1,0){12}}
\put(53.5,36.5){\line(-1,1){12}}
\put(53.5,33.5){\line(-1,-1){12}}
\put(40,50){\circle{3}}
\put(40,20){\circle{3}}
\put(70,35){\circle{3}}
\put(40,12){\makebox(0,0){{\scriptsize 1}}}
\put(40,60){\makebox(0,0)[t]{{\scriptsize 1}}}
\put(58,40){\makebox(0,0)[b]{{\scriptsize 2}}}
\put(70,40){\makebox(0,0)[b]{{\scriptsize 2}}}
\put(30,20){\makebox(0,0){$\alpha_1$}}
\put(25,50){\makebox(0,0){$\alpha_{r+2}$}}
\multiput(73,35)(5,0){7}{.}            
\put(175,38.5){\vector(-1,4){2.5}}     
\put(175,35){\makebox(0,0){$\sigma$}}
\put(175,31){\vector(-1,-4){2.5}}
\put(125,35){\circle{3}}               
\put(123.5,35){\line(-1,0){12}}
\put(126.5,36.5){\line(1,1){12}}
\put(126.5,33.5){\line(1,-1){12}}
\put(140,50){\circle{3}}
\put(140,20){\circle{3}}
\put(110,35){\circle{3}}
\put(140,12){\makebox(0,0){{\scriptsize 1}}}
\put(140,60){\makebox(0,0)[t]{{\scriptsize 1}}}
\put(122,40){\makebox(0,0)[b]{{\scriptsize 2}}}
\put(110,40){\makebox(0,0)[b]{{\scriptsize 2}}}
\put(155,20){\makebox(0,0){$\alpha_{r}$}}
\put(155,50){\makebox(0,0){$\alpha_{r+1}$}}
\end{picture}    
& \begin{picture}(50,70)
  \put(25,35){\makebox(0,0){$\Longrightarrow$}}
  \end{picture}
&
\begin{picture}(155,70)
\thicklines            
\put(35,35){\circle{3}}                
\put(36.5,35){\line(1,0){12}}
\put(33.5,36.5){\line(-1,1){12}}
\put(33.5,33.5){\line(-1,-1){12}}
\put(20,50){\circle{3}}
\put(20,20){\circle{3}}
\put(50,35){\circle{3}}
\put(20,12){\makebox(0,0){{\scriptsize 1}}}   
\put(20,60){\makebox(0,0)[t]{{\scriptsize 1}}}
\put(38,40){\makebox(0,0)[b]{{\scriptsize 2}}}
\put(50,40){\makebox(0,0)[b]{{\scriptsize 2}}}
\put(10,20){\makebox(0,0){$\alpha_1$}}        
\put(5,50){\makebox(0,0){$\alpha_{r+1}$}}    
\multiput(53,35)(5,0){7}{.}            
\put(90,35){\circle{3}}                
\put(103.5,35){\line(-1,0){12}}
\put(105,35){\circle{3}}               
\put(120,35){\circle*{4}}
\put(120,36){\line(-1,0){14}}
\put(120,34){\line(-1,0){14}}
\put(90,40){\makebox(0,0)[b]{{\scriptsize 2}}}
\put(105,40){\makebox(0,0)[b]{{\scriptsize 2}}}
\put(120,40){\makebox(0,0)[b]{{\scriptsize 2}}}
\put(120,20){\makebox(0,0){$\alpha_{r}$}}
\end{picture}
\\ \hline
$D_{r+2}^{(1)} / Z_2$  & & $D_{r+1}^{(2)} \equiv
{\tilde{B}}_r$   \\                                          
\begin{picture}(180,70)
\thicklines               
\put(55,35){\circle{3}}                
\put(56.5,35){\line(1,0){12}}
\put(53.5,36.5){\line(-1,1){12}}
\put(53.5,33.5){\line(-1,-1){12}}
\put(40,50){\circle{3}}
\put(40,20){\circle{3}}
\put(70,35){\circle{3}}
\put(40,12){\makebox(0,0){{\scriptsize 1}}}
\put(40,60){\makebox(0,0)[t]{{\scriptsize 1}}}
\put(58,40){\makebox(0,0)[b]{{\scriptsize 2}}}
\put(70,40){\makebox(0,0)[b]{{\scriptsize 2}}}
\put(30,20){\makebox(0,0){$\alpha_1$}}
\put(25,50){\makebox(0,0){$\alpha_{r+3}$}}
\multiput(73,35)(5,0){7}{.}            
\put(175,35){\makebox(0,0){$Z_2$}}
\multiput(130,35)(5,0){8}{\makebox(0,0){-}}  
\multiput(25,35)(5,0){6}{\makebox(0,0){-}}
\put(125,35){\circle{3}}                      
\put(123.5,35){\line(-1,0){12}}
\put(126.5,36.5){\line(1,1){12}}
\put(126.5,33.5){\line(1,-1){12}}
\put(140,50){\circle{3}}
\put(140,20){\circle{3}}
\put(110,35){\circle{3}}
\put(140,12){\makebox(0,0){{\scriptsize 1}}}
\put(140,60){\makebox(0,0)[t]{{\scriptsize 1}}}
\put(122,40){\makebox(0,0)[b]{{\scriptsize 2}}}
\put(110,40){\makebox(0,0)[b]{{\scriptsize 2}}}
\put(155,20){\makebox(0,0){$\alpha_{r+1}$}}
\put(155,50){\makebox(0,0){$\alpha_{r+2}$}}
\end{picture}    
& \begin{picture}(50,70)
  \put(25,35){\makebox(0,0){$\Longrightarrow$}}
  \end{picture}
&
\begin{picture}(155,70)
\thicklines            
\put(20,35){\circle*{4}}                
\put(20,36){\line(1,0){14}}
\put(20,34){\line(1,0){14}}
\put(17,20){\makebox(0,0){$\alpha_{r+1}$}}
\put(35,20){\makebox(0,0){$\alpha_{1}$}}
\put(35,35){\circle{3}}                 
\put(36.5,35){\line(1,0){12}}
\put(50,35){\circle{3}}
\put(20,40){\makebox(0,0)[b]{{\scriptsize 1}}}
\put(35,40){\makebox(0,0)[b]{{\scriptsize 1}}}
\put(50,40){\makebox(0,0)[b]{{\scriptsize 1}}}
\multiput(53,35)(5,0){7}{.}             
\put(90,35){\circle{3}}                 
\put(103.5,35){\line(-1,0){12}}                     
\put(105,35){\circle{3}}                
\put(120,35){\circle*{4}}                           
\put(120,36){\line(-1,0){14}}                       
\put(120,34){\line(-1,0){14}}                       
\put(90,40){\makebox(0,0)[b]{{\scriptsize 1}}}      
\put(105,40){\makebox(0,0)[b]{{\scriptsize 1}}}     
\put(120,40){\makebox(0,0)[b]{{\scriptsize 1}}}     
\put(120,20){\makebox(0,0){$\alpha_{r}$}}           
\end{picture}
\\ \hline
$A_{2r-1}^{(1)} / Z_2$  & & $C_r^{(1)}$ \\   
\begin{picture}(180,70)
\thicklines            
\multiput(40,20)(15,0){3}{\circle{3}}             
\multiput(41.5,20)(15,0){2}{\line(1,0){12}}
\multiput(40,12)(15,0){3}{\makebox(0,0){{\scriptsize 1}}}
\put(30,20){\makebox(0,0){$\alpha_1$}} 
\put(41.5,20.3){\line(5,1){47}}
\multiput(110,20)(15,0){3}{\circle{3}}             
\multiput(111.5,20)(15,0){2}{\line(1,0){12}}       
\multiput(110,12)(15,0){3}{\makebox(0,0){{\scriptsize1}}}
\put(160,20){\makebox(0,0){$\alpha_{2r-1}$}}                   
\put(138.5,20.3){\line(-5,1){47}}
\multiput(73,20)(5,0){7}{.}            
\put(90,30){\circle{3}}                            
\put(85,35){\makebox(0,0)[b]{{\scriptsize 1}}}
\put(95,35){\makebox(0,0)[lb]{$\alpha_{2r}$}}
\multiput(90,27)(0,-5){5}{\line(0,-1){3}}          
\multiput(90,33)(0,5){6}{\line(0,1){3}}
\put(90,55){\vector(1,0){10}}
\put(90,55){\vector(-1,0){10}}
\put(105,55){\makebox(0,0)[l]{$Z_2$}}
\end{picture}  
& \begin{picture}(50,70)                       
  \put(25,35){\makebox(0,0){$\Longrightarrow$}}
  \end{picture}                                
&                                              
\begin{picture}(155,70)
\thicklines            
\put(20,35){\circle{3}}                
\put(35,36){\line(-1,0){14}}
\put(35,34){\line(-1,0){14}}
\put(17,20){\makebox(0,0){$\alpha_{r+1}$}}
\put(35,20){\makebox(0,0){$\alpha_{1}$}}
\put(35,35){\circle*{4}}                 
\put(36.5,35){\line(1,0){12}}
\put(50,35){\circle*{4}}
\put(20,40){\makebox(0,0)[b]{{\scriptsize 1}}}
\put(35,40){\makebox(0,0)[b]{{\scriptsize 2}}}
\put(50,40){\makebox(0,0)[b]{{\scriptsize 2}}}
\multiput(53,35)(5,0){7}{.}             
\put(90,35){\circle*{4}}                 
\put(103.5,35){\line(-1,0){12}}                     
\put(105,35){\circle*{4}}                
\put(120,35){\circle{3}}                           
\put(105,36){\line(1,0){14}}                       
\put(105,34){\line(1,0){14}}                       
\put(90,40){\makebox(0,0)[b]{{\scriptsize 2}}}      
\put(105,40){\makebox(0,0)[b]{{\scriptsize 2}}}     
\put(120,40){\makebox(0,0)[b]{{\scriptsize 1}}}     
\put(120,20){\makebox(0,0){$\alpha_{r}$}}           
\end{picture}
\\ \hline
$D_{2r}^{(1)} / Z_2$  & &  $A_{2r-1}^{(2)} \equiv 
{\tilde{C}}_r$   \\                                          
\begin{picture}(180,70)
\thicklines               
\put(55,35){\circle{3}}                
\put(56.5,35){\line(1,0){12}}
\put(53.5,36.5){\line(-1,1){12}}
\put(53.5,33.5){\line(-1,-1){12}}
\put(40,50){\circle{3}}
\put(40,20){\circle{3}}
\put(70,35){\circle{3}}
\put(40,12){\makebox(0,0){{\scriptsize 1}}}
\put(40,60){\makebox(0,0)[t]{{\scriptsize 1}}}
\put(58,40){\makebox(0,0)[b]{{\scriptsize 2}}}
\put(70,40){\makebox(0,0)[b]{{\scriptsize 2}}}
\put(30,20){\makebox(0,0){$\alpha_1$}}
\put(25,50){\makebox(0,0){$\alpha_{2r+1}$}}
\multiput(73,35)(5,0){7}{.}            
\multiput(90,30)(0,-5){5}{\line(0,-1){3}}          
\multiput(90,33)(0,5){6}{\line(0,1){3}}
\put(90,55){\vector(1,0){10}}
\put(90,55){\vector(-1,0){10}}
\put(105,55){\makebox(0,0)[l]{$Z_2$}}
\put(125,35){\circle{3}}                      
\put(123.5,35){\line(-1,0){12}}
\put(126.5,36.5){\line(1,1){12}}
\put(126.5,33.5){\line(1,-1){12}}
\put(140,50){\circle{3}}
\put(140,20){\circle{3}}
\put(110,35){\circle{3}}
\put(140,12){\makebox(0,0){{\scriptsize 1}}}
\put(140,60){\makebox(0,0)[t]{{\scriptsize 1}}}
\put(122,40){\makebox(0,0)[b]{{\scriptsize 2}}}
\put(110,40){\makebox(0,0)[b]{{\scriptsize 2}}}
\put(150,20){\makebox(0,0)[l]{$\alpha_{2r-1}$}}
\put(150,50){\makebox(0,0)[l]{$\alpha_{2r}$}}
\end{picture}    
& \begin{picture}(50,70)
  \put(25,35){\makebox(0,0){$\Longrightarrow$}}
  \end{picture}
&
\begin{picture}(155,70)
\thicklines            
\put(35,35){\circle*{4}}                
\put(36.5,35){\line(1,0){12}}
\put(33.5,36.5){\line(-1,1){12}}
\put(33.5,33.5){\line(-1,-1){12}}
\put(20,50){\circle*{4}}
\put(20,20){\circle*{4}}
\put(50,35){\circle*{4}}
\put(20,12){\makebox(0,0){{\scriptsize 1}}}   
\put(20,60){\makebox(0,0)[t]{{\scriptsize 1}}}
\put(38,40){\makebox(0,0)[b]{{\scriptsize 2}}}
\put(50,40){\makebox(0,0)[b]{{\scriptsize 2}}}
\put(10,20){\makebox(0,0){$\alpha_1$}}        
\put(5,50){\makebox(0,0){$\alpha_{r+1}$}}    
\multiput(53,35)(5,0){7}{.}            
\put(90,35){\circle*{4}}                
\put(103.5,35){\line(-1,0){12}}
\put(105,35){\circle*{4}}               
\put(120,35){\circle{3}}
\put(105,36){\line(1,0){14}}
\put(105,34){\line(1,0){14}}
\put(90,40){\makebox(0,0)[b]{{\scriptsize 2}}}
\put(105,40){\makebox(0,0)[b]{{\scriptsize 2}}}
\put(120,40){\makebox(0,0)[b]{{\scriptsize 1}}}
\put(120,20){\makebox(0,0){$\alpha_{r}$}}
\end{picture}
\\ \hline
\begin{picture}(180,40)                                 
\thicklines
\put(80,20){\makebox(0,0){Square length}}
\put(150,20){\circle*{4}}  
\put(179,20){\makebox(0,0)[r]{ $= 1$}}
\end{picture}
&  \begin{picture}(50,40)
   \end{picture}
&
\begin{picture}(155,40)
\put(2,20){\circle{3}}
\put(22,20){\makebox(0,0){ $= 2$}}
\put(100,20){\circle{3}}
\put(100,20){\circle{6}}
\put(120,20){\makebox(0,0){ $= 4$}}
\end{picture}
\\ \hline
\end{tabular}
\end{center}
\begin{center}
{\bf Table A.a:} Foldings of the Dynkin diagrams of the simply-laced algebras: the principal series. Near the roots there are the numbers $q_i$.

\end{center}

\newpage
\begin{center}
\begin{tabular}{|ccc|} \hline
$D_4^{(1)} / \sigma$  &  &    $G_2^{(1)}$  \\             
\begin{picture}(170,70)
\thicklines               
\put(85,35){\circle{3}}
\multiput(70,20)(30,30){2}{\circle{3}}
\multiput(70,50)(30,-30){2}{\circle{3}}
\multiput(71.5,21.5)(15,15){2}{\line(1,1){12}}
\multiput(71.5,48.5)(15,-15){2}{\line(1,-1){12}}
\multiput(70,12)(30,0){2}{\makebox(0,0){{\scriptsize 1}}}
\multiput(70,58)(30,0){2}{\makebox(0,0){{\scriptsize 1}}}
\put(85,27){\makebox(0,0){{\scriptsize 2}}}
\put(103.5,37.5){\vector(-1,4){3}}     
\put(103.5,35){\makebox(0,0){$\sigma$}}
\put(103.5,32){\vector(-1,-4){3}}
\put(85,15.5){\vector(4,1){10}}
\put(85,15.5){\vector(-4,1){10}}
\put(60,50){\makebox(0,0){$\alpha_5$}}
\put(60,20){\makebox(0,0){$\alpha_1$}}
\put(110,50){\makebox(0,0){$\alpha_3$}}
\put(110,20){\makebox(0,0){$\alpha_2$}}
\end{picture}
& \begin{picture}(50,70)
  \put(25,35){\makebox(0,0){$\Longrightarrow$}}
  \end{picture}
&
\begin{picture}(165,70)
\thicklines            
\multiput(67.5,35)(15,0){2}{\circle{3}}
\put(97.5,35){\circle*{4}}
\multiput(69,35)(15,0){2}{\line(1,0){12}}
\put(83,36.5){\line(1,0){14}}
\put(83.5,33.5){\line(1,0){13}}
\put(67.5,40){\makebox(0,0)[b]{{\scriptsize 1}}}
\put(82.5,40){\makebox(0,0)[b]{{\scriptsize 2}}}
\put(97.5,40){\makebox(0,0)[b]{{\scriptsize 3}}}
\put(67.5,20){\makebox(0,0){$\alpha_{3}$}}           
\put(82.5,20){\makebox(0,0){$\alpha_{2}$}} 
\put(97.5,20){\makebox(0,0){$\alpha_{1}$}}           
\end{picture}                          
\\ \hline
$E_6^{(1)} / Z_3$   & & $D_4^{(3)} \equiv {\tilde{G}}_2$  \\     
\begin{picture}(170,70)
\thicklines               
\multiput(55,20)(15,0){5}{\circle{3}}
\multiput(85,35)(0,15){2}{\circle{3}}
\put(85,60){\makebox(0,0){$\alpha_7$}}
\multiput(56.5,20)(15,0){4}{\line(1,0){12}}
\multiput(85,21.5)(0,15){2}{\line(0,1){12}}
\multiput(55,12)(60,0){2}{\makebox(0,0){{\scriptsize 1}}}
\multiput(70,9)(30,0){2}{\makebox(0,0){{\scriptsize 2}}}
\put(85,12){\makebox(0,0){{\scriptsize 3}}}
\put(90,35){\makebox(0,0){{\scriptsize 2}}}
\put(90,50){\makebox(0,0){{\scriptsize 1}}}
\put(100,35){\vector(1,-1){12}}
\put(100,35){\vector(-1,1){12}}
\put(70,35){\vector(1,1){12}}
\put(70,35){\vector(-1,-1){12}}
\put(85,5){\makebox(0,0)[t]{$Z_3$}} 
\put(89,7.5){\vector(2,1){25}}
\put(81,7.5){\vector(-2,1){25}}
\end{picture}
& \begin{picture}(50,70)                       
  \put(25,35){\makebox(0,0){$\Longrightarrow$}}
  \end{picture}                                
&                                              
\begin{picture}(165,70)
\thicklines            
\multiput(67.5,35)(15,0){2}{\circle*{4}}
\put(97.5,35){\circle{3}}
\multiput(69,35)(15,0){2}{\line(1,0){12}}
\put(83,36.5){\line(1,0){14}}
\put(83.5,33.5){\line(1,0){13}}
\put(67.5,40){\makebox(0,0)[b]{{\scriptsize 1}}}
\put(82.5,40){\makebox(0,0)[b]{{\scriptsize 2}}}
\put(97.5,40){\makebox(0,0)[b]{{\scriptsize 1}}}
\put(67.5,20){\makebox(0,0){$\alpha_{3}$}}           
\put(82.5,20){\makebox(0,0){$\alpha_{2}$}} 
\put(97.5,20){\makebox(0,0){$\alpha_{1}$}}           
\end{picture}
\\ \hline
$E_6^{(1)} / Z_2$  & &  $F_4^{(1)}$ \\    
\begin{picture}(170,70)                         
\thicklines               
\multiput(55,20)(15,0){5}{\circle{3}}
\multiput(85,35)(0,15){2}{\circle{3}}
\multiput(56.5,20)(15,0){4}{\line(1,0){12}}
\multiput(85,21.5)(0,15){2}{\line(0,1){12}}
\put(75,50){\makebox(0,0){$\alpha_7$}}
\multiput(55,12)(60,0){2}{\makebox(0,0){{\scriptsize 1}}}
\multiput(70,12)(30,0){2}{\makebox(0,0){{\scriptsize 2}}}
\put(85,12){\makebox(0,0){{\scriptsize 3}}}
\put(90,35){\makebox(0,0){{\scriptsize 2}}}
\put(90,50){\makebox(0,0){{\scriptsize 1}}}
\put(85,18){\line(0,-1){3}}
\put(85,8){\line(0,-1){3}}
\put(85,3){\line(0,-1){3}}
\multiput(85,52)(0,5){3}{\line(0,1){3}}
\put(90,60){\makebox(0,0)[l]{$Z_2$}}
\end{picture}
& \begin{picture}(50,70)                       
  \put(25,35){\makebox(0,0){$\Longrightarrow$}}
  \end{picture}                                
&                                              
\begin{picture}(165,70)
\thicklines            
\multiput(52.5,35)(15,0){3}{\circle{3}}
\multiput(97.5,35)(15,0){2}{\circle*{4}}
\multiput(54,35)(15,0){2}{\line(1,0){12}}
\put(99,35){\line(1,0){12}}
\put(83.5,36.5){\line(1,0){13.5}}
\put(83.5,33.5){\line(1,0){13.5}}
\put(52.5,40){\makebox(0,0)[b]{{\scriptsize 1}}}
\put(67.5,40){\makebox(0,0)[b]{{\scriptsize 2}}}
\put(82.5,40){\makebox(0,0)[b]{{\scriptsize 3}}}
\put(97.5,40){\makebox(0,0)[b]{{\scriptsize 4}}}
\put(112.5,40){\makebox(0,0)[b]{{\scriptsize 2}}}
\put(52.5,20){\makebox(0,0){$\alpha_5$}}
\put(67.5,20){\makebox(0,0){$\alpha_1$}}
\put(82.5,20){\makebox(0,0){$\alpha_2$}}
\put(97.5,20){\makebox(0,0){$\alpha_3$}}
\put(112.5,20){\makebox(0,0){$\alpha_4$}}
\end{picture}
\\ \hline
$E_7^{(1)} / Z_2$  & & $E_6^{(2)} \equiv 
{\tilde{F}}_4$   \\                                       
\begin{picture}(170,70)
\thicklines
\multiput(40,25)(15,0){7}{\circle{3}}
\put(30,25){\makebox(0,0){$\alpha_8$}}
\put(85,40){\circle{3}}
\multiput(41.5,25)(15,0){6}{\line(1,0){12}}
\put(85,26.5){\line(0,1){12}}
\multiput(40,17)(90,0){2}{\makebox(0,0){{\scriptsize 1}}}
\multiput(55,17)(60,0){2}{\makebox(0,0){{\scriptsize 2}}}
\multiput(70,17)(30,0){2}{\makebox(0,0){{\scriptsize 3}}}
\put(85,17){\makebox(0,0){{\scriptsize 4}}}
\put(95,40){\makebox(0,0){{\scriptsize 2}}}
\multiput(85,42)(0,5){5}{\line(0,1){3}}
\multiput(85,12)(0,-5){2}{\line(0,-1){3}}
\put(85,23){\line(0,-1){3}}
\put(90,55){\makebox(0,0)[l]{$Z_2$}}
\end{picture}
& \begin{picture}(50,70)                       
  \put(25,35){\makebox(0,0){$\Longrightarrow$}}
  \end{picture}                                
&                                              
\begin{picture}(165,70)
\thicklines            
\multiput(82.5,35)(15,0){3}{\circle*{4}}
\multiput(52.5,35)(15,0){2}{\circle{3}}
\put(54,35){\line(1,0){12}}
\put(68,36.5){\line(1,0){14}}
\put(68,33.5){\line(1,0){14}}
\put(82.5,35){\line(1,0){30}}
\put(52.5,40){\makebox(0,0)[b]{{\scriptsize 1}}}
\put(67.5,40){\makebox(0,0)[b]{{\scriptsize 2}}}
\put(82.5,40){\makebox(0,0)[b]{{\scriptsize 3}}}
\put(97.5,40){\makebox(0,0)[b]{{\scriptsize 1}}}
\put(112.5,40){\makebox(0,0)[b]{{\scriptsize 2}}}
\put(52.5,20){\makebox(0,0){$\alpha_1$}}
\put(67.5,20){\makebox(0,0){$\alpha_2$}}
\put(82.5,20){\makebox(0,0){$\alpha_3$}}
\put(97.5,20){\makebox(0,0){$\alpha_4$}}
\put(112.5,20){\makebox(0,0){$\alpha_5$}}
\end{picture}
\\ \hline
\begin{picture}(170,40)                                 
\thicklines
\put(80,20){\makebox(0,0){Square length}}
\put(145,20){\circle*{4}}  
\put(165,20){\makebox(0,0){ $= 1$}}
\end{picture}
&  \begin{picture}(50,40)
   \put(12.5,20){\makebox(0,0){ or $2/3$}}
   \end{picture}
&
\begin{picture}(165,40)
\put(15,20){\circle{3}}
\put(35,20){\makebox(0,0){ $= 2$}}
\end{picture}
\\ \hline
\end{tabular}
\end{center}
\begin{center}
{\bf Table A.b:} Foldings of the Dynkin diagrams of the simply-laced algebras: the exceptional series. Near the roots there are the numbers $q_i$ 
entering the Lagrangian (\ref{toda2}). 
\end{center}

\begin{center}  
\begin{tabular}{|c|c|} \hline
$A_{2r}^{(1)}$   &  $A_{2r}^{(2)}$    \\ 
$2M \sin (\frac{\pi i}{2r+1}), \spz 1 \leq i \leq 2r$               &
 $4M \sin (\frac{\pi i}{2r+1}), \spz 1 \leq i \leq r$            \\ \hline
$D_{r+1}^{(1)}$  &  $B_r^{(1)}$       \\ 
$M,M,2M \sin (\frac{\pi i}{2r}), 
			           \hspace{0.4cm} 1 \leq i \leq r-1$  &
 $M,2M \sin (\frac{\pi i}{2r} ), \hspace{0.4cm} 
						1 \leq i \leq r-1$ \\ \hline 
$D_{r+2}^{(1)}$  &  $D_{r+1}^{(2)} \equiv {\tilde{B}}_r$ \\
$M,M,2M \sin (\frac{\pi i}{2r+2} ), 
				     \hspace{0.4cm} 1 \leq i \leq r$  &
 ${\sqrt 2}M \sin (\frac{\pi i}{2r+2} ), 
				\hspace{0.4cm} 1 \leq i \leq r$    \\ \hline
$A_{2r-1}^{(1)}$ &  $C_r^{(1)}$        \\
$2M \sin (\frac{\pi i}{2r} ), \spz 1 \leq i \leq 2r-1$               &
 $2M \sin\left(\frac{\pi i}{2r}\right), \spz 1 \leq i \leq r$              \\ \hline
$D_{2r}^{(1)}$   &  $A_{2r-1}^{(2)} \equiv {\tilde{C}}_r$   \\ 
$M,M,2M \sin (\frac{\pi i}{2(2r-1)} ), 
	\, 1 \leq i \leq 2r-2$                             &
 $\frac{M}{\sqrt 2},{\sqrt 2}M \sin (\frac{\pi i}
 {(2r-1)} ), 
				      \, 1 \leq i \leq r-1$         \\ \hline
$D_4^{(1)}$      &  $G_2^{(1)}$        \\ 
$M,M,M,{\sqrt 3}M$                             &
$M,{\sqrt 3}M$                                               \\ \hline
$E_6^{(1)}$      &  $D_4^{(3)} \equiv {\tilde{G}}_2$   \\
$m_1 =  m_{\overline 1} = M$     &
			$m_3$, $m_4$      \\                      
$m_2 =  m_{\overline 2} = 2M \,\cos(\frac{\pi}{12})$ &
     			\begin{picture}(180,20)
			\put(-6,0){\line(1,0){192}}
			\end{picture}                  \\
$m_3 =  2M \,\cos(\frac{\pi}{4})$ &
			$F_4^{(1)}$             \\                    
$m_4 =  4M \, \cos(\frac{\pi}{12})\cos(\frac{\pi}{4})$ &
			$m_1$, $m_2$, $m_3$, $m_4$ \\          \hline
$E_7^{(1)}$      &    $E_6^{(2)} \equiv {\tilde{F}}_4$   \\
\begin{tabular}{c}
$m_1$ = $M$ \\                                            
$m_2$ = $2 M \cos(\frac{5\pi}{18})$ \\                     
$m_3$ = $2 M \cos(\frac{\pi}{9})$ \\                 
$m_4$ = $2 M \cos(\frac{\pi}{18}) $ \\ 
$m_5$ = $4 M \cos(\frac{5\pi}{18}) \cos(\frac{\pi}{18})$ \\
$m_6$ = $4 M \cos(\frac{\pi}{9}) \cos(\frac{2\pi}{9})$ \\                     
$m_7$ = $4 M \cos(\frac{\pi}{18}) \cos(\frac{\pi}{9})$ \\ 
\end{tabular}								&
$m_2$, $m_4$, $m_5$, $m_6$                                          \\ \hline
\end{tabular}
\end{center}
\begin{center} 
{\bf Table A.c:} Masses of the Toda field theories related by the folding procedure. 
\end{center}

\begin{center}
\begin{tabular}{|c|} \hline
$E_8^{(1)}$                                            \\
$m_1$ = $M$                                            \\
$m_2$ = $2M \cos(\frac{\pi}{5})$                       \\
$m_3$ = $2M \cos(\frac{\pi}{30})$                      \\
$m_4$ = $2 m_2  
\cos(\frac{7\pi}{30})$ \\
$m_5$ = $2m_2  
\cos(\frac{2\pi}{15})$ \\
$m_6$ = $2 m_2  
\cos(\frac{\pi}{30})$  \\
$m_7$ = $4 m_2 \cos2(\frac{\pi}{5}) 
\cos(\frac{7\pi}{30})$ \\
$m_8$ = $4 m_2 \cos(\frac{\pi}{5}) 
\cos(\frac{2\pi}{15})$ \\ \hline
\end{tabular}
\end{center}
\begin{center}
{\bf Table A.d:} Mass spectrum of the Toda field theory $E_8^{(1)}$. 
\end{center}   

\vspace{3mm}
\newpage

\section{Scattering matrix of a contact potential}\label{AppendixA}

We want to study the scattering process of two particles interacting through a density-density potential. It is important to consider separately two different cases, that is to say whether or not the particles are distinguishable.

\vspace{2mm}
\noindent
{\bf Distinguishable particles}. The problem is most easily addressed in first quantization, therefore let $\chi(x,y)$ be the wavefunction, the coordinate $x$ is referred to the first particle of mass $m_a$ and $y$ to the second particle of mass $m_b$. The Hamiltonian referred to these two particles is:
\begin{equation}
H=-\frac{\partial_x^2}{2m_a}-\frac{\partial_y^2}{2m_b}+2\lambda\delta(x-y)\label{ds1}
\end{equation}
The problem is easily solved in the CM frame, therefore we define the coordinates:
\begin{equation}
X=\frac{m_a x+m_b y}{m_a+m_b}, \hspace{3pc}Y=x-y
\end{equation}
In terms of these new variables the Hamiltonian is:
\begin{equation}
H=-\frac{1}{2M}\partial_X^2-\frac{1}{2\mu}\partial_Y^2+2\lambda\delta(Y)\label{sc3}
\end{equation}
where $M$ is the total mass $M=m_a+m_b$ and $\mu$ the reduced mass $\mu=m_am_b/M$. Since the CM coordinate is decoupled from the relative position, we can simply ignore it in the computation of the scattering matrix. Consider now the scattering process in which in the far past the particle $a$ is on the far left of the particle $b$. This choice requires the presence of an ingoing plane-wave of momentum $k$, then we will have a reflected and transmitted plane waves.
The transmission and reflection coefficients $T$ and $R$ are parametrized in terms of the relative velocity of the particles $v=\mu^{-1}k$, that is invariant under Galilean transformation. Their value is found solving the eigenstate problem with boundary conditions:
\begin{equation}
\chi_{\text{CM}}(Y)=\begin{cases}e^{ikY}+R(v)e^{-ikY}\hspace{3pc}Y<0\\
T(v)e^{ikY}\hspace{6pc}\hspace{2pt}Y>0
\end{cases}\hspace{2pc}\text{distinguishable particles}\label{ds4}
\end{equation}
Above, $\chi_{\text{CM}}$ is the reduced wavefunction in the CM reference frame. The solution of the eigenvalue equation requires the continuity of the wavefunction at $Y=0$ and a discontinuity of the first derivative because of the $\delta$ interaction:
\begin{equation}
\chi_{\text{CM}}(0^+)=\chi_{\text{CM}}(0^-),\hspace{4pc}\partial_{Y}\chi_{\text{CM}}(0^+)-\partial_{Y}\chi_{\text{CM}}(0^-)=4\lambda\mu \chi_{\text{CM}}(0^+)\label{ds5}
\end{equation}

Using (\ref{ds4}) in the equations above we fix the reflexion and transmission coefficients $R$ and $T$ as:
\begin{equation}
R(v)=\frac{-i2\lambda}{v+i2\lambda},\hspace{6pc}T(v)=\frac{v}{v+i2\lambda}
\end{equation}

Notice that for any non trivial interaction $\lambda\ne 0$ it is \emph{impossible} to have a purely transmissive scattering, since $R$ is always non zero.

\vspace{2mm}
\noindent
{\bf Indistinguishable particles}. 
If we are rather interested in the scattering event among two \emph{indistinguishable} bosonic particles we cannot distinguish reflexion from transmission, since there is no way to establish which particle is on the left of the other. 
In first quantization the Hamiltonian is still in the form (\ref{ds1}) (for indistinguishable particles we are forced to have mass degeneracy):
\begin{equation}
H=-\frac{\partial_x^2}{2m}-\frac{\partial_y^2}{2m}+2\lambda\delta(x-y)\label{ds7}
\end{equation}

The two body wavefunction $\psi(x,y)$ is now required to be symmetric $\chi(x,y)=\chi(y,x)$, therefore the wavefunction in the CM reference state must be symmetric in $Y$, thus $\chi_{\text{CM}}(Y)=\chi_{\text{CM}}(-Y)$.
This symmetry constrain implies that (\ref{ds4}) is no more consistent and we rather need to impose:
\begin{equation}
\chi_{\text{CM}}(Y)=\begin{cases}e^{ikY}+S(2k)e^{-ikY}\hspace{3pc}Y<0\\
S(2k)e^{ikY}+e^{-ikY}\hspace{3pc}Y>0
\end{cases}\hspace{2pc}\text{indistinguishable particles}
\end{equation}
Where the amplitude is parametrized in terms of the relative momentum (that is simply proportional to the relative velocity for degenerated masses).
Asking the above wavefunction to be an eigenstate of (\ref{ds7}) leads to the same boundary conditions (\ref{ds5}) (for identical particles $\mu=m/2$) whose solution fixes the $S(2k)$ amplitude, that turns out to be the LL scattering matrix, as it should be.
\begin{equation}
S(k)=\frac{k-i2m\lambda}{k+i2m\lambda}
\end{equation}

\vspace{2mm}
\noindent
{\bf Non integrability from the three body problem with different couplings}. 
As we just seen a contact potential is never purely transmissive, thus integrability can be attained only for degenerated masses, exactly as it happens in relativistic models \cite{Mbook}. 
Therefore consider a set of particles of degenerate masses $m$ interacting through density-density potentials, however we allow for an interaction strength dependent on the colliding particles. The two body scattering matrix can be read directly from the previous section (since the masses are degenerated, we can parametrize the scattering in terms of momenta rather than velocities):
\begin{equation}
A_{a}(k_1)A_{b}(k_2)=S_{R}(k_1-k_2,\lambda_{ab})A_{a}(k_2)A_{b}(k_1)+S_{T}(k_1-k_2,\lambda_{ab})A_{b}(k_2)A_{a}(k_1)\label{ds10}
\end{equation}

Where:
\begin{equation}
S_{R}(k,\lambda)=\frac{-i2m\lambda}{k+i2m\lambda},\hspace{5pc}S_{R}(k,\lambda)=\frac{k}{k+i2m\lambda}
\end{equation}

The coupling $\lambda_{ab}$ depends on the kind of particles that interact (of course $\lambda_{ab}=\lambda_{ba}$). Notice that for identical particles (set $a=b$ in (\ref{ds10})) we have a LL scattering:
\begin{equation}
A_{a}(k_1)A_a(k_2)=S_{LL}(k_1-k_2,\lambda_{aa})A_a(k_2)A_a(k_1)=\frac{k_1-k_2-i2m\lambda_{aa}}{k_1-k_2+i2m\lambda_{aa}}A_a(k_2)A_a(k_1)
\end{equation}
\begin{figure}
\begin{center}
\includegraphics[scale=0.3]{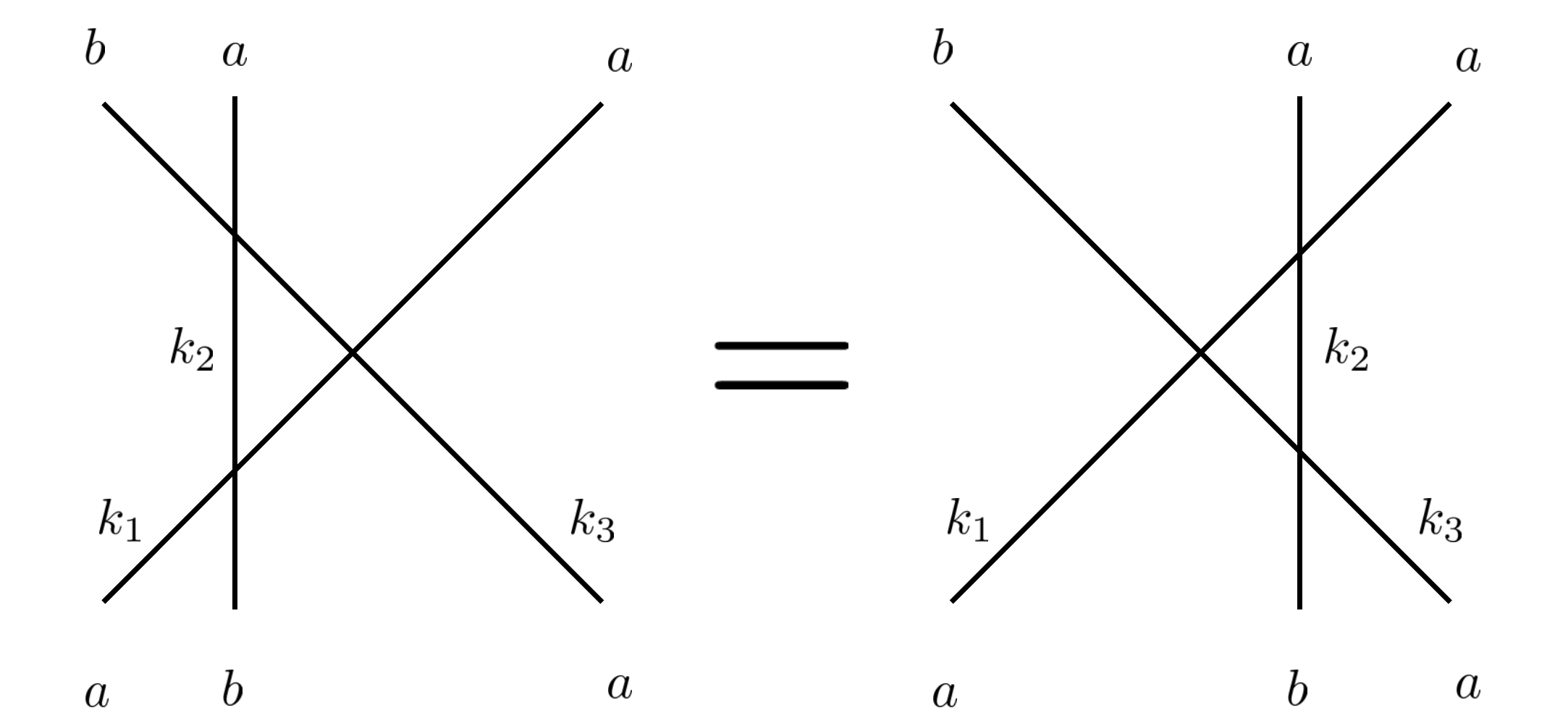}
\caption{\emph{Graphical representation of the Yang Baxter equations we use to test integrability. The left amplitude is reported in (\ref{sd13}), the right amplitude is (\ref{sd14}).}}\label{figapp1}
\end{center}
\end{figure}

In the following we show that integrability requires the couplings $\lambda_{ab}$ to be all equal with each others (apart the trivial case in which different particles do not interact). In order to show it we prove that Yang Baxter equations are violated unless $\lambda_{ab}=\lambda_{aa}=\lambda_{bb}$: since we are assuming more than one type of particle, there are at least two different particles. We will impose the YB equation represented in Figure \ref{figapp1}, the momenta of the in state are labeled $k_1,k_2,k_3$ from the left to the right.
The scattering amplitude of the process on the left is readily written as:
\begin{equation}
S_{T}(k_1-k_2,\lambda_{ab})S_{LL}(k_1-k_3,\lambda_{aa})S_{R}(k_2-k_3,\lambda_{ab})+S_{R}(k_1-k_2,\lambda_{ab})S_{R}(k_1-k_3,\lambda_{ab})S_{T}(k_2-k_3,\lambda_{ab})\label{sd13}
\end{equation}

Instead the amplitude on the right is:
\begin{equation}
S_{R}(k_2-k_3,\lambda_{ab})S_{T}(k_1-k_3,\lambda_{ab})S_{LL}(k_1-k_2,\lambda_{aa})\label{sd14}
\end{equation}

Equating (\ref{sd13}) with (\ref{sd14}) we find only two possible solutions for $\lambda_{ab}=\lambda_{aa}$ (equal interaction) or $\lambda_{ab}=0$ (the particles of species $a$ are decoupled from the particle of species $b$).
In the case $\lambda_{ab}\ne 0$ we can repeat the same argument exchanging the particles $a,b$ and reach the conclusion $\lambda_{bb}=\lambda_{ab}$, therefore a non trivial solution of the YB equations requires $\lambda_{aa}=\lambda_{ab}=\lambda_{bb}$: this case is known to be integrable in the literature \cite{yang,suth}, therefore the YB equations are automatically satisfied.

\newpage 

\section{Non relativistic limit of the dynamics of the Bullogh-Dodd model}
\label{A}

In this Appenix the technical passages to extract the NR limit of the equation of motion for the Bullogh Dodd model are presented. We recall the relativistic equation of motion (\ref{BD34}) in which we already dropped the unessential terms
\begin{equation}
\frac{1}{c^2}\partial_t^2\phi=\partial_x^2\phi-m^2c^2\phi-cm^2\frac{\beta}{2}:\phi^2:-m^2\frac{\beta^2}{2}:\phi^3:\label{A1}
\end{equation}

Actually, it is more instructive to keep arbitrary the interaction couplings and consider:
\begin{equation}
\frac{1}{c^2}\partial_t^2\phi=\partial_x^2\phi-m^2c^2\phi-c \frac{v_3}{2}:\phi^2:-\frac{v_4}{3!}:\phi^3:\label{A2}
\end{equation}

Notice that this equation of motion can be thought to be derived from the relativistic Lagrangian
\begin{equation}
\mathcal{L}=\frac{1}{2}\partial_\mu\phi\partial^\mu\phi-\frac{m^2c^2}{2}\phi^2-c\frac{v_3}{3!}:\phi^3:-\frac{v_4}{4!}:\phi^4:\label{A3}
\end{equation}

Splitting now the field in its modes (\ref{LL11}) , from the equation of motion (\ref{A2}) we write the integral equation for the NR modes:
\begin{eqnarray}
\nonumber&&i\psi^\dagger(t_0+\Delta)=i\psi^{\dagger}(t_0)+\int_{t_0}^{t_0+\Delta}dt\;\left[ e^{-i2mc^2(t-t_0)}i\partial_t\psi+\frac{1}{2m}\partial_{x}^2\left(\psi^\dagger+e^{-i2mc^2(t-t_0)}\psi\right)+\right.\\
\nonumber&&-c\frac{v_3 e^{-imc^2(t-t_0)}}{4m\sqrt{2m}}:\left(e^{imc^2(t-t_0)}\psi^\dagger+e^{-imc^2(t-t_0)}\psi\right)^2:\\
&&\left.-\frac{v_4 e^{-imc^2(t-t_0)}}{3!(2m)^2}:\left(e^{imc^2(t-t_0)}\psi^\dagger+e^{-imc^2(t-t_0)}\psi\right)^3:\right]\label{A4}
\end{eqnarray}

From these equations we want to extract, after having taken the NR limit, the $\mathcal{O}(\Delta)$ term and then get from it the NR equation of motion.
Thanks to the power counting argument (\ref{BD36}) we know that in order to do this we need to solve (\ref{A4}) iteratively up to the second order.
Moreover, as we discussed after (\ref{BD36}), at the second iterative solution the only terms that do not give a vanishing contribution are those of the first iterative solution associated with $v_3$.
In this perspective, we compute the first order solution keeping only the important terms:
\begin{eqnarray}
\nonumber&&\left[i\psi^\dagger(t_0+\Delta)\right]_{\text{first order}}=i\psi^\dagger+\frac{\Delta}{2m}\partial_x^2\psi^\dagger-\frac{\Delta v_4}{8m^2}\psi^\dagger\psi^\dagger\psi-\frac{v_3 c^{-1}}{4m\sqrt{2m}}\left[\frac{e^{ic^2m\Delta}-1}{im}\psi^\dagger\psi^\dagger+\right. \\
&&\left.+2\frac{e^{-ic^2m\Delta}-1}{-im}\psi^\dagger\psi+\frac{e^{-3ic^2m\Delta}-1}{-3im}\psi\psi\right]+...
\end{eqnarray}

where the fields on the right are all computed at time $t_0$. Now we should take the above and plug it in (\ref{A4}) to get the second order solution: this is a tedious but simple calculation. At this point let $c\to\infty$ and drop all the vanishing terms (in particular the linear terms in $v_3$ of the first order solution vanish, but the crucial point is to drop them only after we have computed the second order solution).
Keeping only the non vanishing terms of order $\mathcal{O}(\Delta)$, the result of this operation is:
\begin{equation}
\left[i\psi^\dagger(t_0+\Delta)\right]_{\text{second order}}=i\psi^\dagger+\frac{\Delta}{2m}\partial_x^2\psi^\dagger-\frac{\Delta v_4}{8m^2}\psi^\dagger\psi^\dagger\psi+\Delta\frac{5}{3}\frac{v_3^2}{8m^4}\psi^\dagger\psi^\dagger\psi+...
\end{equation}
Because of the power counting (\ref{BD35}) the second order solution is enough to extract the NR limit. Deriving in $\Delta$ we find the NR equation of motion
\begin{equation}
i\partial_t\psi^\dagger=\frac{1}{2m}\partial_x^2\psi^\dagger-\frac{1}{8m^2}\left(v_4-\frac{5}{3}\frac{v_3^2}{m^2}\right)\psi^\dagger\psi^\dagger\psi
\end{equation}
and from this the non relativistic Hamiltonian
\begin{equation}
H=\int dx \; \frac{\partial_x\psi^\dagger\partial_x\psi}{2m}+\frac{1}{16m^2}\left(v_4-\frac{5}{3}\frac{v_3^2}{m^2}\right)\psi^\dagger\psi^\dagger\psi\psi
\end{equation}

Notice that we can give a physical meaning to the coupling $v_4-\frac{5}{3m^2}v_3^2$, since it can be associated with a scattering amplitude.
In particular, it is equal to the two body scattering amplitude at tree level of the Lagrangian (\ref{A3}) with $c=1$ and at zero rapidity, moreover a similar interpretation of the NR coupling is also present in the more complicated NR limit of the Toda theories (Figure \ref{scattering}). This equality can be traced back to the fact that the iterative solution of the equation of motion (\ref{A2}) can be indeed represented through tree level Feynman diagrams.
Of course, plugging in the above the coefficients $v_3$ and $v_4$ in order to match (\ref{A1}) and (\ref{A2}), we get the NR limit of the Bullogh Dodd model (\ref{BD38}).

\section{The $\Gamma$ propagator}
\label{gammaprop}

We compute the $\Gamma$ propagator defined by eq. (\ref{N72}), i.e. 

\begin{equation}
\Gamma^{-1}\left(k^\mu k_\mu\right)=-\frac{N}{2}\int \frac{d^2 q}{(2\pi)^2}\frac{1}{q^\mu q_\mu-m^2c^2+i\epsilon_1}\frac{1}{\left(k^\mu-q^\mu\right)\left(k_\mu-q_\mu\right)-m^2c^2+i\epsilon_2}\,\,\, .
\end{equation}

First of all we perform a Wick rotation going in the Euclidean space $\bar{q}^\mu=(-iq_0,q_1)$, in this perspective we also consider the Euclidean version of the $k$ momentum $\bar{k}^\mu=(-ik_0,k_1)$ 

\begin{eqnarray}
\nonumber&&\int \frac{d^2 q}{(2\pi)^2}\frac{1}{q^\mu q_\mu-m^2c^2+i\epsilon_1}\frac{1}{\left(k^\mu-q^\mu\right)\left(k_\mu-q_\mu\right)-m^2c^2+i\epsilon_2}\to\\
&&\to i\int \frac{d^2 q}{(2\pi)^2}\frac{1}{\left[\bar{q}^\mu \bar{q}_\mu+m^2c^2\right]\left[\left(\bar{k}^\mu-\bar{q}^\mu\right)\left(\bar{k}_\mu-\bar{q}_\mu\right)+m^2c^2\right]}\label{C3}\,\,\, ,
\end{eqnarray}
where the contractions are performed with the euclidean metric
\begin{equation}
\eta^{E}_{\mu\nu}=\begin{pmatrix} \frac{1}{c^2}&& 0 \\ 0 && 1 \end{pmatrix}\label{C4}\,\,\, .
\end{equation}

The next step is to move in adimensional units and shift $\bar{q}_0\to c^2m\bar{q}_0$ and $\bar{q}_1\to cm\bar{q}_1$, we also introduce the adimensional euclidean vector $s^\mu$ in such a way $\bar{k}^\mu=(c^2ms^0,cms^1)$:
\be
\frac{i}{cm^2}\int \frac{d^2 q}{(2\pi)^2}\frac{1}{\left[\textbf{q}^2+1\left]\left[\left(\textbf{s}-\textbf{q}\right)^2+1\right.\right.\right]}\,\,\, .
\ee

After the rescaling, the scalar products are done with the standard euclidean metric and thus we use the standard notation $\textbf{q}$ to indicate vectorial quantities. Using the Feynman's trick 
\begin{equation}
\frac{1}{AB}=\int_0^1dx\;\frac{1}{\left[xA+(1-x)B\right]^2}
\end{equation}
we rewrite the integral as
\be
\frac{i}{cm^2}\int_0^1 dx\;\int \frac{d^2 q}{(2\pi)^2}\frac{1}{\left[\textbf{q}^2-(1-x)2\textbf{s}\textbf{q}+1+(1-x)\textbf{s}^2\right]^2}\,\,\, .
\ee

The last passages amount to shift the momentum $\textbf{q}$ in order to make the integral symmetric under rotations, at this point the integration in $\textbf{q}$ becomes trivial. The $x$ integration can be done as well and we get
\be
\frac{1}{2\pi}\frac{i}{cm^2}\frac{1}{\sqrt{(4+\textbf{s}^2)\textbf{s}^2}}\log\left[\frac{\sqrt{4+\textbf{s}^2}+\sqrt{\textbf{s}^2}}{\sqrt{4+\textbf{s}^2}-\sqrt{\textbf{s}^2}}\right]\,\,\, .
\ee

Tracking back the euclidean rotation and the rescaling we have $\textbf{s}^2 =-\frac{k_\mu k^\mu}{m^2c^2}$ in the standard Minkowski metric, then $\Gamma$ can be finally written as
\begin{equation}
\Gamma\left(k^\mu k_\mu\right)=i4\pi c N^{-1} m^2\sqrt{\left(\frac{k_\mu k^\mu}{m^2c^2}-4\right)\frac{k_\mu k^\mu}{m^2c^2}}\left(\log\left[\frac{\sqrt{4-\frac{k_\mu k^\mu}{m^2c^2}}+\sqrt{-\frac{k_\mu k^\mu}{m^2c^2}}}{\sqrt{4-\frac{k_\mu k^\mu}{m^2c^2}}-\sqrt{-\frac{k_\mu k^\mu}{m^2c^2}}}\right]\right)^{-1}\,\,\, .
\end{equation}

\end{document}